%% file: pra-submission.tex
\documentclass[preprintnumbers,reprint,pra,amssymb,twocolumn,superscriptaddress,longbibliography]{revtex4}

\usepackage{epsfig}
\usepackage{xcolor}
\usepackage{dcolumn}
\usepackage{graphicx}
\usepackage{hyperref}
\usepackage{graphicx, amsmath, amssymb, times}
\usepackage{url}
\usepackage{todonotes}

\usepackage{nicefrac}

\usepackage[utf8]{inputenc}
\DeclareUnicodeCharacter{2212}{--}

\usepackage{pgfplots}
\pgfplotsset{compat=newest}

\usepgfplotslibrary{external}
\DeclareMathAlphabet{\mathpzc}{OT1}{pzc}{m}{it}

\begin{document}


\title{Fulde-Ferrell states in unequally charged Fermi gases}

\author{A. Levent Suba{\c s}{\i}}\email{alsubasi@itu.edu.tr}
\affiliation{Department of Physics, Istanbul Technical University, Maslak 34469, Istanbul, Turkey}
\affiliation{Center for Nonlinear Studies, Los Alamos National Laboratory, Los Alamos, New Mexico 87545, USA}
\author{Nader Ghazanfari} 
\email{nghazanfari@msgsu.edu.tr}
\affiliation{Department of Physics, Mimar Sinan Fine Arts University, Bomonti 34380, Istanbul, Turkey}

\date{\today}

\begin{abstract}
Atoms with different internal states can exhibit different responses to an artificial magnetic field. An atomic gas mixture of two different components can therefore be interpreted as a mixture of two atomic gases carrying different synthetic charges. 
We consider the superfluid state of such unequally charged Fermi gases coupled to a magnetic field via the orbital effect and trapped in a torus geometry.
The orbital coupling to the magnetic field favors an inhomogeneous superfluid state with optimum finite center-of-mass momentum pairing. 
The resulting population-balanced orbital Fulde-Ferrell (FF) state is robust against the magnetic field and does not undergo pair breaking unlike the conventional BCS and Fulde-Ferrell-Larkin-Ovchinnikov type pairing states under the Zeeman effect. 
We contrast the homogeneous and inhomogeneous cases emphasizing the advantages of the unequally charged systems and 
present their momentum distributions. 
We conclude that an unequally charged atomic Fermi gas system orbitally coupled to an artificial magnetic field provides an ideal candidate for experimental realization of the FF state. 
\end{abstract}


\maketitle
\section{Introduction}\label{sec:introduction}
The Fulde-Ferrell-Larkin-Ovchinnikov (FFLO) state \cite{Ferrell1964, Ovchinnikov1964}, in which Cooper pairs have finite center-of-mass (CM) momenta, has attracted interest for decades in the field of superconductivity~\cite{Sarrao2003, Shimahara2007, Norman1993, Kobayashi2012, Brown2016, Uji2015, Wosnitza2018, Lohneysen2013, Crivelli2013, Lortz2017, Gurevich2011, Hulet2006, Ketterle2006, Ketterle2007, Dukelsky2006, Bakhtiari2008, Mueller2010, Simons2008, Hulet2016, Yanase2011, Sheehy2010, Drut2015, Radzihovsky2011, Iskin2013, Jakubczyk2017, Chen2018, Nardulli2004, Song2019}. The existence of this exotic and elusive state has been investigated in heavy-fermion systems~\cite{Sarrao2003, Shimahara2007, Norman1993}, organic superconductors~\cite{Kobayashi2012, Brown2016, Uji2015, Wosnitza2018}, iron-based superconductors~\cite{Lohneysen2013, Crivelli2013, Lortz2017, Gurevich2011}, ultracold atomic gases~\cite{Hulet2006, Ketterle2006, Ketterle2007, Dukelsky2006, Bakhtiari2008, Mueller2010, Simons2008, Yanase2011, Hulet2016, Sheehy2010, Drut2015, Radzihovsky2011, Iskin2013, Jakubczyk2017, Chen2018} and high energy physics~\cite{Nardulli2004}. There is various observational evidence of FFLO state~\cite{Norman1993, Lohneysen2013, Kobayashi2012,  Crivelli2013, Uji2015, Brown2016, Lortz2017, Wosnitza2018}, the evidence being "circumstantial in organic superconductors, scant in the pnictides, and complex in the heavy fermions"~\cite{Agosta2018}. Despite extensive studies on fermionic superfluids in quantum gases, direct observation of the FFLO states in the field of ultracold atoms is yet to be announced. The experimental and theoretical achievements in the field of ultracold quantum gases~\cite{Hulet2006, Ketterle2006, Ketterle2007, Dukelsky2006, Bakhtiari2008, Mueller2010, Simons2008, Hulet2016, Yanase2011} continue to stimulate the studies of the FFLO state in Fermi mixtures.
\par
\begin{figure}[t]
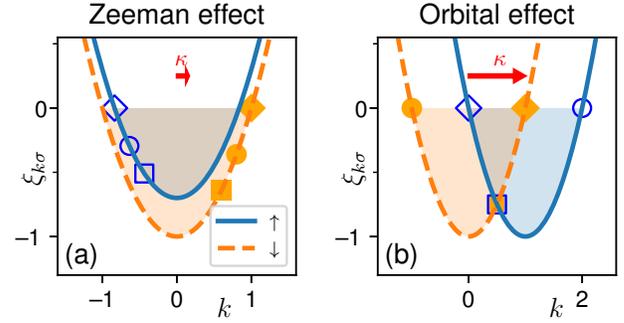

\centering
\resizebox{0.48\textwidth}{!}{%
\InputIfFileExists{epsk.pgf}{}{\huge PGF file missing!}
}
\caption{
\label{fig:dispersion}
Comparison of finite center-of-mass momentum pairing for unequally charged fermions coupled to 
a magnetic field via (a) the Zeeman effect and (b) the orbital effect in the non-interacting limit.
The states forming Cooper pairs are indicated by the same symbols on the dispersion curves of both fermion species.
The red arrows show the center-of-mass momentum $\kappa$.
The orbital effect allows for perfect pairing of degenerate states as shown in (b).
}
\end{figure}

An FFLO state can appear when a population imbalance exists between the two fermion species. 
Such an imbalance causes a mismatch in the Fermi surfaces.
The Fermi surface mismatch in general can arise from various asymmetries between the components~\cite{Simons2008, Mueller2010, Kobayashi2013, Guo2014, Hulet2016, Lebed2018}.
In solid state physics this mismatch can be generated by applying a magnetic field which partially polarizes the system. 
A similar state can be obtained in ultracold atomic gases by introducing an imbalance in the number or mass of different spin species.
With such an asymmetry, involving all possible states in pairing requires non-degenerate pairing
which as a consequence reduces the excitation energy gap.
In such cases, pairing in an FFLO state can favor a finite CM momentum for the Cooper pairs 
involving quantum states that are closer in energy compared to those of zero momentum pairs.
We demonstrate such a scenario for a single CM momentum for all pairs with a generic Zeeman splitting between the components in Fig.~\ref{fig:dispersion}(a).
The partial pairing of states is shown by representative symbols on the non-interacting quadratic dispersion curves. 
These types of non-degenerate Fermi surface mismatch weakens pairing and superfluidity eventually disappears with increasing asymmetry.
\par 
In order to avoid the pair breaking effects of asymmetric pairing, we consider a Fermi surface mismatch with perfect degenerate-state pairing. We propose orbital coupling to the magnetic field and yet allow an imbalance in the electric charge of the fermions forming the Cooper pairs. 
The unequally charged Fermi seas shift differently in momentum space due to the orbital coupling. The shifts are proportional to the magnetic field and respect full pairing of degenerate states provided an optimum CM momentum is chosen (see Fig.~\ref{fig:dispersion}(b)). 
This Fulde-Ferrel (FF) pairing remains robust with respect to changes in both the magnetic field strength and the charge ratio. Considering only the orbital effect due to the magnetic field, such a Fermi surface mismatch does not require any imbalance between particle numbers or particle masses. This ideal pairing can be realized in ultracold atomic Fermi gases coupled to an artificial magnetic field.
\par 
The artificial magnetic fields engineered for neutral ultracold gases make them behave as if they were electrically charged and subject to a magnetic field~\cite{Spielman2009a, Spielman2009b}. The strength of the artificially generated magnetic fields depends on the internal structure of the atoms which determines their coupling to the applied laser field. Therefore, each fermion species can exhibit a different coupling to the magnetic field. Such a system can be interpreted as a mixture of two atomic gases, carrying different synthetic charges~\cite{Oktel2016, Ghazanfari2019}. Using the correspondence between the magnetic field and rotation, this system is analogous to a mixture with component selective rotations. 
\par
The mixture of two unequally charged bosonic or fermionic gases has gathered some attention recently. 
In the presence of a magnetic field, the topological effects in a mixture of two unequally charged condensates~\cite{Babaev2014} and the transfer of angular momentum between two unequally charged bosonic superfluids~\cite{Ghazanfari2019} have been studied. 
The pair breaking process~\cite{Oktel2016} has been investigated for a fermionic superfluid composed of unequally charged Cooper pairs. The realization of unequally charged superfluids is readily possible experimentally and can facilitate the observation of the FF state arising from finite-CM-momentum pairing mentioned above.
\par
The geometry or dimensionality of a system is an important factor in the context of conventional FFLO states. There are different studies stating that these phases are more stable in one dimensional systems than in two or three dimensional systems~\cite{Drummond2007, Torma2008, Bolech2009, Feiguin2010, Ptok2017}. For the orbital FF state, our choice of the ring geometry reflects the advantage of the one dimensional systems. For two- and three-dimensional systems the optimum CM momentum choice for pairing can be much more complex.
\par 
Following the introduction given above the paper is organized as follows: In Sec.~\ref{sec:FF state} we consider the superfluid state of unequally charged interacting Fermi gases under a magnetic field with finite CM momentum pairing. The components interact with a short-range s-wave interaction and are coupled to the magnetic field via the orbital effect. We show that the resulting FF state does not undergo pair breaking, unlike the conventional BCS and FFLO type pairings. Then in Sec.~\ref{sec:stability} we show that the orbital FF state of the population balanced system is robust against phase fluctuations. In Sec.~\ref{sec:experimental-signitures} we discuss how the introduced FF state can be observed via momentum distribution measurements through time-of-flight imaging. We summarize and discuss our results in Sec.~\ref{sec:Conclusion}. A detailed derivation of the superfluid mass density which ensures the local stability of the FF state against the phase fluctuations is given in Appendix.
\section{Fulde-Ferrell state}\label{sec:FF state}
We consider a mixture of two unequally charged Fermi gases, represented by up ($\uparrow$) and down ($\downarrow$) pseudo-spins, with corresponding particle masses $M_\uparrow=M_\downarrow=M$, and synthetic particle charges $q_\uparrow$ and $q_\downarrow$, respectively. The mixture is trapped in a quasi-one-dimensional torus geometry with circumference $L=2\pi R$ ($R$ being the radius of the ring) and cross-section area $S=\pi r^2$, where the tube radius $r \ll R$. The interactions between two gases can be derived from the contact potential model with the coupling constant $g^{1\text{D}} = 4\pi\hbar^2a_s/MS$, so that $g^{1\text{D}}N/L$ gives the mean interaction energy per particle. Here, $a_s$ is the three dimensional \textit{s}-wave scattering length. The system is under a uniform magnetic field (${\bf B}=B_0\hat{e}_z$) along the central axis of the ring, generated by a vector potential ${\cal A}= \frac{B_0\rho}{2}|_{\rho=R}\hat{e}_\theta$ in the symmetric gauge.
\par
The many-body Hamiltonian describing the system of unequally charged Fermi gases under a magnetic field can be written as
\begin{eqnarray}
\begin{aligned}
    \hat{H} &= \sum_\sigma \int d\theta \Psi^\dagger_\sigma(\theta)\xi_\sigma(\theta)\Psi(\theta)\\
     &+ g^{1\text{D}} \int d\theta \Psi^\dagger_\uparrow(\theta)\Psi^\dagger_\downarrow(\theta)\Psi_\downarrow(\theta)\Psi_\uparrow(\theta),
\end{aligned}
\label{eq:MBHam}
\end{eqnarray}
where $\theta$ is the azimuthal angle and $\xi_\sigma(\theta)= (-i\hbar \partial_\theta - q_\sigma{\cal A})^2/2M-\mu_\sigma$ denotes the non-interacting state energies for pseudo-spin $\sigma$ with respect to their chemical potential $\mu_\sigma$.
The reduced Hamiltonian describing this system in the plane wave basis can be written as
\begin{eqnarray}\label{eq:cm-MB-Ham}\nonumber
H &=& \sum_{k} \left[\xi_{k\uparrow} c^{\dagger}_{\frac{\kappa}{2}+k\uparrow} c_{\frac{\kappa}{2}+k\uparrow} + \xi_{-k\downarrow} c^{\dagger}_{\frac{\kappa}{2}-k\downarrow} c_{\frac{\kappa}{2}-k\downarrow}\right] \\ &+& \frac{g}{L}\sum_{\substack{k,k'}}c^{\dagger}_{\frac{\kappa}{2}+k\uparrow}c^{\dagger}_{\frac{\kappa}{2}-k\downarrow}c_{\frac{\kappa}{2}-k'\downarrow}c_{\frac{\kappa}{2}+k'\uparrow}.
\end{eqnarray}
with $\xi_{k\sigma}=\left(\frac{\kappa}{2} + k - Q_\sigma\Phi\right)^2 - \mu_\sigma$ where momenta are referenced from momentum $\kappa/2$ (see Appendix~\ref{app:sufmass} for more details). Note that all the energies are non-dimensionalized via scaling the Hamiltonian. The energy scale is the Fermi energy $E_F=\hbar^2k_F^2/2M$ so that $g=8a_s/r^2k_F^2$ has a length dimension, $Q_\sigma = q_\sigma /2eRk_F$ are the dimensionless synthetic charges, and $\Phi = \pi R^2 B_0/\Phi_0$ is the number of magnetic flux quanta passing through the area enclosed by the torus ($\Phi_0=h/2e$). We consider $N_\uparrow=N_\downarrow=N/2$ with a total density of particles $n=N/L=2k_F/\pi$ in the thermodynamic limit where $k_FR\gg 1$.
\par
The atoms acquire a geometric phase $\gamma(\theta) = Q_\sigma\Phi\theta$ due to the magnetic field. Each species gains a phase according to its electric charge so that a finite CM momentum $\kappa=(Q_\uparrow+Q_\downarrow)\Phi = 2Q_+\Phi$ (see Fig.~\ref{fig:dispersion}(b)) is favoured by the Cooper pairs in the superfluid state. (Here and throughout the momentum referred to is the canonical momentum.) 
We 
adopt a mean-field approximation and introduce the gap function
\begin{equation}\label{eq:pairing-energy}
\Delta = \frac{g}{L}\sum_k\langle c_{\frac{\kappa}{2}-k\downarrow}c_{\frac{\kappa}{2}+k\uparrow} \rangle\, e^{i\kappa\theta}. 
\end{equation}
Note that with this choice of the CM momentum, the paired states are time-reversed partners in the CM frame. As a result, the unequally charged system introduced above does not suffer from the pair breaking effects of partial and non-degenerate pairing.
\begin{figure}[t]
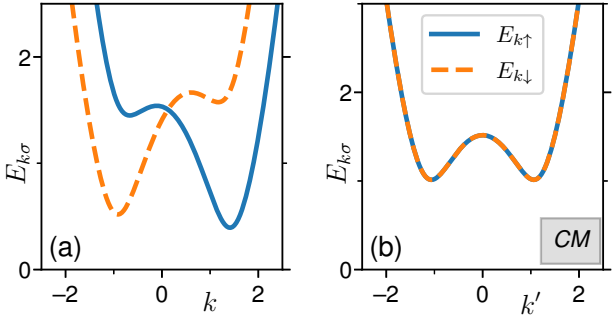

\centering
\resizebox{0.48\textwidth}{!}{%
\InputIfFileExists{Ek.pgf}{}{\huge PGF file missing!}
}
\caption{
\label{fig:int_dispersion}
Excitation spectra for the unequally charged system in a magnetic field for (a) BCS and (b) FFLO states in the superfluid phase. 
The non-degenerate state pairing for the BCS state is reflected in the asymmetry with reduced excitation energy gap in (a).
The degenerate-state pairing for the FFLO has a symmetric excitation spectrum in the CM frame shown in (b) which results in a robust energy gap.
}
\end{figure}
The mean-field approximation leads us to a quadratic Hamiltonian of the form
\begin{eqnarray}\label{eq:Matrix-Ham}\nonumber
H &=& \sum_{k} \psi^{\dagger}_k \left( 
\begin{array}{cc}
\xi_{k\uparrow} & \Delta \\
\Delta^* & -\xi_{-k\downarrow}\\
\end{array} \right) \psi_k + \sum_k \xi_{-k\downarrow} -\frac{\vert \Delta\vert^2 L}{g}
\end{eqnarray} 
where $\psi_k^\dagger = \left(c_{\frac{\kappa}{2}+k\uparrow}^\dagger ~ c_{\frac{\kappa}{2}-k\downarrow}\right)$. The Bogoliubov transformation 
\begin{eqnarray}\label{eq:Bog-transformation}
\begin{aligned}
c_{\frac{\kappa}{2}+k\uparrow}^\dagger = u_k \gamma_{k\uparrow}^\dagger - v_k\gamma_{-k\downarrow} \\
c_{\frac{\kappa}{2}-k\downarrow}= v_k\gamma_{k\uparrow}^\dagger + u_k\gamma_{-k\downarrow}
\end{aligned}
\end{eqnarray}
with $|u_k|^2+|v_k|^2=1$, puts the Hamiltonian in the diagonalized form
\begin{equation}
H = \sum_{k,\sigma} E_{k\sigma} \gamma_{k\sigma}^\dagger\gamma_{k\sigma} + \sum_k \left(\xi_{-k\downarrow}-E_{k\downarrow}\right)- \frac{\vert \Delta\vert^2 L}{g}.
\end{equation}
The operators $\gamma_{k\sigma}^\dagger$ create quasiparticles with excitation energies  
\begin{equation}\label{eq:Bogoliubov-excitations}
E_{k\sigma}= \mathrm{sign}(\sigma) \xi_{k-} + \sqrt{\xi_{k+}^2 + \vert\Delta\vert^2}
\end{equation}
where $\text{sgn}(\uparrow,\downarrow)=(+,-)$ and $\xi_{k\pm}=\left(\xi_{k\uparrow}\pm\xi_{-k\downarrow}\right)/2$ 
introduces the average and difference of the two pairing single particle states' energies. The so-called gap equation
\begin{equation}\label{eq:gap-equation}
-\frac{1}{g}=\frac{1}{L} \sum_k \frac{1-f_{k\uparrow}-f_{k\downarrow}}{2\sqrt{\xi_{k+}^2 + \vert\Delta\vert^2}}
\end{equation}
is obtained from the self-consistency condition of the mean-field approximation using Eq.~\eqref{eq:pairing-energy}.
For a thermal ensemble, the momentum distributions of the particles are given by
\begin{eqnarray}
\!\!\!\!\!\!\! n_{k\sigma} \!=\! \langle c_{k+\frac{\kappa}{2}\sigma}^\dag c_{k+\frac{\kappa}{2}\sigma}\rangle \!=\! f_{k\sigma} + \vert v_k \vert^2 \left[ 1- f_{k\uparrow}- f_{k\downarrow} \right],
\end{eqnarray} 
where 
\begin{equation}
\vert v_k\vert^2 = \frac{1}{2}\left[ 1- \frac{\xi_{k+}}{\sqrt{\xi_{k+}^2 + \vert\Delta\vert^2}} \right]
\end{equation}
and $f_{k\sigma}= 1/[e^{\beta E_{k\sigma}}+1]$ is the Fermi-Dirac distribution function with inverse temperature $\beta = 1/k_BT$. 
The equation for the total number of particles $N = \sum_{k\sigma} n_{k\sigma}$ determines the average chemical potential.
We consider only the zero temperature ground states.

\begin{figure}[t]
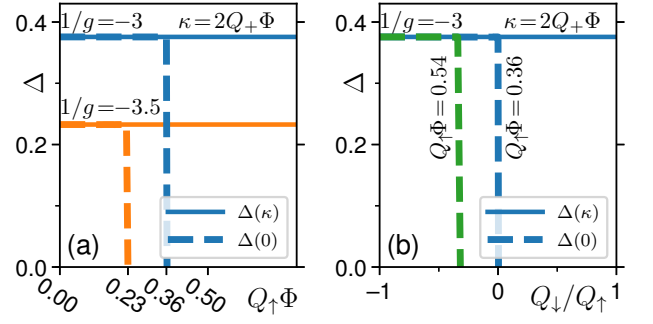

\centering
\resizebox{0.48\textwidth}{!}{%
\InputIfFileExists{gap_Q_B.pgf}{}{\huge PGF file missing!}
}
\caption{
\label{fig:gap}
Gap function $\Delta$ for the FF (solid and dashed lines) and the BCS (dotted and dash-dotted lines) states 
as a function of (a) the magnetic flux $\Phi$ for the charged-uncharged mixture and (b) the charge ratio $Q_\downarrow/Q_\uparrow$. The finite center-of-mass momentum pairing for the FF state is robust against the magnetic field 
whereas the zero center-of-mass momentum BCS states break at critical $\Phi$ and $Q_\downarrow/Q_\uparrow$ values, respectively. Note that $\kappa=0$ in the limit $Q_\uparrow=-Q_\downarrow$.}
\end{figure}
\par
We now contrast the quasiparticle excitation energies of the homogeneous BCS ($\kappa=0$) and inhomogeneous FF ($\kappa\neq 0$) pairing states for a charged-uncharged system in Fig.~\ref{fig:int_dispersion}. For the BCS pairing, the non-zero energy difference of the two pairing states, $\xi_{k-}\neq 0$, leads to an asymmetry in the excitation energies $E_{k\sigma}$ as shown in Fig.~\ref{fig:int_dispersion}(a). The difference $\xi_{k-}$ reduces the energy gap and eventually leads to pair breaking at the so-called Clogston-Chandrasekhar limit when either one of $E_{k\sigma}=0$. Similarly, the excitation spectrum for conventional FF states has an additional linear dependence on momenta~\cite{Radzihovsky2011}. In contrast, the FF pairing has $\xi_{k-}=0$ and the excitation spectra are symmetric in the CM frame as shown in Fig.~\ref{fig:int_dispersion}(b). The magnitude of the ensuing energy gap remains unchanged and equal to its zero field limit.
\par
The inhomogeneous phase considered here is acquired through the orbital coupling to the magnetic field. The pairing involves pairs with kinetic momenta equal in magnitude and opposite in direction in the dispersion relations and  results in a gapped excitation spectrum. In fact, for the FF pairing with optimum CM momentum, we note that the self-consistent equations of systems with the same interactions but different magnetic fields and synthetic charges reduce to identical equations in their CM frame. Because of this invariance the energy gap is robust against the applied magnetic field. 
\par
We demonstrate the robustness of the FF ($\kappa\neq 0$) pairing with respect to that of the BCS ($\kappa=0$) pairing by comparing the gap $\Delta$ in both cases as a function of the system parameters. 
The FF pairing gives a constant energy gap as a function of the magnetic field and charge ratio as shown by the solid horizontal lines in Fig.~\ref{fig:gap}. 
In contrast, a zero CM momentum BCS pairing breaks at a critical magnetic field or critical charge ratio at the Clogston limit. 
This is due to the closing of the energy gap ($E_{k\sigma}=0$), which happens at higher magnetic field value [Fig.~\ref{fig:gap}(a)] 
or a greater charge ratio [Fig.~\ref{fig:gap}(b)] for stronger interactions. 
For example, the BCS state for the charged-uncharged system ($Q_\downarrow=0$) with $1/g=-3$ breaks at $\Phi \sim 0.36$ 
in Fig.~\ref{fig:gap}(a) as indicated by the corresponding vertical line. 
Similarly, as $Q_\downarrow/Q_\uparrow$ ratio is changed at constant $\Phi\sim 0.36$, 
the BCS state is destroyed correspondingly at $Q_\downarrow/Q_\uparrow=0$ as shown in Fig.~\ref{fig:gap}(b). We emphasize that the optimum $\kappa=2Q_+\Phi$ value  also changes with the variation of the parameters in these plots. A state with a fixed value 
of $\kappa$ undergoes a first order transition having the same fate as the state in the $\kappa=0$ limit shown.
In the $Q_\downarrow=-Q_\uparrow$ case of oppositely charged components the homogeneous superfluid state with $\kappa=0$ is favored. Said another way, the equally charged Fermi gas in an anti-parallel magnetic field requires a population imbalance for FFLO states as was recently studied in two-dimensions~\cite{Nishida2019}.
\section{Stability of the Fulde-Ferrell state}\label{sec:stability}
In this section, we analyze the stability of the FF state with respect to phase fluctuations. 
The low-energy phase variations of the gap function can be written as
\begin{equation}
    \Delta(\theta) = \Delta_0 e^{i\kappa\theta+i\varphi(\theta)}
\end{equation}
where $\varphi(\theta)$ denotes the phase fluctuation about the FF order parameter.
The stability requires that fluctuations bring positive contribution to the free energy. Assuming a constant dimensionless superfluid velocity 
$v_s =\frac{1}{2} \frac{d \varphi}{d\theta}$ for the low-energy fluctuations we expand the free energy $F$ about its mean-field value in terms of $v_s$.  
We identify the superfluid mass density~\cite{Samokhin2010, Melo2006, Tanatar2010} 
\begin{equation}\label{eq:sufmass}
\rho_s =
\sum_{k\sigma}  \left[ n_{k\sigma} - \frac{1}{4} \frac{\left(k-Q_- \Phi\right)^2}{\cosh(\frac{\beta E_k}{2})}\right],
\end{equation}
with $Q_-=(Q_\uparrow-Q_\downarrow)/2$, so that $\delta F = \rho_s v_s^2$ in units of $E_F$ (see the Appendix \ref{app:sufmass} for a derivation). 
For a gapped system at zero temperature the second term in $\rho_s$ vanishes so that $\rho_s$ equals the total mass density and 
gives positive phase stiffness. On the other hand, for conventional FFLO states with population or mass imbalance the excitation energies can become gapless which can decrease $\rho_s$ and 
eventually drive the system to instability. 
We conclude that the orbital FF state of population balanced charged system is robust against phase fluctuations.
\section{Experimental Signatures}\label{sec:experimental-signitures}

\begin{figure}[!t]
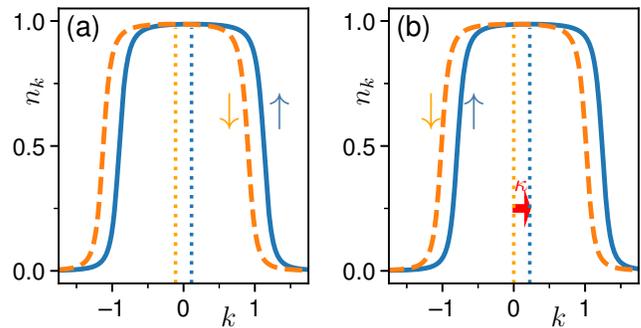

\centering
\resizebox{0.5\textwidth}{!}{%
\InputIfFileExists{mom_dist.pgf}{}{\huge PGF file missing!}
}
\caption{
\label{fig:mom_dist}
Momentum distributions for the charged-uncharged system in a magnetic field for $1/g=-3$ and $\Phi=0.23$. (a) In the BCS state, $n_{k\uparrow}=n_{-k\downarrow}$ such that the total momentum of the system is zero.
(b) In the FF state, $n_{k+\frac{\kappa}{2}\uparrow}=n_{-k+\frac{\kappa}{2}\downarrow}$ and the total momentum is finite with Cooper pairs having center-of-mass momentum $\kappa$. Each component's momentum is indicated by the corresponding vertical dotted line.}
\end{figure}

An observation of the phase of the pairing potential will provide direct evidence for the FF state.
For the unequally charged systems, the effect can be realized 
via momentum distributions, 
which are routinely measured through time-of-flight imaging in ultracold atom experiments.
Figure~\ref{fig:mom_dist} shows the momentum distributions of the charged-uncharged systems in a magnetic field for the weakly interacting limit. 
In the homogeneous BCS state, even though the uncharged $\downarrow$ component is not coupled to the magnetic field,
the pairing enforces occupation of $\pm k$ momentum states so that the total momentum of the system is zero. 
However, due to the magnetic field, the momentum distributions of $\uparrow$ and $\downarrow$ species get shifted in opposite directions such that each component acquires a finite momentum as shown by the vertical dotted lines in Fig.~\ref{fig:mom_dist}(a).
\par
In the inhomogeneous FF state, the momentum distributions in the CM frame are symmetric similar to the zero-field distributions. However, in the laboratory frame, the uncharged component has zero momentum whereas the charged component has finite momentum as seen in  Fig.~\ref{fig:mom_dist}(b). The shift in momentum of the charged component distribution is proportional to the magnetic field strength, whereas the uncharged component distribution is unaffected  and remains symmetric about $k=0$. 
The signature of these states will show as an asymmetry between the distributions of the two components \footnote{For the mechanical momentum distributions, the symmetry properties of the homogeneous and inhomogeneous states will be reversed.}.

The energy gap for the results shown in Figs.~\ref{fig:gap} and \ref{fig:mom_dist} corresponds to a fraction of Fermi temperature $T_F=E_F/k_B$. This is in agreement with the critical temperatures observed in the experiments. On the other hand, we estimate that the model considered here would require torus radii larger than those currently available in experiments with ring geometries~\cite{Hadzibabic2013, Campbell2014, Boshier2014}. The limitation mainly depends on the three-dimensional scattering assumption which requires $R>r>a_s$. A quasi-one-dimensional condensate with larger scattering length can provide stronger interactions which can make the observation of the pairing effects easier.
\section{Conclusion}\label{sec:Conclusion}
We studied the inhomogeneous superfluid FF state with finite center-of-mass momentum pairing for the unequally charged Fermi gases in the presence of a uniform magnetic field. 
The Cooper pairs acquire a geometric phase in the magnetic field.
Their analogous finite center-of-mass momentum makes ideal pairing possible.
Differently from conventional FFLO states, the FF states for orbital coupling fully pair up degenerate states despite the Fermi surface shifts in the momentum space. 
The symmetry of the inhomogeneous FF state in this scenario is manifest in the CM frame. 
These states are robust compared to states with homogeneous pairing which break at a critical magnetic field or critical charge ratio.

Cold atom systems with artificial magnetic fields provide an experimental environment to realize an unequally charged Fermi system.
The momentum distributions, which can be measured in the experiments, will show an asymmetry between the unequally charged components in the FF state with the total angular momentum of the system being non-zero and proportional to the magnetic field. 
\acknowledgements
This work is supported by MSGS\"{U} BAP under Project No. 2019-26. A.L.S. acknowledges the scholarship program by T\"{U}B\.{I}TAK B\.IDEB 2219  
and the hospitality of the Center for Non-linear Studies (CNLS) 
supported by the Laboratory Directed Research and Development program of Los Alamos National Laboratory (LANL) under Project No. 20180045DR. 
The authors acknowledge useful discussions with R. Onur Umucal{\i}lar.
\appendix
\section{Superfluid mass density}\label{app:sufmass}

For the many-body Hamiltonian in Eq.~\eqref{eq:MBHam} we define the mean-field order parameter $\Delta(\theta)=g^{1D}\langle \Psi_\downarrow(\theta)\Psi_\uparrow(\theta) \rangle$ so that the mean-field Hamiltonian can be written in real-space as
\begin{eqnarray}
\!\!\!\!\!\hat{H}_{MF} &=& \sum_\sigma \int d\theta\, \Psi^\dagger_\sigma(\theta)\xi_\sigma(\theta)\Psi_\sigma(\theta) - \int d\theta \frac{|\Delta(\theta)|^2}{g^{1D}} \\ \nonumber
&+&  \int d\theta \left[ \Delta^*(\theta)\Psi_\downarrow(\theta) \Psi_\uparrow(\theta) + \Delta(\theta)\Psi^\dagger_\uparrow(\theta)\Psi^\dagger_\downarrow(\theta)\right]
\end{eqnarray}
The low-energy phase fluctuations of the order parameter in the FF state can be described by $\Delta(\theta) = \Delta_0 e^{i\kappa\theta+i\varphi(\theta)}$ where $\varphi(\theta)$ varies slowly on the scale of the FF momentum $\kappa$~\cite{Samokhin2010}. 
In order to remove the phase fluctuations from the off-diagonal term we apply a unitary transformation $\tilde{\Psi}_\sigma(\theta) = e^{-i\varphi/2}\Psi_\sigma(\theta)$ so that $\xi(\theta)\rightarrow \Tilde{\xi}(\theta)=e^{-i\varphi/2}\xi(\theta)e^{i\varphi/2}=(-i\hbar\frac{ d}{d\theta} + \frac{\hbar}{2}\frac{d\varphi}{d\theta}-q{\cal A})^2/2M$ which gives
\begin{eqnarray}
\hat{\tilde{H}}_{MF} &=& \sum_\sigma \int d\theta\, \tilde{\Psi}^\dagger_\sigma(\theta)\tilde{\xi}_\sigma(\theta)\tilde{\Psi}_\sigma(\theta) 
- \int d\theta \frac{|\Delta(\theta)|^2}{g^{1D}} \\ \nonumber
&+& \!\! \int \!\! d\theta\! \left[\Delta_0 e^{-i\kappa\theta}\tilde{\Psi}_\downarrow(\theta) \tilde{\Psi}_\uparrow(\theta) 
+ \Delta_0e^{i\kappa\theta}\tilde{\Psi}^\dagger_\uparrow(\theta)\tilde{\Psi}^\dagger_\downarrow(\theta)\right].
\end{eqnarray}
We identify the fluctuation terms as
\begin{eqnarray}\nonumber
\!\!\delta\hat{H} &=&   \sum_\sigma \! \!\int\!\! d\theta\, \tilde{\Psi}^\dagger_\sigma(\theta) \left(\!\!-i\hbar\frac{d}{d\theta} - q_\sigma{\cal A} \right)\!\!\left(\frac{\hbar}{2M}\frac{d\varphi}{d\theta}\right) \tilde{\Psi}_\sigma(\theta)\\
&+&   \sum_\sigma \!  \int \!  d\theta\, \tilde{\Psi}^\dagger_\sigma(\theta) 
\frac{1}{2M}\left(\frac{\hbar}{2}\frac{d\varphi}{d\theta}\right)^2  \tilde{\Psi}_\sigma(\theta) \, .
\end{eqnarray}
For slow variations of the phase we assume $v_s = \frac{1}{2}\frac{d\varphi}{d\theta}$ to be constant and 
we expand the field operators in the plane-wave basis
\begin{eqnarray}
 \begin{aligned}
    &\tilde{\Psi}_\uparrow (\theta) = \frac{1}{\sqrt{L}}\sum_k e^{i(\kappa/2+k)\theta}c_{\frac{\kappa}{2}+k\uparrow} \\ 
    &\tilde{\Psi}_\downarrow (\theta) = \frac{1}{\sqrt{L}} \sum_k e^{i(\kappa/2-k)\theta}c_{\frac{\kappa}{2}-k\downarrow}
\end{aligned}
\end{eqnarray}
so that in the units of the Fermi energy we get
\begin{eqnarray}\label{eq:cm-MB-Ham3}\nonumber
\hat{H}_{MF} &=& \sum_{k} \left[\xi_{k\uparrow} c^{\dagger}_{\frac{\kappa}{2}+k\uparrow} c_{\frac{\kappa}{2}+k\uparrow} 
+ \xi_{-k\downarrow} c^{\dagger}_{\frac{\kappa}{2}-k\downarrow} c_{\frac{\kappa}{2}-k\downarrow}\right] \\ 
&+& \sum_{\substack{k}}\Delta_0 \left( c^{\dagger}_{\frac{\kappa}{2}+k\uparrow}c^{\dagger}_{\frac{\kappa}{2}-k\downarrow}
+ c_{\frac{\kappa}{2}-k\downarrow}c_{\frac{\kappa}{2}+k\uparrow}\right) \\ \nonumber
\end{eqnarray}
where $\xi_{k\sigma}=\left[\frac{\kappa}{2} + k - Q_\sigma\Phi\right]^2 - \mu_\sigma$ and 
\begin{eqnarray}
\delta\hat{H} = \sum_k \left[ \varepsilon_{k\uparrow}c^{\dagger}_{\frac{\kappa}{2}+k\uparrow} c_{\frac{\kappa}{2}+k\uparrow}
+ \varepsilon_{-k\downarrow} c^{\dagger}_{\frac{\kappa}{2}-k\downarrow} c_{\frac{\kappa}{2}-k\downarrow} \right]
\end{eqnarray}
with $\varepsilon_{k\sigma}= 2\left(\frac{\kappa}{2}+k - Q_\sigma \Phi\right)v_s+v_s^2$.


Terms linear in $v_s$ vanish, then 
\begin{equation}
    \langle\delta\hat{H}\rangle = \sum_{k\sigma} n_{k\sigma} v_s^2 \, .    
\end{equation}
The change in the free energy is $\delta F =\delta H-T\delta S$. Treating the quasiparticle excitations as independent particles, the entropy can be written as~\cite{Lifshits1980, Legget1998, Rothen2004}
\begin{equation}
    S=-k_B\sum_{k\sigma} \left[ f_{k\sigma}\ln f_{k\sigma} + \left(1-f_{k\sigma}\right)\ln (1-f_{k\sigma})\right],
\end{equation}
so that
\begin{eqnarray}
\begin{aligned}
\delta S &= \sum_{k\sigma} \frac{\partial S}{\partial f_{k\sigma}}\frac{\partial f_{k\sigma}}{\partial E_{k\sigma}}\delta E_{k\sigma}\\
&= k_B \beta^2 \sum_{k\sigma}\frac{E_{k\sigma}e^{\beta E_{k\sigma}}}{\left(e^{\beta E_{k\sigma}}+1\right)^2}\delta E_{k\sigma}.
\end{aligned}
\end{eqnarray}
We introduce $Q_\pm=(Q_\uparrow \pm Q_\downarrow)/2$ and expand the excitation energies as   
\begin{eqnarray}
\!\!\!\!\! E_{k\sigma} (\kappa \!+\! 2v_s)\!\! =\!\! E_{k\sigma} (\kappa) &\!+\!& \frac{\partial E_{k\sigma}}{\partial \kappa} 2v_s 
				\!+\!\frac{1}{2}\frac{\partial^2 E_{k\sigma}}{\partial \kappa^2} 4v_s^2,
\end{eqnarray}
where $E_{k\sigma}$ are given in Eq.~\eqref{eq:Bogoliubov-excitations}. For $\kappa=2Q_+\Phi$, the free energy has an extremum, the first order terms in $v_s$ vanish, and the the quadratic terms become
\begin{equation}
    E_{k\sigma}\delta E_{k\sigma}= \left(k-Q_- \Phi\right)^2 v_s^2.
\end{equation}
The contribution of the phase fluctuations to the free energy can finally be written as
\begin{equation}
    \delta F = \sum_{k\sigma}  \left[ n_{k\sigma} - \frac{\beta}{4} \frac{\left(k-Q_- \Phi\right)^2}{\cosh(\frac{\beta E_k}{2})}\right]v_s^2
\end{equation}
Writing $\delta F =\rho_s v_s^2$, the superfluid mass density $\rho_s$ \cite{Melo2006, Tanatar2010} is given in Eq.~\eqref{eq:sufmass}.

\bibliography{BCS}

\end{document}

%% file: Ek.pgf
\begingroup%
\makeatletter%
\begin{pgfpicture}%
\pgfpathrectangle{\pgfpointorigin}{\pgfqpoint{3.853429in}{2.141145in}}%
\pgfusepath{use as bounding box, clip}%
\begin{pgfscope}%
\pgfsetbuttcap%
\pgfsetmiterjoin%
\definecolor{currentfill}{rgb}{1.000000,1.000000,1.000000}%
\pgfsetfillcolor{currentfill}%
\pgfsetlinewidth{0.000000pt}%
\definecolor{currentstroke}{rgb}{1.000000,1.000000,1.000000}%
\pgfsetstrokecolor{currentstroke}%
\pgfsetdash{}{0pt}%
\pgfpathmoveto{\pgfqpoint{0.000000in}{0.000000in}}%
\pgfpathlineto{\pgfqpoint{3.853429in}{0.000000in}}%
\pgfpathlineto{\pgfqpoint{3.853429in}{2.141145in}}%
\pgfpathlineto{\pgfqpoint{0.000000in}{2.141145in}}%
\pgfpathclose%
\pgfusepath{fill}%
\end{pgfscope}%
\begin{pgfscope}%
\pgfsetbuttcap%
\pgfsetmiterjoin%
\definecolor{currentfill}{rgb}{1.000000,1.000000,1.000000}%
\pgfsetfillcolor{currentfill}%
\pgfsetlinewidth{0.000000pt}%
\definecolor{currentstroke}{rgb}{0.000000,0.000000,0.000000}%
\pgfsetstrokecolor{currentstroke}%
\pgfsetstrokeopacity{0.000000}%
\pgfsetdash{}{0pt}%
\pgfpathmoveto{\pgfqpoint{0.333606in}{0.395708in}}%
\pgfpathlineto{\pgfqpoint{1.779818in}{0.395708in}}%
\pgfpathlineto{\pgfqpoint{1.779818in}{1.992534in}}%
\pgfpathlineto{\pgfqpoint{0.333606in}{1.992534in}}%
\pgfpathclose%
\pgfusepath{fill}%
\end{pgfscope}%
\begin{pgfscope}%
\pgfsetbuttcap%
\pgfsetroundjoin%
\definecolor{currentfill}{rgb}{0.000000,0.000000,0.000000}%
\pgfsetfillcolor{currentfill}%
\pgfsetlinewidth{0.803000pt}%
\definecolor{currentstroke}{rgb}{0.000000,0.000000,0.000000}%
\pgfsetstrokecolor{currentstroke}%
\pgfsetdash{}{0pt}%
\pgfsys@defobject{currentmarker}{\pgfqpoint{0.000000in}{-0.048611in}}{\pgfqpoint{0.000000in}{0.000000in}}{%
\pgfpathmoveto{\pgfqpoint{0.000000in}{0.000000in}}%
\pgfpathlineto{\pgfqpoint{0.000000in}{-0.048611in}}%
\pgfusepath{stroke,fill}%
}%
\begin{pgfscope}%
\pgfsys@transformshift{0.478227in}{0.395708in}%
\pgfsys@useobject{currentmarker}{}%
\end{pgfscope}%
\end{pgfscope}%
\begin{pgfscope}%
\definecolor{textcolor}{rgb}{0.000000,0.000000,0.000000}%
\pgfsetstrokecolor{textcolor}%
\pgfsetfillcolor{textcolor}%
\pgftext[x=0.478227in,y=0.298486in,,top]{\color{textcolor}\sffamily\fontsize{10.000000}{12.000000}\selectfont −2}%
\end{pgfscope}%
\begin{pgfscope}%
\pgfsetbuttcap%
\pgfsetroundjoin%
\definecolor{currentfill}{rgb}{0.000000,0.000000,0.000000}%
\pgfsetfillcolor{currentfill}%
\pgfsetlinewidth{0.803000pt}%
\definecolor{currentstroke}{rgb}{0.000000,0.000000,0.000000}%
\pgfsetstrokecolor{currentstroke}%
\pgfsetdash{}{0pt}%
\pgfsys@defobject{currentmarker}{\pgfqpoint{0.000000in}{-0.048611in}}{\pgfqpoint{0.000000in}{0.000000in}}{%
\pgfpathmoveto{\pgfqpoint{0.000000in}{0.000000in}}%
\pgfpathlineto{\pgfqpoint{0.000000in}{-0.048611in}}%
\pgfusepath{stroke,fill}%
}%
\begin{pgfscope}%
\pgfsys@transformshift{1.056712in}{0.395708in}%
\pgfsys@useobject{currentmarker}{}%
\end{pgfscope}%
\end{pgfscope}%
\begin{pgfscope}%
\definecolor{textcolor}{rgb}{0.000000,0.000000,0.000000}%
\pgfsetstrokecolor{textcolor}%
\pgfsetfillcolor{textcolor}%
\pgftext[x=1.056712in,y=0.298486in,,top]{\color{textcolor}\sffamily\fontsize{10.000000}{12.000000}\selectfont 0}%
\end{pgfscope}%
\begin{pgfscope}%
\pgfsetbuttcap%
\pgfsetroundjoin%
\definecolor{currentfill}{rgb}{0.000000,0.000000,0.000000}%
\pgfsetfillcolor{currentfill}%
\pgfsetlinewidth{0.803000pt}%
\definecolor{currentstroke}{rgb}{0.000000,0.000000,0.000000}%
\pgfsetstrokecolor{currentstroke}%
\pgfsetdash{}{0pt}%
\pgfsys@defobject{currentmarker}{\pgfqpoint{0.000000in}{-0.048611in}}{\pgfqpoint{0.000000in}{0.000000in}}{%
\pgfpathmoveto{\pgfqpoint{0.000000in}{0.000000in}}%
\pgfpathlineto{\pgfqpoint{0.000000in}{-0.048611in}}%
\pgfusepath{stroke,fill}%
}%
\begin{pgfscope}%
\pgfsys@transformshift{1.635197in}{0.395708in}%
\pgfsys@useobject{currentmarker}{}%
\end{pgfscope}%
\end{pgfscope}%
\begin{pgfscope}%
\definecolor{textcolor}{rgb}{0.000000,0.000000,0.000000}%
\pgfsetstrokecolor{textcolor}%
\pgfsetfillcolor{textcolor}%
\pgftext[x=1.635197in,y=0.298486in,,top]{\color{textcolor}\sffamily\fontsize{10.000000}{12.000000}\selectfont 2}%
\end{pgfscope}%
\begin{pgfscope}%
\pgfsetbuttcap%
\pgfsetroundjoin%
\definecolor{currentfill}{rgb}{0.000000,0.000000,0.000000}%
\pgfsetfillcolor{currentfill}%
\pgfsetlinewidth{0.602250pt}%
\definecolor{currentstroke}{rgb}{0.000000,0.000000,0.000000}%
\pgfsetstrokecolor{currentstroke}%
\pgfsetdash{}{0pt}%
\pgfsys@defobject{currentmarker}{\pgfqpoint{0.000000in}{-0.027778in}}{\pgfqpoint{0.000000in}{0.000000in}}{%
\pgfpathmoveto{\pgfqpoint{0.000000in}{0.000000in}}%
\pgfpathlineto{\pgfqpoint{0.000000in}{-0.027778in}}%
\pgfusepath{stroke,fill}%
}%
\begin{pgfscope}%
\pgfsys@transformshift{0.767470in}{0.395708in}%
\pgfsys@useobject{currentmarker}{}%
\end{pgfscope}%
\end{pgfscope}%
\begin{pgfscope}%
\pgfsetbuttcap%
\pgfsetroundjoin%
\definecolor{currentfill}{rgb}{0.000000,0.000000,0.000000}%
\pgfsetfillcolor{currentfill}%
\pgfsetlinewidth{0.602250pt}%
\definecolor{currentstroke}{rgb}{0.000000,0.000000,0.000000}%
\pgfsetstrokecolor{currentstroke}%
\pgfsetdash{}{0pt}%
\pgfsys@defobject{currentmarker}{\pgfqpoint{0.000000in}{-0.027778in}}{\pgfqpoint{0.000000in}{0.000000in}}{%
\pgfpathmoveto{\pgfqpoint{0.000000in}{0.000000in}}%
\pgfpathlineto{\pgfqpoint{0.000000in}{-0.027778in}}%
\pgfusepath{stroke,fill}%
}%
\begin{pgfscope}%
\pgfsys@transformshift{1.345955in}{0.395708in}%
\pgfsys@useobject{currentmarker}{}%
\end{pgfscope}%
\end{pgfscope}%
\begin{pgfscope}%
\definecolor{textcolor}{rgb}{0.000000,0.000000,0.000000}%
\pgfsetstrokecolor{textcolor}%
\pgfsetfillcolor{textcolor}%
\pgftext[x=1.345955in,y=0.261295in,,top]{\color{textcolor}\sffamily\fontsize{12.000000}{14.400000}\selectfont \(\displaystyle k\)}%
\end{pgfscope}%
\begin{pgfscope}%
\pgfsetbuttcap%
\pgfsetroundjoin%
\definecolor{currentfill}{rgb}{0.000000,0.000000,0.000000}%
\pgfsetfillcolor{currentfill}%
\pgfsetlinewidth{0.803000pt}%
\definecolor{currentstroke}{rgb}{0.000000,0.000000,0.000000}%
\pgfsetstrokecolor{currentstroke}%
\pgfsetdash{}{0pt}%
\pgfsys@defobject{currentmarker}{\pgfqpoint{-0.048611in}{0.000000in}}{\pgfqpoint{0.000000in}{0.000000in}}{%
\pgfpathmoveto{\pgfqpoint{0.000000in}{0.000000in}}%
\pgfpathlineto{\pgfqpoint{-0.048611in}{0.000000in}}%
\pgfusepath{stroke,fill}%
}%
\begin{pgfscope}%
\pgfsys@transformshift{0.333606in}{0.395708in}%
\pgfsys@useobject{currentmarker}{}%
\end{pgfscope}%
\end{pgfscope}%
\begin{pgfscope}%
\pgfsetbuttcap%
\pgfsetroundjoin%
\definecolor{currentfill}{rgb}{0.000000,0.000000,0.000000}%
\pgfsetfillcolor{currentfill}%
\pgfsetlinewidth{0.803000pt}%
\definecolor{currentstroke}{rgb}{0.000000,0.000000,0.000000}%
\pgfsetstrokecolor{currentstroke}%
\pgfsetdash{}{0pt}%
\pgfsys@defobject{currentmarker}{\pgfqpoint{0.000000in}{0.000000in}}{\pgfqpoint{0.048611in}{0.000000in}}{%
\pgfpathmoveto{\pgfqpoint{0.000000in}{0.000000in}}%
\pgfpathlineto{\pgfqpoint{0.048611in}{0.000000in}}%
\pgfusepath{stroke,fill}%
}%
\begin{pgfscope}%
\pgfsys@transformshift{1.779818in}{0.395708in}%
\pgfsys@useobject{currentmarker}{}%
\end{pgfscope}%
\end{pgfscope}%
\begin{pgfscope}%
\definecolor{textcolor}{rgb}{0.000000,0.000000,0.000000}%
\pgfsetstrokecolor{textcolor}%
\pgfsetfillcolor{textcolor}%
\pgftext[x=0.196630in,y=0.342947in,left,base]{\color{textcolor}\sffamily\fontsize{10.000000}{12.000000}\selectfont 0}%
\end{pgfscope}%
\begin{pgfscope}%
\pgfsetbuttcap%
\pgfsetroundjoin%
\definecolor{currentfill}{rgb}{0.000000,0.000000,0.000000}%
\pgfsetfillcolor{currentfill}%
\pgfsetlinewidth{0.803000pt}%
\definecolor{currentstroke}{rgb}{0.000000,0.000000,0.000000}%
\pgfsetstrokecolor{currentstroke}%
\pgfsetdash{}{0pt}%
\pgfsys@defobject{currentmarker}{\pgfqpoint{-0.048611in}{0.000000in}}{\pgfqpoint{0.000000in}{0.000000in}}{%
\pgfpathmoveto{\pgfqpoint{0.000000in}{0.000000in}}%
\pgfpathlineto{\pgfqpoint{-0.048611in}{0.000000in}}%
\pgfusepath{stroke,fill}%
}%
\begin{pgfscope}%
\pgfsys@transformshift{0.333606in}{1.673169in}%
\pgfsys@useobject{currentmarker}{}%
\end{pgfscope}%
\end{pgfscope}%
\begin{pgfscope}%
\pgfsetbuttcap%
\pgfsetroundjoin%
\definecolor{currentfill}{rgb}{0.000000,0.000000,0.000000}%
\pgfsetfillcolor{currentfill}%
\pgfsetlinewidth{0.803000pt}%
\definecolor{currentstroke}{rgb}{0.000000,0.000000,0.000000}%
\pgfsetstrokecolor{currentstroke}%
\pgfsetdash{}{0pt}%
\pgfsys@defobject{currentmarker}{\pgfqpoint{0.000000in}{0.000000in}}{\pgfqpoint{0.048611in}{0.000000in}}{%
\pgfpathmoveto{\pgfqpoint{0.000000in}{0.000000in}}%
\pgfpathlineto{\pgfqpoint{0.048611in}{0.000000in}}%
\pgfusepath{stroke,fill}%
}%
\begin{pgfscope}%
\pgfsys@transformshift{1.779818in}{1.673169in}%
\pgfsys@useobject{currentmarker}{}%
\end{pgfscope}%
\end{pgfscope}%
\begin{pgfscope}%
\definecolor{textcolor}{rgb}{0.000000,0.000000,0.000000}%
\pgfsetstrokecolor{textcolor}%
\pgfsetfillcolor{textcolor}%
\pgftext[x=0.196630in,y=1.620407in,left,base]{\color{textcolor}\sffamily\fontsize{10.000000}{12.000000}\selectfont 2}%
\end{pgfscope}%
\begin{pgfscope}%
\pgfsetbuttcap%
\pgfsetroundjoin%
\definecolor{currentfill}{rgb}{0.000000,0.000000,0.000000}%
\pgfsetfillcolor{currentfill}%
\pgfsetlinewidth{0.602250pt}%
\definecolor{currentstroke}{rgb}{0.000000,0.000000,0.000000}%
\pgfsetstrokecolor{currentstroke}%
\pgfsetdash{}{0pt}%
\pgfsys@defobject{currentmarker}{\pgfqpoint{-0.027778in}{0.000000in}}{\pgfqpoint{0.000000in}{0.000000in}}{%
\pgfpathmoveto{\pgfqpoint{0.000000in}{0.000000in}}%
\pgfpathlineto{\pgfqpoint{-0.027778in}{0.000000in}}%
\pgfusepath{stroke,fill}%
}%
\begin{pgfscope}%
\pgfsys@transformshift{0.333606in}{1.034438in}%
\pgfsys@useobject{currentmarker}{}%
\end{pgfscope}%
\end{pgfscope}%
\begin{pgfscope}%
\pgfsetbuttcap%
\pgfsetroundjoin%
\definecolor{currentfill}{rgb}{0.000000,0.000000,0.000000}%
\pgfsetfillcolor{currentfill}%
\pgfsetlinewidth{0.602250pt}%
\definecolor{currentstroke}{rgb}{0.000000,0.000000,0.000000}%
\pgfsetstrokecolor{currentstroke}%
\pgfsetdash{}{0pt}%
\pgfsys@defobject{currentmarker}{\pgfqpoint{0.000000in}{0.000000in}}{\pgfqpoint{0.027778in}{0.000000in}}{%
\pgfpathmoveto{\pgfqpoint{0.000000in}{0.000000in}}%
\pgfpathlineto{\pgfqpoint{0.027778in}{0.000000in}}%
\pgfusepath{stroke,fill}%
}%
\begin{pgfscope}%
\pgfsys@transformshift{1.779818in}{1.034438in}%
\pgfsys@useobject{currentmarker}{}%
\end{pgfscope}%
\end{pgfscope}%
\begin{pgfscope}%
\definecolor{textcolor}{rgb}{0.000000,0.000000,0.000000}%
\pgfsetstrokecolor{textcolor}%
\pgfsetfillcolor{textcolor}%
\pgftext[x=0.261295in,y=1.034438in,,bottom,rotate=90.000000]{\color{textcolor}\sffamily\fontsize{12.000000}{14.400000}\selectfont \(\displaystyle E_{k\sigma}\)}%
\end{pgfscope}%
\begin{pgfscope}%
\pgfpathrectangle{\pgfqpoint{0.333606in}{0.395708in}}{\pgfqpoint{1.446212in}{1.596825in}}%
\pgfusepath{clip}%
\pgfsetrectcap%
\pgfsetroundjoin%
\pgfsetlinewidth{2.007500pt}%
\definecolor{currentstroke}{rgb}{0.121569,0.466667,0.705882}%
\pgfsetstrokecolor{currentstroke}%
\pgfsetdash{}{0pt}%
\pgfpathmoveto{\pgfqpoint{0.663400in}{2.006422in}}%
\pgfpathlineto{\pgfqpoint{0.677354in}{1.911858in}}%
\pgfpathlineto{\pgfqpoint{0.700610in}{1.767469in}}%
\pgfpathlineto{\pgfqpoint{0.723865in}{1.640798in}}%
\pgfpathlineto{\pgfqpoint{0.747121in}{1.534097in}}%
\pgfpathlineto{\pgfqpoint{0.770377in}{1.449364in}}%
\pgfpathlineto{\pgfqpoint{0.782004in}{1.415600in}}%
\pgfpathlineto{\pgfqpoint{0.793632in}{1.387558in}}%
\pgfpathlineto{\pgfqpoint{0.805260in}{1.365064in}}%
\pgfpathlineto{\pgfqpoint{0.816888in}{1.347805in}}%
\pgfpathlineto{\pgfqpoint{0.828516in}{1.335349in}}%
\pgfpathlineto{\pgfqpoint{0.840144in}{1.327171in}}%
\pgfpathlineto{\pgfqpoint{0.851771in}{1.322691in}}%
\pgfpathlineto{\pgfqpoint{0.863399in}{1.321307in}}%
\pgfpathlineto{\pgfqpoint{0.875027in}{1.322431in}}%
\pgfpathlineto{\pgfqpoint{0.898283in}{1.330023in}}%
\pgfpathlineto{\pgfqpoint{0.933166in}{1.347961in}}%
\pgfpathlineto{\pgfqpoint{0.968050in}{1.365687in}}%
\pgfpathlineto{\pgfqpoint{0.991306in}{1.374344in}}%
\pgfpathlineto{\pgfqpoint{1.014561in}{1.379074in}}%
\pgfpathlineto{\pgfqpoint{1.037817in}{1.379063in}}%
\pgfpathlineto{\pgfqpoint{1.061073in}{1.373738in}}%
\pgfpathlineto{\pgfqpoint{1.084328in}{1.362709in}}%
\pgfpathlineto{\pgfqpoint{1.107584in}{1.345720in}}%
\pgfpathlineto{\pgfqpoint{1.130840in}{1.322633in}}%
\pgfpathlineto{\pgfqpoint{1.154095in}{1.293407in}}%
\pgfpathlineto{\pgfqpoint{1.177351in}{1.258097in}}%
\pgfpathlineto{\pgfqpoint{1.200607in}{1.216861in}}%
\pgfpathlineto{\pgfqpoint{1.223862in}{1.169970in}}%
\pgfpathlineto{\pgfqpoint{1.258746in}{1.089994in}}%
\pgfpathlineto{\pgfqpoint{1.293629in}{1.000550in}}%
\pgfpathlineto{\pgfqpoint{1.386652in}{0.752926in}}%
\pgfpathlineto{\pgfqpoint{1.409908in}{0.703455in}}%
\pgfpathlineto{\pgfqpoint{1.421535in}{0.683349in}}%
\pgfpathlineto{\pgfqpoint{1.433163in}{0.667126in}}%
\pgfpathlineto{\pgfqpoint{1.444791in}{0.655353in}}%
\pgfpathlineto{\pgfqpoint{1.456419in}{0.648529in}}%
\pgfpathlineto{\pgfqpoint{1.468047in}{0.647052in}}%
\pgfpathlineto{\pgfqpoint{1.479675in}{0.651192in}}%
\pgfpathlineto{\pgfqpoint{1.491302in}{0.661080in}}%
\pgfpathlineto{\pgfqpoint{1.502930in}{0.676718in}}%
\pgfpathlineto{\pgfqpoint{1.514558in}{0.697991in}}%
\pgfpathlineto{\pgfqpoint{1.526186in}{0.724704in}}%
\pgfpathlineto{\pgfqpoint{1.537814in}{0.756604in}}%
\pgfpathlineto{\pgfqpoint{1.561069in}{0.834835in}}%
\pgfpathlineto{\pgfqpoint{1.584325in}{0.930443in}}%
\pgfpathlineto{\pgfqpoint{1.607581in}{1.041437in}}%
\pgfpathlineto{\pgfqpoint{1.630836in}{1.166206in}}%
\pgfpathlineto{\pgfqpoint{1.665720in}{1.376543in}}%
\pgfpathlineto{\pgfqpoint{1.700604in}{1.612132in}}%
\pgfpathlineto{\pgfqpoint{1.735487in}{1.871000in}}%
\pgfpathlineto{\pgfqpoint{1.752604in}{2.006422in}}%
\pgfpathlineto{\pgfqpoint{1.752604in}{2.006422in}}%
\pgfusepath{stroke}%
\end{pgfscope}%
\begin{pgfscope}%
\pgfpathrectangle{\pgfqpoint{0.333606in}{0.395708in}}{\pgfqpoint{1.446212in}{1.596825in}}%
\pgfusepath{clip}%
\pgfsetbuttcap%
\pgfsetroundjoin%
\pgfsetlinewidth{2.007500pt}%
\definecolor{currentstroke}{rgb}{1.000000,0.498039,0.054902}%
\pgfsetstrokecolor{currentstroke}%
\pgfsetdash{{7.400000pt}{3.200000pt}}{0.000000pt}%
\pgfpathmoveto{\pgfqpoint{0.515448in}{2.006422in}}%
\pgfpathlineto{\pgfqpoint{0.549448in}{1.749297in}}%
\pgfpathlineto{\pgfqpoint{0.584331in}{1.508228in}}%
\pgfpathlineto{\pgfqpoint{0.619215in}{1.291883in}}%
\pgfpathlineto{\pgfqpoint{0.642470in}{1.162684in}}%
\pgfpathlineto{\pgfqpoint{0.665726in}{1.046793in}}%
\pgfpathlineto{\pgfqpoint{0.688982in}{0.945689in}}%
\pgfpathlineto{\pgfqpoint{0.712237in}{0.861239in}}%
\pgfpathlineto{\pgfqpoint{0.723865in}{0.825936in}}%
\pgfpathlineto{\pgfqpoint{0.735493in}{0.795628in}}%
\pgfpathlineto{\pgfqpoint{0.747121in}{0.770590in}}%
\pgfpathlineto{\pgfqpoint{0.758749in}{0.751060in}}%
\pgfpathlineto{\pgfqpoint{0.770377in}{0.737212in}}%
\pgfpathlineto{\pgfqpoint{0.782004in}{0.729126in}}%
\pgfpathlineto{\pgfqpoint{0.793632in}{0.726762in}}%
\pgfpathlineto{\pgfqpoint{0.805260in}{0.729945in}}%
\pgfpathlineto{\pgfqpoint{0.816888in}{0.738363in}}%
\pgfpathlineto{\pgfqpoint{0.828516in}{0.751584in}}%
\pgfpathlineto{\pgfqpoint{0.840144in}{0.769084in}}%
\pgfpathlineto{\pgfqpoint{0.863399in}{0.814576in}}%
\pgfpathlineto{\pgfqpoint{0.886655in}{0.870130in}}%
\pgfpathlineto{\pgfqpoint{0.921538in}{0.963279in}}%
\pgfpathlineto{\pgfqpoint{0.979678in}{1.120489in}}%
\pgfpathlineto{\pgfqpoint{1.014561in}{1.206151in}}%
\pgfpathlineto{\pgfqpoint{1.037817in}{1.257495in}}%
\pgfpathlineto{\pgfqpoint{1.061073in}{1.303526in}}%
\pgfpathlineto{\pgfqpoint{1.084328in}{1.343852in}}%
\pgfpathlineto{\pgfqpoint{1.107584in}{1.378218in}}%
\pgfpathlineto{\pgfqpoint{1.130840in}{1.406486in}}%
\pgfpathlineto{\pgfqpoint{1.154095in}{1.428616in}}%
\pgfpathlineto{\pgfqpoint{1.177351in}{1.444661in}}%
\pgfpathlineto{\pgfqpoint{1.200607in}{1.454780in}}%
\pgfpathlineto{\pgfqpoint{1.223862in}{1.459244in}}%
\pgfpathlineto{\pgfqpoint{1.247118in}{1.458471in}}%
\pgfpathlineto{\pgfqpoint{1.270374in}{1.453067in}}%
\pgfpathlineto{\pgfqpoint{1.293629in}{1.443889in}}%
\pgfpathlineto{\pgfqpoint{1.328513in}{1.425814in}}%
\pgfpathlineto{\pgfqpoint{1.363396in}{1.408326in}}%
\pgfpathlineto{\pgfqpoint{1.386652in}{1.401686in}}%
\pgfpathlineto{\pgfqpoint{1.398280in}{1.401284in}}%
\pgfpathlineto{\pgfqpoint{1.409908in}{1.403571in}}%
\pgfpathlineto{\pgfqpoint{1.421535in}{1.409142in}}%
\pgfpathlineto{\pgfqpoint{1.433163in}{1.418597in}}%
\pgfpathlineto{\pgfqpoint{1.444791in}{1.432501in}}%
\pgfpathlineto{\pgfqpoint{1.456419in}{1.451355in}}%
\pgfpathlineto{\pgfqpoint{1.468047in}{1.475556in}}%
\pgfpathlineto{\pgfqpoint{1.479675in}{1.505373in}}%
\pgfpathlineto{\pgfqpoint{1.491302in}{1.540939in}}%
\pgfpathlineto{\pgfqpoint{1.514558in}{1.629205in}}%
\pgfpathlineto{\pgfqpoint{1.537814in}{1.739173in}}%
\pgfpathlineto{\pgfqpoint{1.561069in}{1.868759in}}%
\pgfpathlineto{\pgfqpoint{1.582893in}{2.006422in}}%
\pgfpathlineto{\pgfqpoint{1.582893in}{2.006422in}}%
\pgfusepath{stroke}%
\end{pgfscope}%
\begin{pgfscope}%
\pgfsetrectcap%
\pgfsetmiterjoin%
\pgfsetlinewidth{0.803000pt}%
\definecolor{currentstroke}{rgb}{0.000000,0.000000,0.000000}%
\pgfsetstrokecolor{currentstroke}%
\pgfsetdash{}{0pt}%
\pgfpathmoveto{\pgfqpoint{0.333606in}{0.395708in}}%
\pgfpathlineto{\pgfqpoint{0.333606in}{1.992534in}}%
\pgfusepath{stroke}%
\end{pgfscope}%
\begin{pgfscope}%
\pgfsetrectcap%
\pgfsetmiterjoin%
\pgfsetlinewidth{0.803000pt}%
\definecolor{currentstroke}{rgb}{0.000000,0.000000,0.000000}%
\pgfsetstrokecolor{currentstroke}%
\pgfsetdash{}{0pt}%
\pgfpathmoveto{\pgfqpoint{1.779818in}{0.395708in}}%
\pgfpathlineto{\pgfqpoint{1.779818in}{1.992534in}}%
\pgfusepath{stroke}%
\end{pgfscope}%
\begin{pgfscope}%
\pgfsetrectcap%
\pgfsetmiterjoin%
\pgfsetlinewidth{0.803000pt}%
\definecolor{currentstroke}{rgb}{0.000000,0.000000,0.000000}%
\pgfsetstrokecolor{currentstroke}%
\pgfsetdash{}{0pt}%
\pgfpathmoveto{\pgfqpoint{0.333606in}{0.395708in}}%
\pgfpathlineto{\pgfqpoint{1.779818in}{0.395708in}}%
\pgfusepath{stroke}%
\end{pgfscope}%
\begin{pgfscope}%
\pgfsetrectcap%
\pgfsetmiterjoin%
\pgfsetlinewidth{0.803000pt}%
\definecolor{currentstroke}{rgb}{0.000000,0.000000,0.000000}%
\pgfsetstrokecolor{currentstroke}%
\pgfsetdash{}{0pt}%
\pgfpathmoveto{\pgfqpoint{0.333606in}{1.992534in}}%
\pgfpathlineto{\pgfqpoint{1.779818in}{1.992534in}}%
\pgfusepath{stroke}%
\end{pgfscope}%
\begin{pgfscope}%
\definecolor{textcolor}{rgb}{0.000000,0.000000,0.000000}%
\pgfsetstrokecolor{textcolor}%
\pgfsetfillcolor{textcolor}%
\pgftext[x=0.376992in,y=0.475549in,left,base]{\color{textcolor}\sffamily\fontsize{12.000000}{14.400000}\selectfont (a)}%
\end{pgfscope}%
\begin{pgfscope}%
\pgfsetbuttcap%
\pgfsetmiterjoin%
\definecolor{currentfill}{rgb}{1.000000,1.000000,1.000000}%
\pgfsetfillcolor{currentfill}%
\pgfsetlinewidth{0.000000pt}%
\definecolor{currentstroke}{rgb}{0.000000,0.000000,0.000000}%
\pgfsetstrokecolor{currentstroke}%
\pgfsetstrokeopacity{0.000000}%
\pgfsetdash{}{0pt}%
\pgfpathmoveto{\pgfqpoint{2.258606in}{0.395708in}}%
\pgfpathlineto{\pgfqpoint{3.704818in}{0.395708in}}%
\pgfpathlineto{\pgfqpoint{3.704818in}{1.992534in}}%
\pgfpathlineto{\pgfqpoint{2.258606in}{1.992534in}}%
\pgfpathclose%
\pgfusepath{fill}%
\end{pgfscope}%
\begin{pgfscope}%
\pgfsetbuttcap%
\pgfsetroundjoin%
\definecolor{currentfill}{rgb}{0.000000,0.000000,0.000000}%
\pgfsetfillcolor{currentfill}%
\pgfsetlinewidth{0.803000pt}%
\definecolor{currentstroke}{rgb}{0.000000,0.000000,0.000000}%
\pgfsetstrokecolor{currentstroke}%
\pgfsetdash{}{0pt}%
\pgfsys@defobject{currentmarker}{\pgfqpoint{0.000000in}{-0.048611in}}{\pgfqpoint{0.000000in}{0.000000in}}{%
\pgfpathmoveto{\pgfqpoint{0.000000in}{0.000000in}}%
\pgfpathlineto{\pgfqpoint{0.000000in}{-0.048611in}}%
\pgfusepath{stroke,fill}%
}%
\begin{pgfscope}%
\pgfsys@transformshift{2.403227in}{0.395708in}%
\pgfsys@useobject{currentmarker}{}%
\end{pgfscope}%
\end{pgfscope}%
\begin{pgfscope}%
\definecolor{textcolor}{rgb}{0.000000,0.000000,0.000000}%
\pgfsetstrokecolor{textcolor}%
\pgfsetfillcolor{textcolor}%
\pgftext[x=2.403227in,y=0.298486in,,top]{\color{textcolor}\sffamily\fontsize{10.000000}{12.000000}\selectfont −2}%
\end{pgfscope}%
\begin{pgfscope}%
\pgfsetbuttcap%
\pgfsetroundjoin%
\definecolor{currentfill}{rgb}{0.000000,0.000000,0.000000}%
\pgfsetfillcolor{currentfill}%
\pgfsetlinewidth{0.803000pt}%
\definecolor{currentstroke}{rgb}{0.000000,0.000000,0.000000}%
\pgfsetstrokecolor{currentstroke}%
\pgfsetdash{}{0pt}%
\pgfsys@defobject{currentmarker}{\pgfqpoint{0.000000in}{-0.048611in}}{\pgfqpoint{0.000000in}{0.000000in}}{%
\pgfpathmoveto{\pgfqpoint{0.000000in}{0.000000in}}%
\pgfpathlineto{\pgfqpoint{0.000000in}{-0.048611in}}%
\pgfusepath{stroke,fill}%
}%
\begin{pgfscope}%
\pgfsys@transformshift{2.981712in}{0.395708in}%
\pgfsys@useobject{currentmarker}{}%
\end{pgfscope}%
\end{pgfscope}%
\begin{pgfscope}%
\definecolor{textcolor}{rgb}{0.000000,0.000000,0.000000}%
\pgfsetstrokecolor{textcolor}%
\pgfsetfillcolor{textcolor}%
\pgftext[x=2.981712in,y=0.298486in,,top]{\color{textcolor}\sffamily\fontsize{10.000000}{12.000000}\selectfont 0}%
\end{pgfscope}%
\begin{pgfscope}%
\pgfsetbuttcap%
\pgfsetroundjoin%
\definecolor{currentfill}{rgb}{0.000000,0.000000,0.000000}%
\pgfsetfillcolor{currentfill}%
\pgfsetlinewidth{0.803000pt}%
\definecolor{currentstroke}{rgb}{0.000000,0.000000,0.000000}%
\pgfsetstrokecolor{currentstroke}%
\pgfsetdash{}{0pt}%
\pgfsys@defobject{currentmarker}{\pgfqpoint{0.000000in}{-0.048611in}}{\pgfqpoint{0.000000in}{0.000000in}}{%
\pgfpathmoveto{\pgfqpoint{0.000000in}{0.000000in}}%
\pgfpathlineto{\pgfqpoint{0.000000in}{-0.048611in}}%
\pgfusepath{stroke,fill}%
}%
\begin{pgfscope}%
\pgfsys@transformshift{3.560197in}{0.395708in}%
\pgfsys@useobject{currentmarker}{}%
\end{pgfscope}%
\end{pgfscope}%
\begin{pgfscope}%
\definecolor{textcolor}{rgb}{0.000000,0.000000,0.000000}%
\pgfsetstrokecolor{textcolor}%
\pgfsetfillcolor{textcolor}%
\pgftext[x=3.560197in,y=0.298486in,,top]{\color{textcolor}\sffamily\fontsize{10.000000}{12.000000}\selectfont 2}%
\end{pgfscope}%
\begin{pgfscope}%
\pgfsetbuttcap%
\pgfsetroundjoin%
\definecolor{currentfill}{rgb}{0.000000,0.000000,0.000000}%
\pgfsetfillcolor{currentfill}%
\pgfsetlinewidth{0.602250pt}%
\definecolor{currentstroke}{rgb}{0.000000,0.000000,0.000000}%
\pgfsetstrokecolor{currentstroke}%
\pgfsetdash{}{0pt}%
\pgfsys@defobject{currentmarker}{\pgfqpoint{0.000000in}{-0.027778in}}{\pgfqpoint{0.000000in}{0.000000in}}{%
\pgfpathmoveto{\pgfqpoint{0.000000in}{0.000000in}}%
\pgfpathlineto{\pgfqpoint{0.000000in}{-0.027778in}}%
\pgfusepath{stroke,fill}%
}%
\begin{pgfscope}%
\pgfsys@transformshift{2.692470in}{0.395708in}%
\pgfsys@useobject{currentmarker}{}%
\end{pgfscope}%
\end{pgfscope}%
\begin{pgfscope}%
\pgfsetbuttcap%
\pgfsetroundjoin%
\definecolor{currentfill}{rgb}{0.000000,0.000000,0.000000}%
\pgfsetfillcolor{currentfill}%
\pgfsetlinewidth{0.602250pt}%
\definecolor{currentstroke}{rgb}{0.000000,0.000000,0.000000}%
\pgfsetstrokecolor{currentstroke}%
\pgfsetdash{}{0pt}%
\pgfsys@defobject{currentmarker}{\pgfqpoint{0.000000in}{-0.027778in}}{\pgfqpoint{0.000000in}{0.000000in}}{%
\pgfpathmoveto{\pgfqpoint{0.000000in}{0.000000in}}%
\pgfpathlineto{\pgfqpoint{0.000000in}{-0.027778in}}%
\pgfusepath{stroke,fill}%
}%
\begin{pgfscope}%
\pgfsys@transformshift{3.270955in}{0.395708in}%
\pgfsys@useobject{currentmarker}{}%
\end{pgfscope}%
\end{pgfscope}%
\begin{pgfscope}%
\definecolor{textcolor}{rgb}{0.000000,0.000000,0.000000}%
\pgfsetstrokecolor{textcolor}%
\pgfsetfillcolor{textcolor}%
\pgftext[x=3.270955in,y=0.261295in,,top]{\color{textcolor}\sffamily\fontsize{12.000000}{14.400000}\selectfont \(\displaystyle k'\)}%
\end{pgfscope}%
\begin{pgfscope}%
\pgfsetbuttcap%
\pgfsetroundjoin%
\definecolor{currentfill}{rgb}{0.000000,0.000000,0.000000}%
\pgfsetfillcolor{currentfill}%
\pgfsetlinewidth{0.803000pt}%
\definecolor{currentstroke}{rgb}{0.000000,0.000000,0.000000}%
\pgfsetstrokecolor{currentstroke}%
\pgfsetdash{}{0pt}%
\pgfsys@defobject{currentmarker}{\pgfqpoint{-0.048611in}{0.000000in}}{\pgfqpoint{0.000000in}{0.000000in}}{%
\pgfpathmoveto{\pgfqpoint{0.000000in}{0.000000in}}%
\pgfpathlineto{\pgfqpoint{-0.048611in}{0.000000in}}%
\pgfusepath{stroke,fill}%
}%
\begin{pgfscope}%
\pgfsys@transformshift{2.258606in}{0.395708in}%
\pgfsys@useobject{currentmarker}{}%
\end{pgfscope}%
\end{pgfscope}%
\begin{pgfscope}%
\pgfsetbuttcap%
\pgfsetroundjoin%
\definecolor{currentfill}{rgb}{0.000000,0.000000,0.000000}%
\pgfsetfillcolor{currentfill}%
\pgfsetlinewidth{0.803000pt}%
\definecolor{currentstroke}{rgb}{0.000000,0.000000,0.000000}%
\pgfsetstrokecolor{currentstroke}%
\pgfsetdash{}{0pt}%
\pgfsys@defobject{currentmarker}{\pgfqpoint{0.000000in}{0.000000in}}{\pgfqpoint{0.048611in}{0.000000in}}{%
\pgfpathmoveto{\pgfqpoint{0.000000in}{0.000000in}}%
\pgfpathlineto{\pgfqpoint{0.048611in}{0.000000in}}%
\pgfusepath{stroke,fill}%
}%
\begin{pgfscope}%
\pgfsys@transformshift{3.704818in}{0.395708in}%
\pgfsys@useobject{currentmarker}{}%
\end{pgfscope}%
\end{pgfscope}%
\begin{pgfscope}%
\definecolor{textcolor}{rgb}{0.000000,0.000000,0.000000}%
\pgfsetstrokecolor{textcolor}%
\pgfsetfillcolor{textcolor}%
\pgftext[x=2.121630in,y=0.342947in,left,base]{\color{textcolor}\sffamily\fontsize{10.000000}{12.000000}\selectfont 0}%
\end{pgfscope}%
\begin{pgfscope}%
\pgfsetbuttcap%
\pgfsetroundjoin%
\definecolor{currentfill}{rgb}{0.000000,0.000000,0.000000}%
\pgfsetfillcolor{currentfill}%
\pgfsetlinewidth{0.803000pt}%
\definecolor{currentstroke}{rgb}{0.000000,0.000000,0.000000}%
\pgfsetstrokecolor{currentstroke}%
\pgfsetdash{}{0pt}%
\pgfsys@defobject{currentmarker}{\pgfqpoint{-0.048611in}{0.000000in}}{\pgfqpoint{0.000000in}{0.000000in}}{%
\pgfpathmoveto{\pgfqpoint{0.000000in}{0.000000in}}%
\pgfpathlineto{\pgfqpoint{-0.048611in}{0.000000in}}%
\pgfusepath{stroke,fill}%
}%
\begin{pgfscope}%
\pgfsys@transformshift{2.258606in}{1.460258in}%
\pgfsys@useobject{currentmarker}{}%
\end{pgfscope}%
\end{pgfscope}%
\begin{pgfscope}%
\pgfsetbuttcap%
\pgfsetroundjoin%
\definecolor{currentfill}{rgb}{0.000000,0.000000,0.000000}%
\pgfsetfillcolor{currentfill}%
\pgfsetlinewidth{0.803000pt}%
\definecolor{currentstroke}{rgb}{0.000000,0.000000,0.000000}%
\pgfsetstrokecolor{currentstroke}%
\pgfsetdash{}{0pt}%
\pgfsys@defobject{currentmarker}{\pgfqpoint{0.000000in}{0.000000in}}{\pgfqpoint{0.048611in}{0.000000in}}{%
\pgfpathmoveto{\pgfqpoint{0.000000in}{0.000000in}}%
\pgfpathlineto{\pgfqpoint{0.048611in}{0.000000in}}%
\pgfusepath{stroke,fill}%
}%
\begin{pgfscope}%
\pgfsys@transformshift{3.704818in}{1.460258in}%
\pgfsys@useobject{currentmarker}{}%
\end{pgfscope}%
\end{pgfscope}%
\begin{pgfscope}%
\definecolor{textcolor}{rgb}{0.000000,0.000000,0.000000}%
\pgfsetstrokecolor{textcolor}%
\pgfsetfillcolor{textcolor}%
\pgftext[x=2.121630in,y=1.407497in,left,base]{\color{textcolor}\sffamily\fontsize{10.000000}{12.000000}\selectfont 2}%
\end{pgfscope}%
\begin{pgfscope}%
\pgfsetbuttcap%
\pgfsetroundjoin%
\definecolor{currentfill}{rgb}{0.000000,0.000000,0.000000}%
\pgfsetfillcolor{currentfill}%
\pgfsetlinewidth{0.602250pt}%
\definecolor{currentstroke}{rgb}{0.000000,0.000000,0.000000}%
\pgfsetstrokecolor{currentstroke}%
\pgfsetdash{}{0pt}%
\pgfsys@defobject{currentmarker}{\pgfqpoint{-0.027778in}{0.000000in}}{\pgfqpoint{0.000000in}{0.000000in}}{%
\pgfpathmoveto{\pgfqpoint{0.000000in}{0.000000in}}%
\pgfpathlineto{\pgfqpoint{-0.027778in}{0.000000in}}%
\pgfusepath{stroke,fill}%
}%
\begin{pgfscope}%
\pgfsys@transformshift{2.258606in}{0.927983in}%
\pgfsys@useobject{currentmarker}{}%
\end{pgfscope}%
\end{pgfscope}%
\begin{pgfscope}%
\pgfsetbuttcap%
\pgfsetroundjoin%
\definecolor{currentfill}{rgb}{0.000000,0.000000,0.000000}%
\pgfsetfillcolor{currentfill}%
\pgfsetlinewidth{0.602250pt}%
\definecolor{currentstroke}{rgb}{0.000000,0.000000,0.000000}%
\pgfsetstrokecolor{currentstroke}%
\pgfsetdash{}{0pt}%
\pgfsys@defobject{currentmarker}{\pgfqpoint{0.000000in}{0.000000in}}{\pgfqpoint{0.027778in}{0.000000in}}{%
\pgfpathmoveto{\pgfqpoint{0.000000in}{0.000000in}}%
\pgfpathlineto{\pgfqpoint{0.027778in}{0.000000in}}%
\pgfusepath{stroke,fill}%
}%
\begin{pgfscope}%
\pgfsys@transformshift{3.704818in}{0.927983in}%
\pgfsys@useobject{currentmarker}{}%
\end{pgfscope}%
\end{pgfscope}%
\begin{pgfscope}%
\pgfsetbuttcap%
\pgfsetroundjoin%
\definecolor{currentfill}{rgb}{0.000000,0.000000,0.000000}%
\pgfsetfillcolor{currentfill}%
\pgfsetlinewidth{0.602250pt}%
\definecolor{currentstroke}{rgb}{0.000000,0.000000,0.000000}%
\pgfsetstrokecolor{currentstroke}%
\pgfsetdash{}{0pt}%
\pgfsys@defobject{currentmarker}{\pgfqpoint{-0.027778in}{0.000000in}}{\pgfqpoint{0.000000in}{0.000000in}}{%
\pgfpathmoveto{\pgfqpoint{0.000000in}{0.000000in}}%
\pgfpathlineto{\pgfqpoint{-0.027778in}{0.000000in}}%
\pgfusepath{stroke,fill}%
}%
\begin{pgfscope}%
\pgfsys@transformshift{2.258606in}{1.992534in}%
\pgfsys@useobject{currentmarker}{}%
\end{pgfscope}%
\end{pgfscope}%
\begin{pgfscope}%
\pgfsetbuttcap%
\pgfsetroundjoin%
\definecolor{currentfill}{rgb}{0.000000,0.000000,0.000000}%
\pgfsetfillcolor{currentfill}%
\pgfsetlinewidth{0.602250pt}%
\definecolor{currentstroke}{rgb}{0.000000,0.000000,0.000000}%
\pgfsetstrokecolor{currentstroke}%
\pgfsetdash{}{0pt}%
\pgfsys@defobject{currentmarker}{\pgfqpoint{0.000000in}{0.000000in}}{\pgfqpoint{0.027778in}{0.000000in}}{%
\pgfpathmoveto{\pgfqpoint{0.000000in}{0.000000in}}%
\pgfpathlineto{\pgfqpoint{0.027778in}{0.000000in}}%
\pgfusepath{stroke,fill}%
}%
\begin{pgfscope}%
\pgfsys@transformshift{3.704818in}{1.992534in}%
\pgfsys@useobject{currentmarker}{}%
\end{pgfscope}%
\end{pgfscope}%
\begin{pgfscope}%
\definecolor{textcolor}{rgb}{0.000000,0.000000,0.000000}%
\pgfsetstrokecolor{textcolor}%
\pgfsetfillcolor{textcolor}%
\pgftext[x=2.186295in,y=1.034438in,,bottom,rotate=90.000000]{\color{textcolor}\sffamily\fontsize{12.000000}{14.400000}\selectfont \(\displaystyle E_{k\sigma}\)}%
\end{pgfscope}%
\begin{pgfscope}%
\pgfpathrectangle{\pgfqpoint{2.258606in}{0.395708in}}{\pgfqpoint{1.446212in}{1.596825in}}%
\pgfusepath{clip}%
\pgfsetrectcap%
\pgfsetroundjoin%
\pgfsetlinewidth{2.007500pt}%
\definecolor{currentstroke}{rgb}{0.121569,0.466667,0.705882}%
\pgfsetstrokecolor{currentstroke}%
\pgfsetdash{}{0pt}%
\pgfpathmoveto{\pgfqpoint{2.404629in}{2.006422in}}%
\pgfpathlineto{\pgfqpoint{2.427936in}{1.849483in}}%
\pgfpathlineto{\pgfqpoint{2.462820in}{1.631552in}}%
\pgfpathlineto{\pgfqpoint{2.497703in}{1.435811in}}%
\pgfpathlineto{\pgfqpoint{2.520959in}{1.319119in}}%
\pgfpathlineto{\pgfqpoint{2.544215in}{1.214904in}}%
\pgfpathlineto{\pgfqpoint{2.567470in}{1.124756in}}%
\pgfpathlineto{\pgfqpoint{2.590726in}{1.050514in}}%
\pgfpathlineto{\pgfqpoint{2.602354in}{1.019939in}}%
\pgfpathlineto{\pgfqpoint{2.613982in}{0.993990in}}%
\pgfpathlineto{\pgfqpoint{2.625610in}{0.972799in}}%
\pgfpathlineto{\pgfqpoint{2.637237in}{0.956411in}}%
\pgfpathlineto{\pgfqpoint{2.648865in}{0.944770in}}%
\pgfpathlineto{\pgfqpoint{2.660493in}{0.937704in}}%
\pgfpathlineto{\pgfqpoint{2.672121in}{0.934930in}}%
\pgfpathlineto{\pgfqpoint{2.683749in}{0.936068in}}%
\pgfpathlineto{\pgfqpoint{2.695377in}{0.940669in}}%
\pgfpathlineto{\pgfqpoint{2.707004in}{0.948242in}}%
\pgfpathlineto{\pgfqpoint{2.730260in}{0.970316in}}%
\pgfpathlineto{\pgfqpoint{2.753516in}{0.998549in}}%
\pgfpathlineto{\pgfqpoint{2.846538in}{1.120734in}}%
\pgfpathlineto{\pgfqpoint{2.869794in}{1.145671in}}%
\pgfpathlineto{\pgfqpoint{2.893050in}{1.166595in}}%
\pgfpathlineto{\pgfqpoint{2.916306in}{1.183050in}}%
\pgfpathlineto{\pgfqpoint{2.939561in}{1.194722in}}%
\pgfpathlineto{\pgfqpoint{2.962817in}{1.201411in}}%
\pgfpathlineto{\pgfqpoint{2.986073in}{1.203010in}}%
\pgfpathlineto{\pgfqpoint{3.009328in}{1.199495in}}%
\pgfpathlineto{\pgfqpoint{3.032584in}{1.190920in}}%
\pgfpathlineto{\pgfqpoint{3.055840in}{1.177426in}}%
\pgfpathlineto{\pgfqpoint{3.079095in}{1.159251in}}%
\pgfpathlineto{\pgfqpoint{3.102351in}{1.136759in}}%
\pgfpathlineto{\pgfqpoint{3.125607in}{1.110475in}}%
\pgfpathlineto{\pgfqpoint{3.160490in}{1.065640in}}%
\pgfpathlineto{\pgfqpoint{3.218629in}{0.987454in}}%
\pgfpathlineto{\pgfqpoint{3.241885in}{0.961126in}}%
\pgfpathlineto{\pgfqpoint{3.253513in}{0.950540in}}%
\pgfpathlineto{\pgfqpoint{3.265141in}{0.942303in}}%
\pgfpathlineto{\pgfqpoint{3.276768in}{0.936912in}}%
\pgfpathlineto{\pgfqpoint{3.288396in}{0.934864in}}%
\pgfpathlineto{\pgfqpoint{3.300024in}{0.936622in}}%
\pgfpathlineto{\pgfqpoint{3.311652in}{0.942584in}}%
\pgfpathlineto{\pgfqpoint{3.323280in}{0.953061in}}%
\pgfpathlineto{\pgfqpoint{3.334908in}{0.968252in}}%
\pgfpathlineto{\pgfqpoint{3.346535in}{0.988243in}}%
\pgfpathlineto{\pgfqpoint{3.358163in}{1.013011in}}%
\pgfpathlineto{\pgfqpoint{3.369791in}{1.042446in}}%
\pgfpathlineto{\pgfqpoint{3.393047in}{1.114567in}}%
\pgfpathlineto{\pgfqpoint{3.416302in}{1.202830in}}%
\pgfpathlineto{\pgfqpoint{3.439558in}{1.305380in}}%
\pgfpathlineto{\pgfqpoint{3.462814in}{1.420588in}}%
\pgfpathlineto{\pgfqpoint{3.497697in}{1.614358in}}%
\pgfpathlineto{\pgfqpoint{3.532581in}{1.830530in}}%
\pgfpathlineto{\pgfqpoint{3.558767in}{2.006422in}}%
\pgfpathlineto{\pgfqpoint{3.558767in}{2.006422in}}%
\pgfusepath{stroke}%
\end{pgfscope}%
\begin{pgfscope}%
\pgfpathrectangle{\pgfqpoint{2.258606in}{0.395708in}}{\pgfqpoint{1.446212in}{1.596825in}}%
\pgfusepath{clip}%
\pgfsetbuttcap%
\pgfsetroundjoin%
\pgfsetlinewidth{2.007500pt}%
\definecolor{currentstroke}{rgb}{1.000000,0.498039,0.054902}%
\pgfsetstrokecolor{currentstroke}%
\pgfsetdash{{7.400000pt}{3.200000pt}}{0.000000pt}%
\pgfpathmoveto{\pgfqpoint{2.404629in}{2.006422in}}%
\pgfpathlineto{\pgfqpoint{2.427936in}{1.849483in}}%
\pgfpathlineto{\pgfqpoint{2.462820in}{1.631552in}}%
\pgfpathlineto{\pgfqpoint{2.497703in}{1.435811in}}%
\pgfpathlineto{\pgfqpoint{2.520959in}{1.319119in}}%
\pgfpathlineto{\pgfqpoint{2.544215in}{1.214904in}}%
\pgfpathlineto{\pgfqpoint{2.567470in}{1.124756in}}%
\pgfpathlineto{\pgfqpoint{2.590726in}{1.050514in}}%
\pgfpathlineto{\pgfqpoint{2.602354in}{1.019939in}}%
\pgfpathlineto{\pgfqpoint{2.613982in}{0.993990in}}%
\pgfpathlineto{\pgfqpoint{2.625610in}{0.972799in}}%
\pgfpathlineto{\pgfqpoint{2.637237in}{0.956411in}}%
\pgfpathlineto{\pgfqpoint{2.648865in}{0.944770in}}%
\pgfpathlineto{\pgfqpoint{2.660493in}{0.937704in}}%
\pgfpathlineto{\pgfqpoint{2.672121in}{0.934930in}}%
\pgfpathlineto{\pgfqpoint{2.683749in}{0.936068in}}%
\pgfpathlineto{\pgfqpoint{2.695377in}{0.940669in}}%
\pgfpathlineto{\pgfqpoint{2.707004in}{0.948242in}}%
\pgfpathlineto{\pgfqpoint{2.730260in}{0.970316in}}%
\pgfpathlineto{\pgfqpoint{2.753516in}{0.998549in}}%
\pgfpathlineto{\pgfqpoint{2.846538in}{1.120734in}}%
\pgfpathlineto{\pgfqpoint{2.869794in}{1.145671in}}%
\pgfpathlineto{\pgfqpoint{2.893050in}{1.166595in}}%
\pgfpathlineto{\pgfqpoint{2.916306in}{1.183050in}}%
\pgfpathlineto{\pgfqpoint{2.939561in}{1.194722in}}%
\pgfpathlineto{\pgfqpoint{2.962817in}{1.201411in}}%
\pgfpathlineto{\pgfqpoint{2.986073in}{1.203010in}}%
\pgfpathlineto{\pgfqpoint{3.009328in}{1.199495in}}%
\pgfpathlineto{\pgfqpoint{3.032584in}{1.190920in}}%
\pgfpathlineto{\pgfqpoint{3.055840in}{1.177426in}}%
\pgfpathlineto{\pgfqpoint{3.079095in}{1.159251in}}%
\pgfpathlineto{\pgfqpoint{3.102351in}{1.136759in}}%
\pgfpathlineto{\pgfqpoint{3.125607in}{1.110475in}}%
\pgfpathlineto{\pgfqpoint{3.160490in}{1.065640in}}%
\pgfpathlineto{\pgfqpoint{3.218629in}{0.987454in}}%
\pgfpathlineto{\pgfqpoint{3.241885in}{0.961126in}}%
\pgfpathlineto{\pgfqpoint{3.253513in}{0.950540in}}%
\pgfpathlineto{\pgfqpoint{3.265141in}{0.942303in}}%
\pgfpathlineto{\pgfqpoint{3.276768in}{0.936912in}}%
\pgfpathlineto{\pgfqpoint{3.288396in}{0.934864in}}%
\pgfpathlineto{\pgfqpoint{3.300024in}{0.936622in}}%
\pgfpathlineto{\pgfqpoint{3.311652in}{0.942584in}}%
\pgfpathlineto{\pgfqpoint{3.323280in}{0.953061in}}%
\pgfpathlineto{\pgfqpoint{3.334908in}{0.968252in}}%
\pgfpathlineto{\pgfqpoint{3.346535in}{0.988243in}}%
\pgfpathlineto{\pgfqpoint{3.358163in}{1.013011in}}%
\pgfpathlineto{\pgfqpoint{3.369791in}{1.042446in}}%
\pgfpathlineto{\pgfqpoint{3.393047in}{1.114567in}}%
\pgfpathlineto{\pgfqpoint{3.416302in}{1.202830in}}%
\pgfpathlineto{\pgfqpoint{3.439558in}{1.305380in}}%
\pgfpathlineto{\pgfqpoint{3.462814in}{1.420588in}}%
\pgfpathlineto{\pgfqpoint{3.497697in}{1.614358in}}%
\pgfpathlineto{\pgfqpoint{3.532581in}{1.830530in}}%
\pgfpathlineto{\pgfqpoint{3.558767in}{2.006422in}}%
\pgfpathlineto{\pgfqpoint{3.558767in}{2.006422in}}%
\pgfusepath{stroke}%
\end{pgfscope}%
\begin{pgfscope}%
\pgfsetrectcap%
\pgfsetmiterjoin%
\pgfsetlinewidth{0.803000pt}%
\definecolor{currentstroke}{rgb}{0.000000,0.000000,0.000000}%
\pgfsetstrokecolor{currentstroke}%
\pgfsetdash{}{0pt}%
\pgfpathmoveto{\pgfqpoint{2.258606in}{0.395708in}}%
\pgfpathlineto{\pgfqpoint{2.258606in}{1.992534in}}%
\pgfusepath{stroke}%
\end{pgfscope}%
\begin{pgfscope}%
\pgfsetrectcap%
\pgfsetmiterjoin%
\pgfsetlinewidth{0.803000pt}%
\definecolor{currentstroke}{rgb}{0.000000,0.000000,0.000000}%
\pgfsetstrokecolor{currentstroke}%
\pgfsetdash{}{0pt}%
\pgfpathmoveto{\pgfqpoint{3.704818in}{0.395708in}}%
\pgfpathlineto{\pgfqpoint{3.704818in}{1.992534in}}%
\pgfusepath{stroke}%
\end{pgfscope}%
\begin{pgfscope}%
\pgfsetrectcap%
\pgfsetmiterjoin%
\pgfsetlinewidth{0.803000pt}%
\definecolor{currentstroke}{rgb}{0.000000,0.000000,0.000000}%
\pgfsetstrokecolor{currentstroke}%
\pgfsetdash{}{0pt}%
\pgfpathmoveto{\pgfqpoint{2.258606in}{0.395708in}}%
\pgfpathlineto{\pgfqpoint{3.704818in}{0.395708in}}%
\pgfusepath{stroke}%
\end{pgfscope}%
\begin{pgfscope}%
\pgfsetrectcap%
\pgfsetmiterjoin%
\pgfsetlinewidth{0.803000pt}%
\definecolor{currentstroke}{rgb}{0.000000,0.000000,0.000000}%
\pgfsetstrokecolor{currentstroke}%
\pgfsetdash{}{0pt}%
\pgfpathmoveto{\pgfqpoint{2.258606in}{1.992534in}}%
\pgfpathlineto{\pgfqpoint{3.704818in}{1.992534in}}%
\pgfusepath{stroke}%
\end{pgfscope}%
\begin{pgfscope}%
\definecolor{textcolor}{rgb}{0.000000,0.000000,0.000000}%
\pgfsetstrokecolor{textcolor}%
\pgfsetfillcolor{textcolor}%
\pgftext[x=2.301992in,y=0.475549in,left,base]{\color{textcolor}\sffamily\fontsize{12.000000}{14.400000}\selectfont (b)}%
\end{pgfscope}%
\begin{pgfscope}%
\pgfsetbuttcap%
\pgfsetmiterjoin%
\definecolor{currentfill}{rgb}{0.501961,0.501961,0.501961}%
\pgfsetfillcolor{currentfill}%
\pgfsetfillopacity{0.250000}%
\pgfsetlinewidth{1.003750pt}%
\definecolor{currentstroke}{rgb}{0.000000,0.000000,0.000000}%
\pgfsetstrokecolor{currentstroke}%
\pgfsetstrokeopacity{0.250000}%
\pgfsetdash{}{0pt}%
\pgfpathmoveto{\pgfqpoint{3.331791in}{0.430443in}}%
\pgfpathlineto{\pgfqpoint{3.687490in}{0.430443in}}%
\pgfpathlineto{\pgfqpoint{3.687490in}{0.703744in}}%
\pgfpathlineto{\pgfqpoint{3.331791in}{0.703744in}}%
\pgfpathclose%
\pgfusepath{stroke,fill}%
\end{pgfscope}%
\begin{pgfscope}%
\definecolor{textcolor}{rgb}{0.000000,0.000000,0.000000}%
\pgfsetstrokecolor{textcolor}%
\pgfsetfillcolor{textcolor}%
\pgftext[x=3.618045in,y=0.528777in,right,base]{\color{textcolor}\sffamily\fontsize{10.000000}{12.000000}\itshape\selectfont CM}%
\end{pgfscope}%
\begin{pgfscope}%
\pgfsetbuttcap%
\pgfsetmiterjoin%
\definecolor{currentfill}{rgb}{1.000000,1.000000,1.000000}%
\pgfsetfillcolor{currentfill}%
\pgfsetfillopacity{0.800000}%
\pgfsetlinewidth{1.003750pt}%
\definecolor{currentstroke}{rgb}{0.800000,0.800000,0.800000}%
\pgfsetstrokecolor{currentstroke}%
\pgfsetstrokeopacity{0.800000}%
\pgfsetdash{}{0pt}%
\pgfpathmoveto{\pgfqpoint{2.645606in}{1.452013in}}%
\pgfpathlineto{\pgfqpoint{3.317819in}{1.452013in}}%
\pgfpathquadraticcurveto{\pgfqpoint{3.345596in}{1.452013in}}{\pgfqpoint{3.345596in}{1.479791in}}%
\pgfpathlineto{\pgfqpoint{3.345596in}{1.895311in}}%
\pgfpathquadraticcurveto{\pgfqpoint{3.345596in}{1.923089in}}{\pgfqpoint{3.317819in}{1.923089in}}%
\pgfpathlineto{\pgfqpoint{2.645606in}{1.923089in}}%
\pgfpathquadraticcurveto{\pgfqpoint{2.617828in}{1.923089in}}{\pgfqpoint{2.617828in}{1.895311in}}%
\pgfpathlineto{\pgfqpoint{2.617828in}{1.479791in}}%
\pgfpathquadraticcurveto{\pgfqpoint{2.617828in}{1.452013in}}{\pgfqpoint{2.645606in}{1.452013in}}%
\pgfpathclose%
\pgfusepath{stroke,fill}%
\end{pgfscope}%
\begin{pgfscope}%
\pgfsetrectcap%
\pgfsetroundjoin%
\pgfsetlinewidth{2.007500pt}%
\definecolor{currentstroke}{rgb}{0.121569,0.466667,0.705882}%
\pgfsetstrokecolor{currentstroke}%
\pgfsetdash{}{0pt}%
\pgfpathmoveto{\pgfqpoint{2.673383in}{1.810622in}}%
\pgfpathlineto{\pgfqpoint{2.951161in}{1.810622in}}%
\pgfusepath{stroke}%
\end{pgfscope}%
\begin{pgfscope}%
\definecolor{textcolor}{rgb}{0.000000,0.000000,0.000000}%
\pgfsetstrokecolor{textcolor}%
\pgfsetfillcolor{textcolor}%
\pgftext[x=3.062272in,y=1.762011in,left,base]{\color{textcolor}\sffamily\fontsize{10.000000}{12.000000}\selectfont \(\displaystyle E_{k\uparrow}\)}%
\end{pgfscope}%
\begin{pgfscope}%
\pgfsetbuttcap%
\pgfsetroundjoin%
\pgfsetlinewidth{2.007500pt}%
\definecolor{currentstroke}{rgb}{1.000000,0.498039,0.054902}%
\pgfsetstrokecolor{currentstroke}%
\pgfsetdash{{7.400000pt}{3.200000pt}}{0.000000pt}%
\pgfpathmoveto{\pgfqpoint{2.673383in}{1.595917in}}%
\pgfpathlineto{\pgfqpoint{2.951161in}{1.595917in}}%
\pgfusepath{stroke}%
\end{pgfscope}%
\begin{pgfscope}%
\definecolor{textcolor}{rgb}{0.000000,0.000000,0.000000}%
\pgfsetstrokecolor{textcolor}%
\pgfsetfillcolor{textcolor}%
\pgftext[x=3.062272in,y=1.547306in,left,base]{\color{textcolor}\sffamily\fontsize{10.000000}{12.000000}\selectfont \(\displaystyle E_{k\downarrow}\)}%
\end{pgfscope}%
\end{pgfpicture}%
\makeatother%
\endgroup%

%% file: gap_Q_B.pgf
\begingroup%
\makeatletter%
\begin{pgfpicture}%
\pgfpathrectangle{\pgfpointorigin}{\pgfqpoint{3.890299in}{2.181779in}}%
\pgfusepath{use as bounding box, clip}%
\begin{pgfscope}%
\pgfsetbuttcap%
\pgfsetmiterjoin%
\definecolor{currentfill}{rgb}{1.000000,1.000000,1.000000}%
\pgfsetfillcolor{currentfill}%
\pgfsetlinewidth{0.000000pt}%
\definecolor{currentstroke}{rgb}{1.000000,1.000000,1.000000}%
\pgfsetstrokecolor{currentstroke}%
\pgfsetdash{}{0pt}%
\pgfpathmoveto{\pgfqpoint{0.000000in}{0.000000in}}%
\pgfpathlineto{\pgfqpoint{3.890299in}{0.000000in}}%
\pgfpathlineto{\pgfqpoint{3.890299in}{2.181779in}}%
\pgfpathlineto{\pgfqpoint{0.000000in}{2.181779in}}%
\pgfpathclose%
\pgfusepath{fill}%
\end{pgfscope}%
\begin{pgfscope}%
\pgfsetbuttcap%
\pgfsetmiterjoin%
\definecolor{currentfill}{rgb}{1.000000,1.000000,1.000000}%
\pgfsetfillcolor{currentfill}%
\pgfsetlinewidth{0.000000pt}%
\definecolor{currentstroke}{rgb}{0.000000,0.000000,0.000000}%
\pgfsetstrokecolor{currentstroke}%
\pgfsetstrokeopacity{0.000000}%
\pgfsetdash{}{0pt}%
\pgfpathmoveto{\pgfqpoint{0.369491in}{0.436091in}}%
\pgfpathlineto{\pgfqpoint{1.804307in}{0.436091in}}%
\pgfpathlineto{\pgfqpoint{1.804307in}{2.033168in}}%
\pgfpathlineto{\pgfqpoint{0.369491in}{2.033168in}}%
\pgfpathclose%
\pgfusepath{fill}%
\end{pgfscope}%
\begin{pgfscope}%
\pgfsetbuttcap%
\pgfsetroundjoin%
\definecolor{currentfill}{rgb}{0.000000,0.000000,0.000000}%
\pgfsetfillcolor{currentfill}%
\pgfsetlinewidth{0.803000pt}%
\definecolor{currentstroke}{rgb}{0.000000,0.000000,0.000000}%
\pgfsetstrokecolor{currentstroke}%
\pgfsetdash{}{0pt}%
\pgfsys@defobject{currentmarker}{\pgfqpoint{0.000000in}{-0.048611in}}{\pgfqpoint{0.000000in}{0.000000in}}{%
\pgfpathmoveto{\pgfqpoint{0.000000in}{0.000000in}}%
\pgfpathlineto{\pgfqpoint{0.000000in}{-0.048611in}}%
\pgfusepath{stroke,fill}%
}%
\begin{pgfscope}%
\pgfsys@transformshift{0.369491in}{0.436091in}%
\pgfsys@useobject{currentmarker}{}%
\end{pgfscope}%
\end{pgfscope}%
\begin{pgfscope}%
\definecolor{textcolor}{rgb}{0.000000,0.000000,0.000000}%
\pgfsetstrokecolor{textcolor}%
\pgfsetfillcolor{textcolor}%
\pgftext[x=0.220854in,y=0.301041in,left,base,rotate=325.000000]{\color{textcolor}\sffamily\fontsize{10.000000}{12.000000}\selectfont 0.00}%
\end{pgfscope}%
\begin{pgfscope}%
\pgfsetbuttcap%
\pgfsetroundjoin%
\definecolor{currentfill}{rgb}{0.000000,0.000000,0.000000}%
\pgfsetfillcolor{currentfill}%
\pgfsetlinewidth{0.803000pt}%
\definecolor{currentstroke}{rgb}{0.000000,0.000000,0.000000}%
\pgfsetstrokecolor{currentstroke}%
\pgfsetdash{}{0pt}%
\pgfsys@defobject{currentmarker}{\pgfqpoint{0.000000in}{-0.048611in}}{\pgfqpoint{0.000000in}{0.000000in}}{%
\pgfpathmoveto{\pgfqpoint{0.000000in}{0.000000in}}%
\pgfpathlineto{\pgfqpoint{0.000000in}{-0.048611in}}%
\pgfusepath{stroke,fill}%
}%
\begin{pgfscope}%
\pgfsys@transformshift{0.782000in}{0.436091in}%
\pgfsys@useobject{currentmarker}{}%
\end{pgfscope}%
\end{pgfscope}%
\begin{pgfscope}%
\definecolor{textcolor}{rgb}{0.000000,0.000000,0.000000}%
\pgfsetstrokecolor{textcolor}%
\pgfsetfillcolor{textcolor}%
\pgftext[x=0.633363in,y=0.301041in,left,base,rotate=325.000000]{\color{textcolor}\sffamily\fontsize{10.000000}{12.000000}\selectfont 0.23}%
\end{pgfscope}%
\begin{pgfscope}%
\pgfsetbuttcap%
\pgfsetroundjoin%
\definecolor{currentfill}{rgb}{0.000000,0.000000,0.000000}%
\pgfsetfillcolor{currentfill}%
\pgfsetlinewidth{0.803000pt}%
\definecolor{currentstroke}{rgb}{0.000000,0.000000,0.000000}%
\pgfsetstrokecolor{currentstroke}%
\pgfsetdash{}{0pt}%
\pgfsys@defobject{currentmarker}{\pgfqpoint{0.000000in}{-0.048611in}}{\pgfqpoint{0.000000in}{0.000000in}}{%
\pgfpathmoveto{\pgfqpoint{0.000000in}{0.000000in}}%
\pgfpathlineto{\pgfqpoint{0.000000in}{-0.048611in}}%
\pgfusepath{stroke,fill}%
}%
\begin{pgfscope}%
\pgfsys@transformshift{1.015158in}{0.436091in}%
\pgfsys@useobject{currentmarker}{}%
\end{pgfscope}%
\end{pgfscope}%
\begin{pgfscope}%
\definecolor{textcolor}{rgb}{0.000000,0.000000,0.000000}%
\pgfsetstrokecolor{textcolor}%
\pgfsetfillcolor{textcolor}%
\pgftext[x=0.866521in,y=0.301041in,left,base,rotate=325.000000]{\color{textcolor}\sffamily\fontsize{10.000000}{12.000000}\selectfont 0.36}%
\end{pgfscope}%
\begin{pgfscope}%
\pgfsetbuttcap%
\pgfsetroundjoin%
\definecolor{currentfill}{rgb}{0.000000,0.000000,0.000000}%
\pgfsetfillcolor{currentfill}%
\pgfsetlinewidth{0.803000pt}%
\definecolor{currentstroke}{rgb}{0.000000,0.000000,0.000000}%
\pgfsetstrokecolor{currentstroke}%
\pgfsetdash{}{0pt}%
\pgfsys@defobject{currentmarker}{\pgfqpoint{0.000000in}{-0.048611in}}{\pgfqpoint{0.000000in}{0.000000in}}{%
\pgfpathmoveto{\pgfqpoint{0.000000in}{0.000000in}}%
\pgfpathlineto{\pgfqpoint{0.000000in}{-0.048611in}}%
\pgfusepath{stroke,fill}%
}%
\begin{pgfscope}%
\pgfsys@transformshift{1.266251in}{0.436091in}%
\pgfsys@useobject{currentmarker}{}%
\end{pgfscope}%
\end{pgfscope}%
\begin{pgfscope}%
\definecolor{textcolor}{rgb}{0.000000,0.000000,0.000000}%
\pgfsetstrokecolor{textcolor}%
\pgfsetfillcolor{textcolor}%
\pgftext[x=1.117614in,y=0.301041in,left,base,rotate=325.000000]{\color{textcolor}\sffamily\fontsize{10.000000}{12.000000}\selectfont 0.50}%
\end{pgfscope}%
\begin{pgfscope}%
\definecolor{textcolor}{rgb}{0.000000,0.000000,0.000000}%
\pgfsetstrokecolor{textcolor}%
\pgfsetfillcolor{textcolor}%
\pgftext[x=1.660825in,y=0.308333in,,top]{\color{textcolor}\sffamily\fontsize{12.000000}{14.400000}\selectfont \(\displaystyle Q_\uparrow\Phi\)}%
\end{pgfscope}%
\begin{pgfscope}%
\pgfsetbuttcap%
\pgfsetroundjoin%
\definecolor{currentfill}{rgb}{0.000000,0.000000,0.000000}%
\pgfsetfillcolor{currentfill}%
\pgfsetlinewidth{0.803000pt}%
\definecolor{currentstroke}{rgb}{0.000000,0.000000,0.000000}%
\pgfsetstrokecolor{currentstroke}%
\pgfsetdash{}{0pt}%
\pgfsys@defobject{currentmarker}{\pgfqpoint{-0.048611in}{0.000000in}}{\pgfqpoint{0.000000in}{0.000000in}}{%
\pgfpathmoveto{\pgfqpoint{0.000000in}{0.000000in}}%
\pgfpathlineto{\pgfqpoint{-0.048611in}{0.000000in}}%
\pgfusepath{stroke,fill}%
}%
\begin{pgfscope}%
\pgfsys@transformshift{0.369491in}{0.436091in}%
\pgfsys@useobject{currentmarker}{}%
\end{pgfscope}%
\end{pgfscope}%
\begin{pgfscope}%
\pgfsetbuttcap%
\pgfsetroundjoin%
\definecolor{currentfill}{rgb}{0.000000,0.000000,0.000000}%
\pgfsetfillcolor{currentfill}%
\pgfsetlinewidth{0.803000pt}%
\definecolor{currentstroke}{rgb}{0.000000,0.000000,0.000000}%
\pgfsetstrokecolor{currentstroke}%
\pgfsetdash{}{0pt}%
\pgfsys@defobject{currentmarker}{\pgfqpoint{0.000000in}{0.000000in}}{\pgfqpoint{0.048611in}{0.000000in}}{%
\pgfpathmoveto{\pgfqpoint{0.000000in}{0.000000in}}%
\pgfpathlineto{\pgfqpoint{0.048611in}{0.000000in}}%
\pgfusepath{stroke,fill}%
}%
\begin{pgfscope}%
\pgfsys@transformshift{1.804307in}{0.436091in}%
\pgfsys@useobject{currentmarker}{}%
\end{pgfscope}%
\end{pgfscope}%
\begin{pgfscope}%
\definecolor{textcolor}{rgb}{0.000000,0.000000,0.000000}%
\pgfsetstrokecolor{textcolor}%
\pgfsetfillcolor{textcolor}%
\pgftext[x=0.100000in,y=0.383330in,left,base]{\color{textcolor}\sffamily\fontsize{10.000000}{12.000000}\selectfont 0.0}%
\end{pgfscope}%
\begin{pgfscope}%
\pgfsetbuttcap%
\pgfsetroundjoin%
\definecolor{currentfill}{rgb}{0.000000,0.000000,0.000000}%
\pgfsetfillcolor{currentfill}%
\pgfsetlinewidth{0.803000pt}%
\definecolor{currentstroke}{rgb}{0.000000,0.000000,0.000000}%
\pgfsetstrokecolor{currentstroke}%
\pgfsetdash{}{0pt}%
\pgfsys@defobject{currentmarker}{\pgfqpoint{-0.048611in}{0.000000in}}{\pgfqpoint{0.000000in}{0.000000in}}{%
\pgfpathmoveto{\pgfqpoint{0.000000in}{0.000000in}}%
\pgfpathlineto{\pgfqpoint{-0.048611in}{0.000000in}}%
\pgfusepath{stroke,fill}%
}%
\begin{pgfscope}%
\pgfsys@transformshift{0.369491in}{1.178918in}%
\pgfsys@useobject{currentmarker}{}%
\end{pgfscope}%
\end{pgfscope}%
\begin{pgfscope}%
\pgfsetbuttcap%
\pgfsetroundjoin%
\definecolor{currentfill}{rgb}{0.000000,0.000000,0.000000}%
\pgfsetfillcolor{currentfill}%
\pgfsetlinewidth{0.803000pt}%
\definecolor{currentstroke}{rgb}{0.000000,0.000000,0.000000}%
\pgfsetstrokecolor{currentstroke}%
\pgfsetdash{}{0pt}%
\pgfsys@defobject{currentmarker}{\pgfqpoint{0.000000in}{0.000000in}}{\pgfqpoint{0.048611in}{0.000000in}}{%
\pgfpathmoveto{\pgfqpoint{0.000000in}{0.000000in}}%
\pgfpathlineto{\pgfqpoint{0.048611in}{0.000000in}}%
\pgfusepath{stroke,fill}%
}%
\begin{pgfscope}%
\pgfsys@transformshift{1.804307in}{1.178918in}%
\pgfsys@useobject{currentmarker}{}%
\end{pgfscope}%
\end{pgfscope}%
\begin{pgfscope}%
\definecolor{textcolor}{rgb}{0.000000,0.000000,0.000000}%
\pgfsetstrokecolor{textcolor}%
\pgfsetfillcolor{textcolor}%
\pgftext[x=0.100000in,y=1.126156in,left,base]{\color{textcolor}\sffamily\fontsize{10.000000}{12.000000}\selectfont 0.2}%
\end{pgfscope}%
\begin{pgfscope}%
\pgfsetbuttcap%
\pgfsetroundjoin%
\definecolor{currentfill}{rgb}{0.000000,0.000000,0.000000}%
\pgfsetfillcolor{currentfill}%
\pgfsetlinewidth{0.803000pt}%
\definecolor{currentstroke}{rgb}{0.000000,0.000000,0.000000}%
\pgfsetstrokecolor{currentstroke}%
\pgfsetdash{}{0pt}%
\pgfsys@defobject{currentmarker}{\pgfqpoint{-0.048611in}{0.000000in}}{\pgfqpoint{0.000000in}{0.000000in}}{%
\pgfpathmoveto{\pgfqpoint{0.000000in}{0.000000in}}%
\pgfpathlineto{\pgfqpoint{-0.048611in}{0.000000in}}%
\pgfusepath{stroke,fill}%
}%
\begin{pgfscope}%
\pgfsys@transformshift{0.369491in}{1.921744in}%
\pgfsys@useobject{currentmarker}{}%
\end{pgfscope}%
\end{pgfscope}%
\begin{pgfscope}%
\pgfsetbuttcap%
\pgfsetroundjoin%
\definecolor{currentfill}{rgb}{0.000000,0.000000,0.000000}%
\pgfsetfillcolor{currentfill}%
\pgfsetlinewidth{0.803000pt}%
\definecolor{currentstroke}{rgb}{0.000000,0.000000,0.000000}%
\pgfsetstrokecolor{currentstroke}%
\pgfsetdash{}{0pt}%
\pgfsys@defobject{currentmarker}{\pgfqpoint{0.000000in}{0.000000in}}{\pgfqpoint{0.048611in}{0.000000in}}{%
\pgfpathmoveto{\pgfqpoint{0.000000in}{0.000000in}}%
\pgfpathlineto{\pgfqpoint{0.048611in}{0.000000in}}%
\pgfusepath{stroke,fill}%
}%
\begin{pgfscope}%
\pgfsys@transformshift{1.804307in}{1.921744in}%
\pgfsys@useobject{currentmarker}{}%
\end{pgfscope}%
\end{pgfscope}%
\begin{pgfscope}%
\definecolor{textcolor}{rgb}{0.000000,0.000000,0.000000}%
\pgfsetstrokecolor{textcolor}%
\pgfsetfillcolor{textcolor}%
\pgftext[x=0.100000in,y=1.868982in,left,base]{\color{textcolor}\sffamily\fontsize{10.000000}{12.000000}\selectfont 0.4}%
\end{pgfscope}%
\begin{pgfscope}%
\pgfsetbuttcap%
\pgfsetroundjoin%
\definecolor{currentfill}{rgb}{0.000000,0.000000,0.000000}%
\pgfsetfillcolor{currentfill}%
\pgfsetlinewidth{0.602250pt}%
\definecolor{currentstroke}{rgb}{0.000000,0.000000,0.000000}%
\pgfsetstrokecolor{currentstroke}%
\pgfsetdash{}{0pt}%
\pgfsys@defobject{currentmarker}{\pgfqpoint{-0.027778in}{0.000000in}}{\pgfqpoint{0.000000in}{0.000000in}}{%
\pgfpathmoveto{\pgfqpoint{0.000000in}{0.000000in}}%
\pgfpathlineto{\pgfqpoint{-0.027778in}{0.000000in}}%
\pgfusepath{stroke,fill}%
}%
\begin{pgfscope}%
\pgfsys@transformshift{0.369491in}{0.436091in}%
\pgfsys@useobject{currentmarker}{}%
\end{pgfscope}%
\end{pgfscope}%
\begin{pgfscope}%
\pgfsetbuttcap%
\pgfsetroundjoin%
\definecolor{currentfill}{rgb}{0.000000,0.000000,0.000000}%
\pgfsetfillcolor{currentfill}%
\pgfsetlinewidth{0.602250pt}%
\definecolor{currentstroke}{rgb}{0.000000,0.000000,0.000000}%
\pgfsetstrokecolor{currentstroke}%
\pgfsetdash{}{0pt}%
\pgfsys@defobject{currentmarker}{\pgfqpoint{-0.027778in}{0.000000in}}{\pgfqpoint{0.000000in}{0.000000in}}{%
\pgfpathmoveto{\pgfqpoint{0.000000in}{0.000000in}}%
\pgfpathlineto{\pgfqpoint{-0.027778in}{0.000000in}}%
\pgfusepath{stroke,fill}%
}%
\begin{pgfscope}%
\pgfsys@transformshift{0.369491in}{0.807504in}%
\pgfsys@useobject{currentmarker}{}%
\end{pgfscope}%
\end{pgfscope}%
\begin{pgfscope}%
\pgfsetbuttcap%
\pgfsetroundjoin%
\definecolor{currentfill}{rgb}{0.000000,0.000000,0.000000}%
\pgfsetfillcolor{currentfill}%
\pgfsetlinewidth{0.602250pt}%
\definecolor{currentstroke}{rgb}{0.000000,0.000000,0.000000}%
\pgfsetstrokecolor{currentstroke}%
\pgfsetdash{}{0pt}%
\pgfsys@defobject{currentmarker}{\pgfqpoint{-0.027778in}{0.000000in}}{\pgfqpoint{0.000000in}{0.000000in}}{%
\pgfpathmoveto{\pgfqpoint{0.000000in}{0.000000in}}%
\pgfpathlineto{\pgfqpoint{-0.027778in}{0.000000in}}%
\pgfusepath{stroke,fill}%
}%
\begin{pgfscope}%
\pgfsys@transformshift{0.369491in}{1.550331in}%
\pgfsys@useobject{currentmarker}{}%
\end{pgfscope}%
\end{pgfscope}%
\begin{pgfscope}%
\definecolor{textcolor}{rgb}{0.000000,0.000000,0.000000}%
\pgfsetstrokecolor{textcolor}%
\pgfsetfillcolor{textcolor}%
\pgftext[x=0.269053in,y=1.554045in,,bottom,rotate=90.000000]{\color{textcolor}\sffamily\fontsize{12.000000}{14.400000}\selectfont \(\displaystyle \Delta\)}%
\end{pgfscope}%
\begin{pgfscope}%
\pgfpathrectangle{\pgfqpoint{0.369491in}{0.436091in}}{\pgfqpoint{1.434816in}{1.597077in}}%
\pgfusepath{clip}%
\pgfsetrectcap%
\pgfsetroundjoin%
\pgfsetlinewidth{2.007500pt}%
\definecolor{currentstroke}{rgb}{0.121569,0.466667,0.705882}%
\pgfsetstrokecolor{currentstroke}%
\pgfsetdash{}{0pt}%
\pgfpathmoveto{\pgfqpoint{0.369491in}{1.831528in}}%
\pgfpathlineto{\pgfqpoint{1.818195in}{1.831528in}}%
\pgfpathlineto{\pgfqpoint{1.818195in}{1.831528in}}%
\pgfusepath{stroke}%
\end{pgfscope}%
\begin{pgfscope}%
\pgfpathrectangle{\pgfqpoint{0.369491in}{0.436091in}}{\pgfqpoint{1.434816in}{1.597077in}}%
\pgfusepath{clip}%
\pgfsetrectcap%
\pgfsetroundjoin%
\pgfsetlinewidth{2.007500pt}%
\definecolor{currentstroke}{rgb}{1.000000,0.498039,0.054902}%
\pgfsetstrokecolor{currentstroke}%
\pgfsetdash{}{0pt}%
\pgfpathmoveto{\pgfqpoint{0.369491in}{1.300122in}}%
\pgfpathlineto{\pgfqpoint{1.818195in}{1.300122in}}%
\pgfpathlineto{\pgfqpoint{1.818195in}{1.300122in}}%
\pgfusepath{stroke}%
\end{pgfscope}%
\begin{pgfscope}%
\pgfpathrectangle{\pgfqpoint{0.369491in}{0.436091in}}{\pgfqpoint{1.434816in}{1.597077in}}%
\pgfusepath{clip}%
\pgfsetbuttcap%
\pgfsetroundjoin%
\pgfsetlinewidth{3.011250pt}%
\definecolor{currentstroke}{rgb}{0.121569,0.466667,0.705882}%
\pgfsetstrokecolor{currentstroke}%
\pgfsetdash{{11.100000pt}{4.800000pt}}{0.000000pt}%
\pgfpathmoveto{\pgfqpoint{0.369491in}{1.831528in}}%
\pgfpathlineto{\pgfqpoint{1.015158in}{1.831233in}}%
\pgfpathlineto{\pgfqpoint{1.018745in}{0.436463in}}%
\pgfpathlineto{\pgfqpoint{1.018745in}{0.436463in}}%
\pgfusepath{stroke}%
\end{pgfscope}%
\begin{pgfscope}%
\pgfpathrectangle{\pgfqpoint{0.369491in}{0.436091in}}{\pgfqpoint{1.434816in}{1.597077in}}%
\pgfusepath{clip}%
\pgfsetbuttcap%
\pgfsetroundjoin%
\pgfsetlinewidth{3.011250pt}%
\definecolor{currentstroke}{rgb}{1.000000,0.498039,0.054902}%
\pgfsetstrokecolor{currentstroke}%
\pgfsetdash{{11.100000pt}{4.800000pt}}{0.000000pt}%
\pgfpathmoveto{\pgfqpoint{0.369491in}{1.300122in}}%
\pgfpathlineto{\pgfqpoint{0.378458in}{1.300122in}}%
\pgfpathlineto{\pgfqpoint{0.387426in}{1.300122in}}%
\pgfpathlineto{\pgfqpoint{0.396393in}{1.300122in}}%
\pgfpathlineto{\pgfqpoint{0.405361in}{1.300122in}}%
\pgfpathlineto{\pgfqpoint{0.414329in}{1.300122in}}%
\pgfpathlineto{\pgfqpoint{0.423296in}{1.300122in}}%
\pgfpathlineto{\pgfqpoint{0.432264in}{1.300122in}}%
\pgfpathlineto{\pgfqpoint{0.441231in}{1.300122in}}%
\pgfpathlineto{\pgfqpoint{0.450199in}{1.300122in}}%
\pgfpathlineto{\pgfqpoint{0.459167in}{1.300122in}}%
\pgfpathlineto{\pgfqpoint{0.468134in}{1.300122in}}%
\pgfpathlineto{\pgfqpoint{0.477102in}{1.300122in}}%
\pgfpathlineto{\pgfqpoint{0.486069in}{1.300122in}}%
\pgfpathlineto{\pgfqpoint{0.495037in}{1.300122in}}%
\pgfpathlineto{\pgfqpoint{0.504005in}{1.300122in}}%
\pgfpathlineto{\pgfqpoint{0.512972in}{1.300122in}}%
\pgfpathlineto{\pgfqpoint{0.521940in}{1.300122in}}%
\pgfpathlineto{\pgfqpoint{0.530907in}{1.300120in}}%
\pgfpathlineto{\pgfqpoint{0.539875in}{1.300122in}}%
\pgfpathlineto{\pgfqpoint{0.548843in}{1.300122in}}%
\pgfpathlineto{\pgfqpoint{0.557810in}{1.299916in}}%
\pgfpathlineto{\pgfqpoint{0.566778in}{1.300122in}}%
\pgfpathlineto{\pgfqpoint{0.575745in}{1.300074in}}%
\pgfpathlineto{\pgfqpoint{0.584713in}{1.300122in}}%
\pgfpathlineto{\pgfqpoint{0.593681in}{1.299802in}}%
\pgfpathlineto{\pgfqpoint{0.602648in}{1.299866in}}%
\pgfpathlineto{\pgfqpoint{0.611616in}{1.299890in}}%
\pgfpathlineto{\pgfqpoint{0.620583in}{1.299810in}}%
\pgfpathlineto{\pgfqpoint{0.629551in}{1.299984in}}%
\pgfpathlineto{\pgfqpoint{0.638519in}{1.300122in}}%
\pgfpathlineto{\pgfqpoint{0.647486in}{1.300122in}}%
\pgfpathlineto{\pgfqpoint{0.656454in}{1.300122in}}%
\pgfpathlineto{\pgfqpoint{0.665421in}{1.300122in}}%
\pgfpathlineto{\pgfqpoint{0.674389in}{1.300122in}}%
\pgfpathlineto{\pgfqpoint{0.683357in}{1.300122in}}%
\pgfpathlineto{\pgfqpoint{0.692324in}{1.300122in}}%
\pgfpathlineto{\pgfqpoint{0.701292in}{1.299814in}}%
\pgfpathlineto{\pgfqpoint{0.710259in}{1.300112in}}%
\pgfpathlineto{\pgfqpoint{0.719227in}{1.300122in}}%
\pgfpathlineto{\pgfqpoint{0.728195in}{1.300122in}}%
\pgfpathlineto{\pgfqpoint{0.737162in}{1.300122in}}%
\pgfpathlineto{\pgfqpoint{0.746130in}{1.300122in}}%
\pgfpathlineto{\pgfqpoint{0.755097in}{1.300122in}}%
\pgfpathlineto{\pgfqpoint{0.764065in}{1.300122in}}%
\pgfpathlineto{\pgfqpoint{0.773033in}{1.300122in}}%
\pgfpathlineto{\pgfqpoint{0.782000in}{0.436463in}}%
\pgfusepath{stroke}%
\end{pgfscope}%
\begin{pgfscope}%
\pgfsetrectcap%
\pgfsetmiterjoin%
\pgfsetlinewidth{0.803000pt}%
\definecolor{currentstroke}{rgb}{0.000000,0.000000,0.000000}%
\pgfsetstrokecolor{currentstroke}%
\pgfsetdash{}{0pt}%
\pgfpathmoveto{\pgfqpoint{0.369491in}{0.436091in}}%
\pgfpathlineto{\pgfqpoint{0.369491in}{2.033168in}}%
\pgfusepath{stroke}%
\end{pgfscope}%
\begin{pgfscope}%
\pgfsetrectcap%
\pgfsetmiterjoin%
\pgfsetlinewidth{0.803000pt}%
\definecolor{currentstroke}{rgb}{0.000000,0.000000,0.000000}%
\pgfsetstrokecolor{currentstroke}%
\pgfsetdash{}{0pt}%
\pgfpathmoveto{\pgfqpoint{1.804307in}{0.436091in}}%
\pgfpathlineto{\pgfqpoint{1.804307in}{2.033168in}}%
\pgfusepath{stroke}%
\end{pgfscope}%
\begin{pgfscope}%
\pgfsetrectcap%
\pgfsetmiterjoin%
\pgfsetlinewidth{0.803000pt}%
\definecolor{currentstroke}{rgb}{0.000000,0.000000,0.000000}%
\pgfsetstrokecolor{currentstroke}%
\pgfsetdash{}{0pt}%
\pgfpathmoveto{\pgfqpoint{0.369491in}{0.436091in}}%
\pgfpathlineto{\pgfqpoint{1.804307in}{0.436091in}}%
\pgfusepath{stroke}%
\end{pgfscope}%
\begin{pgfscope}%
\pgfsetrectcap%
\pgfsetmiterjoin%
\pgfsetlinewidth{0.803000pt}%
\definecolor{currentstroke}{rgb}{0.000000,0.000000,0.000000}%
\pgfsetstrokecolor{currentstroke}%
\pgfsetdash{}{0pt}%
\pgfpathmoveto{\pgfqpoint{0.369491in}{2.033168in}}%
\pgfpathlineto{\pgfqpoint{1.804307in}{2.033168in}}%
\pgfusepath{stroke}%
\end{pgfscope}%
\begin{pgfscope}%
\definecolor{textcolor}{rgb}{0.000000,0.000000,0.000000}%
\pgfsetstrokecolor{textcolor}%
\pgfsetfillcolor{textcolor}%
\pgftext[x=0.412535in,y=0.515945in,left,base]{\color{textcolor}\sffamily\fontsize{12.000000}{14.400000}\selectfont (a)}%
\end{pgfscope}%
\begin{pgfscope}%
\definecolor{textcolor}{rgb}{0.000000,0.000000,0.000000}%
\pgfsetstrokecolor{textcolor}%
\pgfsetfillcolor{textcolor}%
\pgftext[x=0.369491in,y=1.364624in,left,base]{\color{textcolor}\sffamily\fontsize{10.000000}{12.000000}\selectfont \(\displaystyle 1/g\!=\!\!-3.5\)}%
\end{pgfscope}%
\begin{pgfscope}%
\definecolor{textcolor}{rgb}{0.000000,0.000000,0.000000}%
\pgfsetstrokecolor{textcolor}%
\pgfsetfillcolor{textcolor}%
\pgftext[x=0.369491in,y=1.884603in,left,base]{\color{textcolor}\sffamily\fontsize{10.000000}{12.000000}\selectfont \(\displaystyle 1/g\!=\!\!-3\)}%
\end{pgfscope}%
\begin{pgfscope}%
\definecolor{textcolor}{rgb}{0.000000,0.000000,0.000000}%
\pgfsetstrokecolor{textcolor}%
\pgfsetfillcolor{textcolor}%
\pgftext[x=1.086899in,y=1.884603in,left,base]{\color{textcolor}\sffamily\fontsize{10.000000}{12.000000}\selectfont \(\displaystyle \kappa\!=\!2Q_+\Phi\)}%
\end{pgfscope}%
\begin{pgfscope}%
\pgfsetbuttcap%
\pgfsetmiterjoin%
\definecolor{currentfill}{rgb}{1.000000,1.000000,1.000000}%
\pgfsetfillcolor{currentfill}%
\pgfsetfillopacity{0.800000}%
\pgfsetlinewidth{1.003750pt}%
\definecolor{currentstroke}{rgb}{0.800000,0.800000,0.800000}%
\pgfsetstrokecolor{currentstroke}%
\pgfsetstrokeopacity{0.800000}%
\pgfsetdash{}{0pt}%
\pgfpathmoveto{\pgfqpoint{1.001628in}{0.491647in}}%
\pgfpathlineto{\pgfqpoint{1.726529in}{0.491647in}}%
\pgfpathquadraticcurveto{\pgfqpoint{1.748751in}{0.491647in}}{\pgfqpoint{1.748751in}{0.513869in}}%
\pgfpathlineto{\pgfqpoint{1.748751in}{0.838261in}}%
\pgfpathquadraticcurveto{\pgfqpoint{1.748751in}{0.860484in}}{\pgfqpoint{1.726529in}{0.860484in}}%
\pgfpathlineto{\pgfqpoint{1.001628in}{0.860484in}}%
\pgfpathquadraticcurveto{\pgfqpoint{0.979406in}{0.860484in}}{\pgfqpoint{0.979406in}{0.838261in}}%
\pgfpathlineto{\pgfqpoint{0.979406in}{0.513869in}}%
\pgfpathquadraticcurveto{\pgfqpoint{0.979406in}{0.491647in}}{\pgfqpoint{1.001628in}{0.491647in}}%
\pgfpathclose%
\pgfusepath{stroke,fill}%
\end{pgfscope}%
\begin{pgfscope}%
\pgfsetrectcap%
\pgfsetroundjoin%
\pgfsetlinewidth{2.007500pt}%
\definecolor{currentstroke}{rgb}{0.121569,0.466667,0.705882}%
\pgfsetstrokecolor{currentstroke}%
\pgfsetdash{}{0pt}%
\pgfpathmoveto{\pgfqpoint{1.023850in}{0.770510in}}%
\pgfpathlineto{\pgfqpoint{1.357184in}{0.770510in}}%
\pgfusepath{stroke}%
\end{pgfscope}%
\begin{pgfscope}%
\definecolor{textcolor}{rgb}{0.000000,0.000000,0.000000}%
\pgfsetstrokecolor{textcolor}%
\pgfsetfillcolor{textcolor}%
\pgftext[x=1.446072in,y=0.731621in,left,base]{\color{textcolor}\sffamily\fontsize{8.000000}{9.600000}\selectfont \(\displaystyle \Delta(\kappa)\)}%
\end{pgfscope}%
\begin{pgfscope}%
\pgfsetbuttcap%
\pgfsetroundjoin%
\pgfsetlinewidth{3.011250pt}%
\definecolor{currentstroke}{rgb}{0.121569,0.466667,0.705882}%
\pgfsetstrokecolor{currentstroke}%
\pgfsetdash{{11.100000pt}{4.800000pt}}{0.000000pt}%
\pgfpathmoveto{\pgfqpoint{1.023850in}{0.602758in}}%
\pgfpathlineto{\pgfqpoint{1.357184in}{0.602758in}}%
\pgfusepath{stroke}%
\end{pgfscope}%
\begin{pgfscope}%
\definecolor{textcolor}{rgb}{0.000000,0.000000,0.000000}%
\pgfsetstrokecolor{textcolor}%
\pgfsetfillcolor{textcolor}%
\pgftext[x=1.446072in,y=0.563869in,left,base]{\color{textcolor}\sffamily\fontsize{8.000000}{9.600000}\selectfont \(\displaystyle \Delta(0)\)}%
\end{pgfscope}%
\begin{pgfscope}%
\pgfsetbuttcap%
\pgfsetmiterjoin%
\definecolor{currentfill}{rgb}{1.000000,1.000000,1.000000}%
\pgfsetfillcolor{currentfill}%
\pgfsetlinewidth{0.000000pt}%
\definecolor{currentstroke}{rgb}{0.000000,0.000000,0.000000}%
\pgfsetstrokecolor{currentstroke}%
\pgfsetstrokeopacity{0.000000}%
\pgfsetdash{}{0pt}%
\pgfpathmoveto{\pgfqpoint{2.306872in}{0.436091in}}%
\pgfpathlineto{\pgfqpoint{3.741688in}{0.436091in}}%
\pgfpathlineto{\pgfqpoint{3.741688in}{2.033168in}}%
\pgfpathlineto{\pgfqpoint{2.306872in}{2.033168in}}%
\pgfpathclose%
\pgfusepath{fill}%
\end{pgfscope}%
\begin{pgfscope}%
\pgfsetbuttcap%
\pgfsetroundjoin%
\definecolor{currentfill}{rgb}{0.000000,0.000000,0.000000}%
\pgfsetfillcolor{currentfill}%
\pgfsetlinewidth{0.803000pt}%
\definecolor{currentstroke}{rgb}{0.000000,0.000000,0.000000}%
\pgfsetstrokecolor{currentstroke}%
\pgfsetdash{}{0pt}%
\pgfsys@defobject{currentmarker}{\pgfqpoint{0.000000in}{-0.048611in}}{\pgfqpoint{0.000000in}{0.000000in}}{%
\pgfpathmoveto{\pgfqpoint{0.000000in}{0.000000in}}%
\pgfpathlineto{\pgfqpoint{0.000000in}{-0.048611in}}%
\pgfusepath{stroke,fill}%
}%
\begin{pgfscope}%
\pgfsys@transformshift{2.306872in}{0.436091in}%
\pgfsys@useobject{currentmarker}{}%
\end{pgfscope}%
\end{pgfscope}%
\begin{pgfscope}%
\definecolor{textcolor}{rgb}{0.000000,0.000000,0.000000}%
\pgfsetstrokecolor{textcolor}%
\pgfsetfillcolor{textcolor}%
\pgftext[x=2.306872in,y=0.338869in,,top]{\color{textcolor}\sffamily\fontsize{10.000000}{12.000000}\selectfont −1}%
\end{pgfscope}%
\begin{pgfscope}%
\pgfsetbuttcap%
\pgfsetroundjoin%
\definecolor{currentfill}{rgb}{0.000000,0.000000,0.000000}%
\pgfsetfillcolor{currentfill}%
\pgfsetlinewidth{0.803000pt}%
\definecolor{currentstroke}{rgb}{0.000000,0.000000,0.000000}%
\pgfsetstrokecolor{currentstroke}%
\pgfsetdash{}{0pt}%
\pgfsys@defobject{currentmarker}{\pgfqpoint{0.000000in}{-0.048611in}}{\pgfqpoint{0.000000in}{0.000000in}}{%
\pgfpathmoveto{\pgfqpoint{0.000000in}{0.000000in}}%
\pgfpathlineto{\pgfqpoint{0.000000in}{-0.048611in}}%
\pgfusepath{stroke,fill}%
}%
\begin{pgfscope}%
\pgfsys@transformshift{3.024280in}{0.436091in}%
\pgfsys@useobject{currentmarker}{}%
\end{pgfscope}%
\end{pgfscope}%
\begin{pgfscope}%
\definecolor{textcolor}{rgb}{0.000000,0.000000,0.000000}%
\pgfsetstrokecolor{textcolor}%
\pgfsetfillcolor{textcolor}%
\pgftext[x=3.024280in,y=0.338869in,,top]{\color{textcolor}\sffamily\fontsize{10.000000}{12.000000}\selectfont 0}%
\end{pgfscope}%
\begin{pgfscope}%
\pgfsetbuttcap%
\pgfsetroundjoin%
\definecolor{currentfill}{rgb}{0.000000,0.000000,0.000000}%
\pgfsetfillcolor{currentfill}%
\pgfsetlinewidth{0.803000pt}%
\definecolor{currentstroke}{rgb}{0.000000,0.000000,0.000000}%
\pgfsetstrokecolor{currentstroke}%
\pgfsetdash{}{0pt}%
\pgfsys@defobject{currentmarker}{\pgfqpoint{0.000000in}{-0.048611in}}{\pgfqpoint{0.000000in}{0.000000in}}{%
\pgfpathmoveto{\pgfqpoint{0.000000in}{0.000000in}}%
\pgfpathlineto{\pgfqpoint{0.000000in}{-0.048611in}}%
\pgfusepath{stroke,fill}%
}%
\begin{pgfscope}%
\pgfsys@transformshift{3.741688in}{0.436091in}%
\pgfsys@useobject{currentmarker}{}%
\end{pgfscope}%
\end{pgfscope}%
\begin{pgfscope}%
\definecolor{textcolor}{rgb}{0.000000,0.000000,0.000000}%
\pgfsetstrokecolor{textcolor}%
\pgfsetfillcolor{textcolor}%
\pgftext[x=3.741688in,y=0.338869in,,top]{\color{textcolor}\sffamily\fontsize{10.000000}{12.000000}\selectfont 1}%
\end{pgfscope}%
\begin{pgfscope}%
\pgfsetbuttcap%
\pgfsetroundjoin%
\definecolor{currentfill}{rgb}{0.000000,0.000000,0.000000}%
\pgfsetfillcolor{currentfill}%
\pgfsetlinewidth{0.602250pt}%
\definecolor{currentstroke}{rgb}{0.000000,0.000000,0.000000}%
\pgfsetstrokecolor{currentstroke}%
\pgfsetdash{}{0pt}%
\pgfsys@defobject{currentmarker}{\pgfqpoint{0.000000in}{-0.027778in}}{\pgfqpoint{0.000000in}{0.000000in}}{%
\pgfpathmoveto{\pgfqpoint{0.000000in}{0.000000in}}%
\pgfpathlineto{\pgfqpoint{0.000000in}{-0.027778in}}%
\pgfusepath{stroke,fill}%
}%
\begin{pgfscope}%
\pgfsys@transformshift{2.665576in}{0.436091in}%
\pgfsys@useobject{currentmarker}{}%
\end{pgfscope}%
\end{pgfscope}%
\begin{pgfscope}%
\pgfsetbuttcap%
\pgfsetroundjoin%
\definecolor{currentfill}{rgb}{0.000000,0.000000,0.000000}%
\pgfsetfillcolor{currentfill}%
\pgfsetlinewidth{0.602250pt}%
\definecolor{currentstroke}{rgb}{0.000000,0.000000,0.000000}%
\pgfsetstrokecolor{currentstroke}%
\pgfsetdash{}{0pt}%
\pgfsys@defobject{currentmarker}{\pgfqpoint{0.000000in}{-0.027778in}}{\pgfqpoint{0.000000in}{0.000000in}}{%
\pgfpathmoveto{\pgfqpoint{0.000000in}{0.000000in}}%
\pgfpathlineto{\pgfqpoint{0.000000in}{-0.027778in}}%
\pgfusepath{stroke,fill}%
}%
\begin{pgfscope}%
\pgfsys@transformshift{3.382984in}{0.436091in}%
\pgfsys@useobject{currentmarker}{}%
\end{pgfscope}%
\end{pgfscope}%
\begin{pgfscope}%
\definecolor{textcolor}{rgb}{0.000000,0.000000,0.000000}%
\pgfsetstrokecolor{textcolor}%
\pgfsetfillcolor{textcolor}%
\pgftext[x=3.454724in,y=0.329456in,,top]{\color{textcolor}\sffamily\fontsize{12.000000}{14.400000}\selectfont \(\displaystyle Q_\downarrow/Q_\uparrow\)}%
\end{pgfscope}%
\begin{pgfscope}%
\pgfsetbuttcap%
\pgfsetroundjoin%
\definecolor{currentfill}{rgb}{0.000000,0.000000,0.000000}%
\pgfsetfillcolor{currentfill}%
\pgfsetlinewidth{0.803000pt}%
\definecolor{currentstroke}{rgb}{0.000000,0.000000,0.000000}%
\pgfsetstrokecolor{currentstroke}%
\pgfsetdash{}{0pt}%
\pgfsys@defobject{currentmarker}{\pgfqpoint{-0.048611in}{0.000000in}}{\pgfqpoint{0.000000in}{0.000000in}}{%
\pgfpathmoveto{\pgfqpoint{0.000000in}{0.000000in}}%
\pgfpathlineto{\pgfqpoint{-0.048611in}{0.000000in}}%
\pgfusepath{stroke,fill}%
}%
\begin{pgfscope}%
\pgfsys@transformshift{2.306872in}{0.436091in}%
\pgfsys@useobject{currentmarker}{}%
\end{pgfscope}%
\end{pgfscope}%
\begin{pgfscope}%
\pgfsetbuttcap%
\pgfsetroundjoin%
\definecolor{currentfill}{rgb}{0.000000,0.000000,0.000000}%
\pgfsetfillcolor{currentfill}%
\pgfsetlinewidth{0.803000pt}%
\definecolor{currentstroke}{rgb}{0.000000,0.000000,0.000000}%
\pgfsetstrokecolor{currentstroke}%
\pgfsetdash{}{0pt}%
\pgfsys@defobject{currentmarker}{\pgfqpoint{0.000000in}{0.000000in}}{\pgfqpoint{0.048611in}{0.000000in}}{%
\pgfpathmoveto{\pgfqpoint{0.000000in}{0.000000in}}%
\pgfpathlineto{\pgfqpoint{0.048611in}{0.000000in}}%
\pgfusepath{stroke,fill}%
}%
\begin{pgfscope}%
\pgfsys@transformshift{3.741688in}{0.436091in}%
\pgfsys@useobject{currentmarker}{}%
\end{pgfscope}%
\end{pgfscope}%
\begin{pgfscope}%
\definecolor{textcolor}{rgb}{0.000000,0.000000,0.000000}%
\pgfsetstrokecolor{textcolor}%
\pgfsetfillcolor{textcolor}%
\pgftext[x=2.037381in,y=0.383330in,left,base]{\color{textcolor}\sffamily\fontsize{10.000000}{12.000000}\selectfont 0.0}%
\end{pgfscope}%
\begin{pgfscope}%
\pgfsetbuttcap%
\pgfsetroundjoin%
\definecolor{currentfill}{rgb}{0.000000,0.000000,0.000000}%
\pgfsetfillcolor{currentfill}%
\pgfsetlinewidth{0.803000pt}%
\definecolor{currentstroke}{rgb}{0.000000,0.000000,0.000000}%
\pgfsetstrokecolor{currentstroke}%
\pgfsetdash{}{0pt}%
\pgfsys@defobject{currentmarker}{\pgfqpoint{-0.048611in}{0.000000in}}{\pgfqpoint{0.000000in}{0.000000in}}{%
\pgfpathmoveto{\pgfqpoint{0.000000in}{0.000000in}}%
\pgfpathlineto{\pgfqpoint{-0.048611in}{0.000000in}}%
\pgfusepath{stroke,fill}%
}%
\begin{pgfscope}%
\pgfsys@transformshift{2.306872in}{1.178918in}%
\pgfsys@useobject{currentmarker}{}%
\end{pgfscope}%
\end{pgfscope}%
\begin{pgfscope}%
\pgfsetbuttcap%
\pgfsetroundjoin%
\definecolor{currentfill}{rgb}{0.000000,0.000000,0.000000}%
\pgfsetfillcolor{currentfill}%
\pgfsetlinewidth{0.803000pt}%
\definecolor{currentstroke}{rgb}{0.000000,0.000000,0.000000}%
\pgfsetstrokecolor{currentstroke}%
\pgfsetdash{}{0pt}%
\pgfsys@defobject{currentmarker}{\pgfqpoint{0.000000in}{0.000000in}}{\pgfqpoint{0.048611in}{0.000000in}}{%
\pgfpathmoveto{\pgfqpoint{0.000000in}{0.000000in}}%
\pgfpathlineto{\pgfqpoint{0.048611in}{0.000000in}}%
\pgfusepath{stroke,fill}%
}%
\begin{pgfscope}%
\pgfsys@transformshift{3.741688in}{1.178918in}%
\pgfsys@useobject{currentmarker}{}%
\end{pgfscope}%
\end{pgfscope}%
\begin{pgfscope}%
\definecolor{textcolor}{rgb}{0.000000,0.000000,0.000000}%
\pgfsetstrokecolor{textcolor}%
\pgfsetfillcolor{textcolor}%
\pgftext[x=2.037381in,y=1.126156in,left,base]{\color{textcolor}\sffamily\fontsize{10.000000}{12.000000}\selectfont 0.2}%
\end{pgfscope}%
\begin{pgfscope}%
\pgfsetbuttcap%
\pgfsetroundjoin%
\definecolor{currentfill}{rgb}{0.000000,0.000000,0.000000}%
\pgfsetfillcolor{currentfill}%
\pgfsetlinewidth{0.803000pt}%
\definecolor{currentstroke}{rgb}{0.000000,0.000000,0.000000}%
\pgfsetstrokecolor{currentstroke}%
\pgfsetdash{}{0pt}%
\pgfsys@defobject{currentmarker}{\pgfqpoint{-0.048611in}{0.000000in}}{\pgfqpoint{0.000000in}{0.000000in}}{%
\pgfpathmoveto{\pgfqpoint{0.000000in}{0.000000in}}%
\pgfpathlineto{\pgfqpoint{-0.048611in}{0.000000in}}%
\pgfusepath{stroke,fill}%
}%
\begin{pgfscope}%
\pgfsys@transformshift{2.306872in}{1.921744in}%
\pgfsys@useobject{currentmarker}{}%
\end{pgfscope}%
\end{pgfscope}%
\begin{pgfscope}%
\pgfsetbuttcap%
\pgfsetroundjoin%
\definecolor{currentfill}{rgb}{0.000000,0.000000,0.000000}%
\pgfsetfillcolor{currentfill}%
\pgfsetlinewidth{0.803000pt}%
\definecolor{currentstroke}{rgb}{0.000000,0.000000,0.000000}%
\pgfsetstrokecolor{currentstroke}%
\pgfsetdash{}{0pt}%
\pgfsys@defobject{currentmarker}{\pgfqpoint{0.000000in}{0.000000in}}{\pgfqpoint{0.048611in}{0.000000in}}{%
\pgfpathmoveto{\pgfqpoint{0.000000in}{0.000000in}}%
\pgfpathlineto{\pgfqpoint{0.048611in}{0.000000in}}%
\pgfusepath{stroke,fill}%
}%
\begin{pgfscope}%
\pgfsys@transformshift{3.741688in}{1.921744in}%
\pgfsys@useobject{currentmarker}{}%
\end{pgfscope}%
\end{pgfscope}%
\begin{pgfscope}%
\definecolor{textcolor}{rgb}{0.000000,0.000000,0.000000}%
\pgfsetstrokecolor{textcolor}%
\pgfsetfillcolor{textcolor}%
\pgftext[x=2.037381in,y=1.868982in,left,base]{\color{textcolor}\sffamily\fontsize{10.000000}{12.000000}\selectfont 0.4}%
\end{pgfscope}%
\begin{pgfscope}%
\pgfsetbuttcap%
\pgfsetroundjoin%
\definecolor{currentfill}{rgb}{0.000000,0.000000,0.000000}%
\pgfsetfillcolor{currentfill}%
\pgfsetlinewidth{0.602250pt}%
\definecolor{currentstroke}{rgb}{0.000000,0.000000,0.000000}%
\pgfsetstrokecolor{currentstroke}%
\pgfsetdash{}{0pt}%
\pgfsys@defobject{currentmarker}{\pgfqpoint{-0.027778in}{0.000000in}}{\pgfqpoint{0.000000in}{0.000000in}}{%
\pgfpathmoveto{\pgfqpoint{0.000000in}{0.000000in}}%
\pgfpathlineto{\pgfqpoint{-0.027778in}{0.000000in}}%
\pgfusepath{stroke,fill}%
}%
\begin{pgfscope}%
\pgfsys@transformshift{2.306872in}{0.807504in}%
\pgfsys@useobject{currentmarker}{}%
\end{pgfscope}%
\end{pgfscope}%
\begin{pgfscope}%
\pgfsetbuttcap%
\pgfsetroundjoin%
\definecolor{currentfill}{rgb}{0.000000,0.000000,0.000000}%
\pgfsetfillcolor{currentfill}%
\pgfsetlinewidth{0.602250pt}%
\definecolor{currentstroke}{rgb}{0.000000,0.000000,0.000000}%
\pgfsetstrokecolor{currentstroke}%
\pgfsetdash{}{0pt}%
\pgfsys@defobject{currentmarker}{\pgfqpoint{-0.027778in}{0.000000in}}{\pgfqpoint{0.000000in}{0.000000in}}{%
\pgfpathmoveto{\pgfqpoint{0.000000in}{0.000000in}}%
\pgfpathlineto{\pgfqpoint{-0.027778in}{0.000000in}}%
\pgfusepath{stroke,fill}%
}%
\begin{pgfscope}%
\pgfsys@transformshift{2.306872in}{1.550331in}%
\pgfsys@useobject{currentmarker}{}%
\end{pgfscope}%
\end{pgfscope}%
\begin{pgfscope}%
\definecolor{textcolor}{rgb}{0.000000,0.000000,0.000000}%
\pgfsetstrokecolor{textcolor}%
\pgfsetfillcolor{textcolor}%
\pgftext[x=2.206434in,y=1.554045in,,bottom,rotate=90.000000]{\color{textcolor}\sffamily\fontsize{12.000000}{14.400000}\selectfont \(\displaystyle \Delta\)}%
\end{pgfscope}%
\begin{pgfscope}%
\pgfpathrectangle{\pgfqpoint{2.306872in}{0.436091in}}{\pgfqpoint{1.434816in}{1.597077in}}%
\pgfusepath{clip}%
\pgfsetrectcap%
\pgfsetroundjoin%
\pgfsetlinewidth{2.007500pt}%
\definecolor{currentstroke}{rgb}{0.121569,0.466667,0.705882}%
\pgfsetstrokecolor{currentstroke}%
\pgfsetdash{}{0pt}%
\pgfpathmoveto{\pgfqpoint{2.306872in}{1.831528in}}%
\pgfpathlineto{\pgfqpoint{2.335568in}{1.831528in}}%
\pgfpathlineto{\pgfqpoint{2.364264in}{1.831528in}}%
\pgfpathlineto{\pgfqpoint{2.392961in}{1.831528in}}%
\pgfpathlineto{\pgfqpoint{2.421657in}{1.831528in}}%
\pgfpathlineto{\pgfqpoint{2.450353in}{1.831528in}}%
\pgfpathlineto{\pgfqpoint{2.479049in}{1.831528in}}%
\pgfpathlineto{\pgfqpoint{2.507746in}{1.831528in}}%
\pgfpathlineto{\pgfqpoint{2.536442in}{1.831528in}}%
\pgfpathlineto{\pgfqpoint{2.565138in}{1.831528in}}%
\pgfpathlineto{\pgfqpoint{2.593835in}{1.831528in}}%
\pgfpathlineto{\pgfqpoint{2.622531in}{1.831528in}}%
\pgfpathlineto{\pgfqpoint{2.651227in}{1.831528in}}%
\pgfpathlineto{\pgfqpoint{2.679924in}{1.831528in}}%
\pgfpathlineto{\pgfqpoint{2.708620in}{1.831528in}}%
\pgfpathlineto{\pgfqpoint{2.737316in}{1.831528in}}%
\pgfpathlineto{\pgfqpoint{2.766013in}{1.831528in}}%
\pgfpathlineto{\pgfqpoint{2.794709in}{1.831528in}}%
\pgfpathlineto{\pgfqpoint{2.823405in}{1.831528in}}%
\pgfpathlineto{\pgfqpoint{2.852102in}{1.831528in}}%
\pgfpathlineto{\pgfqpoint{2.880798in}{1.831528in}}%
\pgfpathlineto{\pgfqpoint{2.909494in}{1.831528in}}%
\pgfpathlineto{\pgfqpoint{2.938191in}{1.831528in}}%
\pgfpathlineto{\pgfqpoint{2.966887in}{1.831528in}}%
\pgfpathlineto{\pgfqpoint{2.995583in}{1.831528in}}%
\pgfpathlineto{\pgfqpoint{3.024280in}{1.831528in}}%
\pgfpathlineto{\pgfqpoint{3.052976in}{1.831528in}}%
\pgfpathlineto{\pgfqpoint{3.081672in}{1.831528in}}%
\pgfpathlineto{\pgfqpoint{3.110369in}{1.831528in}}%
\pgfpathlineto{\pgfqpoint{3.139065in}{1.831528in}}%
\pgfpathlineto{\pgfqpoint{3.167761in}{1.831528in}}%
\pgfpathlineto{\pgfqpoint{3.196457in}{1.831528in}}%
\pgfpathlineto{\pgfqpoint{3.225154in}{1.831528in}}%
\pgfpathlineto{\pgfqpoint{3.253850in}{1.831528in}}%
\pgfpathlineto{\pgfqpoint{3.282546in}{1.831528in}}%
\pgfpathlineto{\pgfqpoint{3.311243in}{1.831528in}}%
\pgfpathlineto{\pgfqpoint{3.339939in}{1.831528in}}%
\pgfpathlineto{\pgfqpoint{3.368635in}{1.831528in}}%
\pgfpathlineto{\pgfqpoint{3.397332in}{1.831528in}}%
\pgfpathlineto{\pgfqpoint{3.426028in}{1.831528in}}%
\pgfpathlineto{\pgfqpoint{3.454724in}{1.831528in}}%
\pgfpathlineto{\pgfqpoint{3.483421in}{1.831528in}}%
\pgfpathlineto{\pgfqpoint{3.512117in}{1.831528in}}%
\pgfpathlineto{\pgfqpoint{3.540813in}{1.831528in}}%
\pgfpathlineto{\pgfqpoint{3.569510in}{1.831528in}}%
\pgfpathlineto{\pgfqpoint{3.598206in}{1.831528in}}%
\pgfpathlineto{\pgfqpoint{3.626902in}{1.831528in}}%
\pgfpathlineto{\pgfqpoint{3.655599in}{1.831528in}}%
\pgfpathlineto{\pgfqpoint{3.684295in}{1.831528in}}%
\pgfpathlineto{\pgfqpoint{3.712991in}{1.831528in}}%
\pgfpathlineto{\pgfqpoint{3.741688in}{1.831528in}}%
\pgfusepath{stroke}%
\end{pgfscope}%
\begin{pgfscope}%
\pgfpathrectangle{\pgfqpoint{2.306872in}{0.436091in}}{\pgfqpoint{1.434816in}{1.597077in}}%
\pgfusepath{clip}%
\pgfsetbuttcap%
\pgfsetroundjoin%
\pgfsetlinewidth{3.011250pt}%
\definecolor{currentstroke}{rgb}{0.121569,0.466667,0.705882}%
\pgfsetstrokecolor{currentstroke}%
\pgfsetdash{{11.100000pt}{4.800000pt}}{0.000000pt}%
\pgfpathmoveto{\pgfqpoint{2.306872in}{1.831528in}}%
\pgfpathlineto{\pgfqpoint{2.992714in}{1.830390in}}%
\pgfpathlineto{\pgfqpoint{2.995583in}{1.829665in}}%
\pgfpathlineto{\pgfqpoint{3.001322in}{1.831528in}}%
\pgfpathlineto{\pgfqpoint{3.024280in}{1.831492in}}%
\pgfpathlineto{\pgfqpoint{3.027149in}{0.436463in}}%
\pgfpathlineto{\pgfqpoint{3.027149in}{0.436463in}}%
\pgfusepath{stroke}%
\end{pgfscope}%
\begin{pgfscope}%
\pgfpathrectangle{\pgfqpoint{2.306872in}{0.436091in}}{\pgfqpoint{1.434816in}{1.597077in}}%
\pgfusepath{clip}%
\pgfsetbuttcap%
\pgfsetroundjoin%
\pgfsetlinewidth{3.011250pt}%
\definecolor{currentstroke}{rgb}{0.172549,0.627451,0.172549}%
\pgfsetstrokecolor{currentstroke}%
\pgfsetdash{{11.100000pt}{4.800000pt}}{0.000000pt}%
\pgfpathmoveto{\pgfqpoint{2.306872in}{1.831528in}}%
\pgfpathlineto{\pgfqpoint{2.322655in}{1.831528in}}%
\pgfpathlineto{\pgfqpoint{2.338438in}{1.831528in}}%
\pgfpathlineto{\pgfqpoint{2.354220in}{1.831528in}}%
\pgfpathlineto{\pgfqpoint{2.370003in}{1.831528in}}%
\pgfpathlineto{\pgfqpoint{2.385786in}{1.831528in}}%
\pgfpathlineto{\pgfqpoint{2.401569in}{1.831528in}}%
\pgfpathlineto{\pgfqpoint{2.417352in}{1.831528in}}%
\pgfpathlineto{\pgfqpoint{2.433135in}{1.831528in}}%
\pgfpathlineto{\pgfqpoint{2.448918in}{1.831528in}}%
\pgfpathlineto{\pgfqpoint{2.464701in}{1.831528in}}%
\pgfpathlineto{\pgfqpoint{2.480484in}{1.831528in}}%
\pgfpathlineto{\pgfqpoint{2.496267in}{1.831528in}}%
\pgfpathlineto{\pgfqpoint{2.512050in}{1.831528in}}%
\pgfpathlineto{\pgfqpoint{2.527833in}{1.831528in}}%
\pgfpathlineto{\pgfqpoint{2.543616in}{1.831528in}}%
\pgfpathlineto{\pgfqpoint{2.559399in}{1.831528in}}%
\pgfpathlineto{\pgfqpoint{2.575182in}{1.831528in}}%
\pgfpathlineto{\pgfqpoint{2.590965in}{1.831528in}}%
\pgfpathlineto{\pgfqpoint{2.606748in}{1.831528in}}%
\pgfpathlineto{\pgfqpoint{2.622531in}{1.831528in}}%
\pgfpathlineto{\pgfqpoint{2.638314in}{1.831528in}}%
\pgfpathlineto{\pgfqpoint{2.654097in}{1.831528in}}%
\pgfpathlineto{\pgfqpoint{2.669880in}{1.831528in}}%
\pgfpathlineto{\pgfqpoint{2.685663in}{1.831528in}}%
\pgfpathlineto{\pgfqpoint{2.701446in}{1.831528in}}%
\pgfpathlineto{\pgfqpoint{2.717229in}{1.831528in}}%
\pgfpathlineto{\pgfqpoint{2.733012in}{1.831528in}}%
\pgfpathlineto{\pgfqpoint{2.748795in}{1.831528in}}%
\pgfpathlineto{\pgfqpoint{2.764578in}{1.831528in}}%
\pgfpathlineto{\pgfqpoint{2.780361in}{1.831528in}}%
\pgfpathlineto{\pgfqpoint{2.796144in}{0.436463in}}%
\pgfusepath{stroke}%
\end{pgfscope}%
\begin{pgfscope}%
\pgfsetrectcap%
\pgfsetmiterjoin%
\pgfsetlinewidth{0.803000pt}%
\definecolor{currentstroke}{rgb}{0.000000,0.000000,0.000000}%
\pgfsetstrokecolor{currentstroke}%
\pgfsetdash{}{0pt}%
\pgfpathmoveto{\pgfqpoint{2.306872in}{0.436091in}}%
\pgfpathlineto{\pgfqpoint{2.306872in}{2.033168in}}%
\pgfusepath{stroke}%
\end{pgfscope}%
\begin{pgfscope}%
\pgfsetrectcap%
\pgfsetmiterjoin%
\pgfsetlinewidth{0.803000pt}%
\definecolor{currentstroke}{rgb}{0.000000,0.000000,0.000000}%
\pgfsetstrokecolor{currentstroke}%
\pgfsetdash{}{0pt}%
\pgfpathmoveto{\pgfqpoint{3.741688in}{0.436091in}}%
\pgfpathlineto{\pgfqpoint{3.741688in}{2.033168in}}%
\pgfusepath{stroke}%
\end{pgfscope}%
\begin{pgfscope}%
\pgfsetrectcap%
\pgfsetmiterjoin%
\pgfsetlinewidth{0.803000pt}%
\definecolor{currentstroke}{rgb}{0.000000,0.000000,0.000000}%
\pgfsetstrokecolor{currentstroke}%
\pgfsetdash{}{0pt}%
\pgfpathmoveto{\pgfqpoint{2.306872in}{0.436091in}}%
\pgfpathlineto{\pgfqpoint{3.741688in}{0.436091in}}%
\pgfusepath{stroke}%
\end{pgfscope}%
\begin{pgfscope}%
\pgfsetrectcap%
\pgfsetmiterjoin%
\pgfsetlinewidth{0.803000pt}%
\definecolor{currentstroke}{rgb}{0.000000,0.000000,0.000000}%
\pgfsetstrokecolor{currentstroke}%
\pgfsetdash{}{0pt}%
\pgfpathmoveto{\pgfqpoint{2.306872in}{2.033168in}}%
\pgfpathlineto{\pgfqpoint{3.741688in}{2.033168in}}%
\pgfusepath{stroke}%
\end{pgfscope}%
\begin{pgfscope}%
\definecolor{textcolor}{rgb}{0.000000,0.000000,0.000000}%
\pgfsetstrokecolor{textcolor}%
\pgfsetfillcolor{textcolor}%
\pgftext[x=2.349916in,y=0.515945in,left,base]{\color{textcolor}\sffamily\fontsize{12.000000}{14.400000}\selectfont (b)}%
\end{pgfscope}%
\begin{pgfscope}%
\definecolor{textcolor}{rgb}{0.000000,0.000000,0.000000}%
\pgfsetstrokecolor{textcolor}%
\pgfsetfillcolor{textcolor}%
\pgftext[x=2.306872in,y=1.884603in,left,base]{\color{textcolor}\sffamily\fontsize{10.000000}{12.000000}\selectfont \(\displaystyle 1/g\!=\!\!-3\)}%
\end{pgfscope}%
\begin{pgfscope}%
\definecolor{textcolor}{rgb}{0.000000,0.000000,0.000000}%
\pgfsetstrokecolor{textcolor}%
\pgfsetfillcolor{textcolor}%
\pgftext[x=3.060150in,y=1.884603in,left,base]{\color{textcolor}\sffamily\fontsize{10.000000}{12.000000}\selectfont \(\displaystyle \kappa\!=\!2Q_+\Phi\)}%
\end{pgfscope}%
\begin{pgfscope}%
\definecolor{textcolor}{rgb}{0.000000,0.000000,0.000000}%
\pgfsetstrokecolor{textcolor}%
\pgfsetfillcolor{textcolor}%
\pgftext[x=2.710264in,y=1.082907in,left,base,rotate=90.000000]{\color{textcolor}\sffamily\fontsize{10.000000}{12.000000}\selectfont \(\displaystyle Q_\uparrow\!\Phi\!=\!0.54\)}%
\end{pgfscope}%
\begin{pgfscope}%
\definecolor{textcolor}{rgb}{0.000000,0.000000,0.000000}%
\pgfsetstrokecolor{textcolor}%
\pgfsetfillcolor{textcolor}%
\pgftext[x=3.169405in,y=1.082907in,left,base,rotate=90.000000]{\color{textcolor}\sffamily\fontsize{10.000000}{12.000000}\selectfont \(\displaystyle Q_\uparrow\!\Phi\!=\!0.36\)}%
\end{pgfscope}%
\begin{pgfscope}%
\pgfsetbuttcap%
\pgfsetmiterjoin%
\definecolor{currentfill}{rgb}{1.000000,1.000000,1.000000}%
\pgfsetfillcolor{currentfill}%
\pgfsetfillopacity{0.800000}%
\pgfsetlinewidth{1.003750pt}%
\definecolor{currentstroke}{rgb}{0.800000,0.800000,0.800000}%
\pgfsetstrokecolor{currentstroke}%
\pgfsetstrokeopacity{0.800000}%
\pgfsetdash{}{0pt}%
\pgfpathmoveto{\pgfqpoint{2.939009in}{0.491647in}}%
\pgfpathlineto{\pgfqpoint{3.663910in}{0.491647in}}%
\pgfpathquadraticcurveto{\pgfqpoint{3.686132in}{0.491647in}}{\pgfqpoint{3.686132in}{0.513869in}}%
\pgfpathlineto{\pgfqpoint{3.686132in}{0.838261in}}%
\pgfpathquadraticcurveto{\pgfqpoint{3.686132in}{0.860484in}}{\pgfqpoint{3.663910in}{0.860484in}}%
\pgfpathlineto{\pgfqpoint{2.939009in}{0.860484in}}%
\pgfpathquadraticcurveto{\pgfqpoint{2.916787in}{0.860484in}}{\pgfqpoint{2.916787in}{0.838261in}}%
\pgfpathlineto{\pgfqpoint{2.916787in}{0.513869in}}%
\pgfpathquadraticcurveto{\pgfqpoint{2.916787in}{0.491647in}}{\pgfqpoint{2.939009in}{0.491647in}}%
\pgfpathclose%
\pgfusepath{stroke,fill}%
\end{pgfscope}%
\begin{pgfscope}%
\pgfsetrectcap%
\pgfsetroundjoin%
\pgfsetlinewidth{2.007500pt}%
\definecolor{currentstroke}{rgb}{0.121569,0.466667,0.705882}%
\pgfsetstrokecolor{currentstroke}%
\pgfsetdash{}{0pt}%
\pgfpathmoveto{\pgfqpoint{2.961231in}{0.770510in}}%
\pgfpathlineto{\pgfqpoint{3.294565in}{0.770510in}}%
\pgfusepath{stroke}%
\end{pgfscope}%
\begin{pgfscope}%
\definecolor{textcolor}{rgb}{0.000000,0.000000,0.000000}%
\pgfsetstrokecolor{textcolor}%
\pgfsetfillcolor{textcolor}%
\pgftext[x=3.383453in,y=0.731621in,left,base]{\color{textcolor}\sffamily\fontsize{8.000000}{9.600000}\selectfont \(\displaystyle \Delta(\kappa)\)}%
\end{pgfscope}%
\begin{pgfscope}%
\pgfsetbuttcap%
\pgfsetroundjoin%
\pgfsetlinewidth{3.011250pt}%
\definecolor{currentstroke}{rgb}{0.121569,0.466667,0.705882}%
\pgfsetstrokecolor{currentstroke}%
\pgfsetdash{{11.100000pt}{4.800000pt}}{0.000000pt}%
\pgfpathmoveto{\pgfqpoint{2.961231in}{0.602758in}}%
\pgfpathlineto{\pgfqpoint{3.294565in}{0.602758in}}%
\pgfusepath{stroke}%
\end{pgfscope}%
\begin{pgfscope}%
\definecolor{textcolor}{rgb}{0.000000,0.000000,0.000000}%
\pgfsetstrokecolor{textcolor}%
\pgfsetfillcolor{textcolor}%
\pgftext[x=3.383453in,y=0.563869in,left,base]{\color{textcolor}\sffamily\fontsize{8.000000}{9.600000}\selectfont \(\displaystyle \Delta(0)\)}%
\end{pgfscope}%
\end{pgfpicture}%
\makeatother%
\endgroup%

%% file: mom_dist.pgf
\begingroup%
\makeatletter%
\begin{pgfpicture}%
\pgfpathrectangle{\pgfpointorigin}{\pgfqpoint{3.899004in}{2.148089in}}%
\pgfusepath{use as bounding box, clip}%
\begin{pgfscope}%
\pgfsetbuttcap%
\pgfsetmiterjoin%
\definecolor{currentfill}{rgb}{1.000000,1.000000,1.000000}%
\pgfsetfillcolor{currentfill}%
\pgfsetlinewidth{0.000000pt}%
\definecolor{currentstroke}{rgb}{1.000000,1.000000,1.000000}%
\pgfsetstrokecolor{currentstroke}%
\pgfsetdash{}{0pt}%
\pgfpathmoveto{\pgfqpoint{0.000000in}{0.000000in}}%
\pgfpathlineto{\pgfqpoint{3.899004in}{0.000000in}}%
\pgfpathlineto{\pgfqpoint{3.899004in}{2.148089in}}%
\pgfpathlineto{\pgfqpoint{0.000000in}{2.148089in}}%
\pgfpathclose%
\pgfusepath{fill}%
\end{pgfscope}%
\begin{pgfscope}%
\pgfsetbuttcap%
\pgfsetmiterjoin%
\definecolor{currentfill}{rgb}{1.000000,1.000000,1.000000}%
\pgfsetfillcolor{currentfill}%
\pgfsetlinewidth{0.000000pt}%
\definecolor{currentstroke}{rgb}{0.000000,0.000000,0.000000}%
\pgfsetstrokecolor{currentstroke}%
\pgfsetstrokeopacity{0.000000}%
\pgfsetdash{}{0pt}%
\pgfpathmoveto{\pgfqpoint{0.369491in}{0.395708in}}%
\pgfpathlineto{\pgfqpoint{1.825393in}{0.395708in}}%
\pgfpathlineto{\pgfqpoint{1.825393in}{1.999478in}}%
\pgfpathlineto{\pgfqpoint{0.369491in}{1.999478in}}%
\pgfpathclose%
\pgfusepath{fill}%
\end{pgfscope}%
\begin{pgfscope}%
\pgfsetbuttcap%
\pgfsetroundjoin%
\definecolor{currentfill}{rgb}{0.000000,0.000000,0.000000}%
\pgfsetfillcolor{currentfill}%
\pgfsetlinewidth{0.803000pt}%
\definecolor{currentstroke}{rgb}{0.000000,0.000000,0.000000}%
\pgfsetstrokecolor{currentstroke}%
\pgfsetdash{}{0pt}%
\pgfsys@defobject{currentmarker}{\pgfqpoint{0.000000in}{-0.048611in}}{\pgfqpoint{0.000000in}{0.000000in}}{%
\pgfpathmoveto{\pgfqpoint{0.000000in}{0.000000in}}%
\pgfpathlineto{\pgfqpoint{0.000000in}{-0.048611in}}%
\pgfusepath{stroke,fill}%
}%
\begin{pgfscope}%
\pgfsys@transformshift{0.681470in}{0.395708in}%
\pgfsys@useobject{currentmarker}{}%
\end{pgfscope}%
\end{pgfscope}%
\begin{pgfscope}%
\definecolor{textcolor}{rgb}{0.000000,0.000000,0.000000}%
\pgfsetstrokecolor{textcolor}%
\pgfsetfillcolor{textcolor}%
\pgftext[x=0.681470in,y=0.298486in,,top]{\color{textcolor}\sffamily\fontsize{10.000000}{12.000000}\selectfont −1}%
\end{pgfscope}%
\begin{pgfscope}%
\pgfsetbuttcap%
\pgfsetroundjoin%
\definecolor{currentfill}{rgb}{0.000000,0.000000,0.000000}%
\pgfsetfillcolor{currentfill}%
\pgfsetlinewidth{0.803000pt}%
\definecolor{currentstroke}{rgb}{0.000000,0.000000,0.000000}%
\pgfsetstrokecolor{currentstroke}%
\pgfsetdash{}{0pt}%
\pgfsys@defobject{currentmarker}{\pgfqpoint{0.000000in}{-0.048611in}}{\pgfqpoint{0.000000in}{0.000000in}}{%
\pgfpathmoveto{\pgfqpoint{0.000000in}{0.000000in}}%
\pgfpathlineto{\pgfqpoint{0.000000in}{-0.048611in}}%
\pgfusepath{stroke,fill}%
}%
\begin{pgfscope}%
\pgfsys@transformshift{1.097442in}{0.395708in}%
\pgfsys@useobject{currentmarker}{}%
\end{pgfscope}%
\end{pgfscope}%
\begin{pgfscope}%
\definecolor{textcolor}{rgb}{0.000000,0.000000,0.000000}%
\pgfsetstrokecolor{textcolor}%
\pgfsetfillcolor{textcolor}%
\pgftext[x=1.097442in,y=0.298486in,,top]{\color{textcolor}\sffamily\fontsize{10.000000}{12.000000}\selectfont 0}%
\end{pgfscope}%
\begin{pgfscope}%
\pgfsetbuttcap%
\pgfsetroundjoin%
\definecolor{currentfill}{rgb}{0.000000,0.000000,0.000000}%
\pgfsetfillcolor{currentfill}%
\pgfsetlinewidth{0.803000pt}%
\definecolor{currentstroke}{rgb}{0.000000,0.000000,0.000000}%
\pgfsetstrokecolor{currentstroke}%
\pgfsetdash{}{0pt}%
\pgfsys@defobject{currentmarker}{\pgfqpoint{0.000000in}{-0.048611in}}{\pgfqpoint{0.000000in}{0.000000in}}{%
\pgfpathmoveto{\pgfqpoint{0.000000in}{0.000000in}}%
\pgfpathlineto{\pgfqpoint{0.000000in}{-0.048611in}}%
\pgfusepath{stroke,fill}%
}%
\begin{pgfscope}%
\pgfsys@transformshift{1.513414in}{0.395708in}%
\pgfsys@useobject{currentmarker}{}%
\end{pgfscope}%
\end{pgfscope}%
\begin{pgfscope}%
\definecolor{textcolor}{rgb}{0.000000,0.000000,0.000000}%
\pgfsetstrokecolor{textcolor}%
\pgfsetfillcolor{textcolor}%
\pgftext[x=1.513414in,y=0.298486in,,top]{\color{textcolor}\sffamily\fontsize{10.000000}{12.000000}\selectfont 1}%
\end{pgfscope}%
\begin{pgfscope}%
\pgfsetbuttcap%
\pgfsetroundjoin%
\definecolor{currentfill}{rgb}{0.000000,0.000000,0.000000}%
\pgfsetfillcolor{currentfill}%
\pgfsetlinewidth{0.602250pt}%
\definecolor{currentstroke}{rgb}{0.000000,0.000000,0.000000}%
\pgfsetstrokecolor{currentstroke}%
\pgfsetdash{}{0pt}%
\pgfsys@defobject{currentmarker}{\pgfqpoint{0.000000in}{-0.027778in}}{\pgfqpoint{0.000000in}{0.000000in}}{%
\pgfpathmoveto{\pgfqpoint{0.000000in}{0.000000in}}%
\pgfpathlineto{\pgfqpoint{0.000000in}{-0.027778in}}%
\pgfusepath{stroke,fill}%
}%
\begin{pgfscope}%
\pgfsys@transformshift{0.473484in}{0.395708in}%
\pgfsys@useobject{currentmarker}{}%
\end{pgfscope}%
\end{pgfscope}%
\begin{pgfscope}%
\pgfsetbuttcap%
\pgfsetroundjoin%
\definecolor{currentfill}{rgb}{0.000000,0.000000,0.000000}%
\pgfsetfillcolor{currentfill}%
\pgfsetlinewidth{0.602250pt}%
\definecolor{currentstroke}{rgb}{0.000000,0.000000,0.000000}%
\pgfsetstrokecolor{currentstroke}%
\pgfsetdash{}{0pt}%
\pgfsys@defobject{currentmarker}{\pgfqpoint{0.000000in}{-0.027778in}}{\pgfqpoint{0.000000in}{0.000000in}}{%
\pgfpathmoveto{\pgfqpoint{0.000000in}{0.000000in}}%
\pgfpathlineto{\pgfqpoint{0.000000in}{-0.027778in}}%
\pgfusepath{stroke,fill}%
}%
\begin{pgfscope}%
\pgfsys@transformshift{0.889456in}{0.395708in}%
\pgfsys@useobject{currentmarker}{}%
\end{pgfscope}%
\end{pgfscope}%
\begin{pgfscope}%
\pgfsetbuttcap%
\pgfsetroundjoin%
\definecolor{currentfill}{rgb}{0.000000,0.000000,0.000000}%
\pgfsetfillcolor{currentfill}%
\pgfsetlinewidth{0.602250pt}%
\definecolor{currentstroke}{rgb}{0.000000,0.000000,0.000000}%
\pgfsetstrokecolor{currentstroke}%
\pgfsetdash{}{0pt}%
\pgfsys@defobject{currentmarker}{\pgfqpoint{0.000000in}{-0.027778in}}{\pgfqpoint{0.000000in}{0.000000in}}{%
\pgfpathmoveto{\pgfqpoint{0.000000in}{0.000000in}}%
\pgfpathlineto{\pgfqpoint{0.000000in}{-0.027778in}}%
\pgfusepath{stroke,fill}%
}%
\begin{pgfscope}%
\pgfsys@transformshift{1.305428in}{0.395708in}%
\pgfsys@useobject{currentmarker}{}%
\end{pgfscope}%
\end{pgfscope}%
\begin{pgfscope}%
\pgfsetbuttcap%
\pgfsetroundjoin%
\definecolor{currentfill}{rgb}{0.000000,0.000000,0.000000}%
\pgfsetfillcolor{currentfill}%
\pgfsetlinewidth{0.602250pt}%
\definecolor{currentstroke}{rgb}{0.000000,0.000000,0.000000}%
\pgfsetstrokecolor{currentstroke}%
\pgfsetdash{}{0pt}%
\pgfsys@defobject{currentmarker}{\pgfqpoint{0.000000in}{-0.027778in}}{\pgfqpoint{0.000000in}{0.000000in}}{%
\pgfpathmoveto{\pgfqpoint{0.000000in}{0.000000in}}%
\pgfpathlineto{\pgfqpoint{0.000000in}{-0.027778in}}%
\pgfusepath{stroke,fill}%
}%
\begin{pgfscope}%
\pgfsys@transformshift{1.721400in}{0.395708in}%
\pgfsys@useobject{currentmarker}{}%
\end{pgfscope}%
\end{pgfscope}%
\begin{pgfscope}%
\definecolor{textcolor}{rgb}{0.000000,0.000000,0.000000}%
\pgfsetstrokecolor{textcolor}%
\pgfsetfillcolor{textcolor}%
\pgftext[x=1.359504in,y=0.261295in,,top]{\color{textcolor}\sffamily\fontsize{12.000000}{14.400000}\selectfont \(\displaystyle k\)}%
\end{pgfscope}%
\begin{pgfscope}%
\pgfsetbuttcap%
\pgfsetroundjoin%
\definecolor{currentfill}{rgb}{0.000000,0.000000,0.000000}%
\pgfsetfillcolor{currentfill}%
\pgfsetlinewidth{0.803000pt}%
\definecolor{currentstroke}{rgb}{0.000000,0.000000,0.000000}%
\pgfsetstrokecolor{currentstroke}%
\pgfsetdash{}{0pt}%
\pgfsys@defobject{currentmarker}{\pgfqpoint{-0.048611in}{0.000000in}}{\pgfqpoint{0.000000in}{0.000000in}}{%
\pgfpathmoveto{\pgfqpoint{0.000000in}{0.000000in}}%
\pgfpathlineto{\pgfqpoint{-0.048611in}{0.000000in}}%
\pgfusepath{stroke,fill}%
}%
\begin{pgfscope}%
\pgfsys@transformshift{0.369491in}{0.468607in}%
\pgfsys@useobject{currentmarker}{}%
\end{pgfscope}%
\end{pgfscope}%
\begin{pgfscope}%
\definecolor{textcolor}{rgb}{0.000000,0.000000,0.000000}%
\pgfsetstrokecolor{textcolor}%
\pgfsetfillcolor{textcolor}%
\pgftext[x=0.100000in,y=0.415845in,left,base]{\color{textcolor}\sffamily\fontsize{10.000000}{12.000000}\selectfont 0.0}%
\end{pgfscope}%
\begin{pgfscope}%
\pgfsetbuttcap%
\pgfsetroundjoin%
\definecolor{currentfill}{rgb}{0.000000,0.000000,0.000000}%
\pgfsetfillcolor{currentfill}%
\pgfsetlinewidth{0.803000pt}%
\definecolor{currentstroke}{rgb}{0.000000,0.000000,0.000000}%
\pgfsetstrokecolor{currentstroke}%
\pgfsetdash{}{0pt}%
\pgfsys@defobject{currentmarker}{\pgfqpoint{-0.048611in}{0.000000in}}{\pgfqpoint{0.000000in}{0.000000in}}{%
\pgfpathmoveto{\pgfqpoint{0.000000in}{0.000000in}}%
\pgfpathlineto{\pgfqpoint{-0.048611in}{0.000000in}}%
\pgfusepath{stroke,fill}%
}%
\begin{pgfscope}%
\pgfsys@transformshift{0.369491in}{1.197593in}%
\pgfsys@useobject{currentmarker}{}%
\end{pgfscope}%
\end{pgfscope}%
\begin{pgfscope}%
\definecolor{textcolor}{rgb}{0.000000,0.000000,0.000000}%
\pgfsetstrokecolor{textcolor}%
\pgfsetfillcolor{textcolor}%
\pgftext[x=0.100000in,y=1.144832in,left,base]{\color{textcolor}\sffamily\fontsize{10.000000}{12.000000}\selectfont 0.5}%
\end{pgfscope}%
\begin{pgfscope}%
\pgfsetbuttcap%
\pgfsetroundjoin%
\definecolor{currentfill}{rgb}{0.000000,0.000000,0.000000}%
\pgfsetfillcolor{currentfill}%
\pgfsetlinewidth{0.803000pt}%
\definecolor{currentstroke}{rgb}{0.000000,0.000000,0.000000}%
\pgfsetstrokecolor{currentstroke}%
\pgfsetdash{}{0pt}%
\pgfsys@defobject{currentmarker}{\pgfqpoint{-0.048611in}{0.000000in}}{\pgfqpoint{0.000000in}{0.000000in}}{%
\pgfpathmoveto{\pgfqpoint{0.000000in}{0.000000in}}%
\pgfpathlineto{\pgfqpoint{-0.048611in}{0.000000in}}%
\pgfusepath{stroke,fill}%
}%
\begin{pgfscope}%
\pgfsys@transformshift{0.369491in}{1.926579in}%
\pgfsys@useobject{currentmarker}{}%
\end{pgfscope}%
\end{pgfscope}%
\begin{pgfscope}%
\definecolor{textcolor}{rgb}{0.000000,0.000000,0.000000}%
\pgfsetstrokecolor{textcolor}%
\pgfsetfillcolor{textcolor}%
\pgftext[x=0.100000in,y=1.873818in,left,base]{\color{textcolor}\sffamily\fontsize{10.000000}{12.000000}\selectfont 1.0}%
\end{pgfscope}%
\begin{pgfscope}%
\pgfsetbuttcap%
\pgfsetroundjoin%
\definecolor{currentfill}{rgb}{0.000000,0.000000,0.000000}%
\pgfsetfillcolor{currentfill}%
\pgfsetlinewidth{0.602250pt}%
\definecolor{currentstroke}{rgb}{0.000000,0.000000,0.000000}%
\pgfsetstrokecolor{currentstroke}%
\pgfsetdash{}{0pt}%
\pgfsys@defobject{currentmarker}{\pgfqpoint{-0.027778in}{0.000000in}}{\pgfqpoint{0.000000in}{0.000000in}}{%
\pgfpathmoveto{\pgfqpoint{0.000000in}{0.000000in}}%
\pgfpathlineto{\pgfqpoint{-0.027778in}{0.000000in}}%
\pgfusepath{stroke,fill}%
}%
\begin{pgfscope}%
\pgfsys@transformshift{0.369491in}{0.833100in}%
\pgfsys@useobject{currentmarker}{}%
\end{pgfscope}%
\end{pgfscope}%
\begin{pgfscope}%
\pgfsetbuttcap%
\pgfsetroundjoin%
\definecolor{currentfill}{rgb}{0.000000,0.000000,0.000000}%
\pgfsetfillcolor{currentfill}%
\pgfsetlinewidth{0.602250pt}%
\definecolor{currentstroke}{rgb}{0.000000,0.000000,0.000000}%
\pgfsetstrokecolor{currentstroke}%
\pgfsetdash{}{0pt}%
\pgfsys@defobject{currentmarker}{\pgfqpoint{-0.027778in}{0.000000in}}{\pgfqpoint{0.000000in}{0.000000in}}{%
\pgfpathmoveto{\pgfqpoint{0.000000in}{0.000000in}}%
\pgfpathlineto{\pgfqpoint{-0.027778in}{0.000000in}}%
\pgfusepath{stroke,fill}%
}%
\begin{pgfscope}%
\pgfsys@transformshift{0.369491in}{1.562086in}%
\pgfsys@useobject{currentmarker}{}%
\end{pgfscope}%
\end{pgfscope}%
\begin{pgfscope}%
\definecolor{textcolor}{rgb}{0.000000,0.000000,0.000000}%
\pgfsetstrokecolor{textcolor}%
\pgfsetfillcolor{textcolor}%
\pgftext[x=0.280556in,y=1.518347in,,bottom,rotate=90.000000]{\color{textcolor}\sffamily\fontsize{12.000000}{14.400000}\selectfont \(\displaystyle n_k\)}%
\end{pgfscope}%
\begin{pgfscope}%
\pgfpathrectangle{\pgfqpoint{0.369491in}{0.395708in}}{\pgfqpoint{1.455903in}{1.603770in}}%
\pgfusepath{clip}%
\pgfsetrectcap%
\pgfsetroundjoin%
\pgfsetlinewidth{2.007500pt}%
\definecolor{currentstroke}{rgb}{0.121569,0.466667,0.705882}%
\pgfsetstrokecolor{currentstroke}%
\pgfsetdash{}{0pt}%
\pgfpathmoveto{\pgfqpoint{0.355602in}{0.471580in}}%
\pgfpathlineto{\pgfqpoint{0.453154in}{0.475083in}}%
\pgfpathlineto{\pgfqpoint{0.507366in}{0.479710in}}%
\pgfpathlineto{\pgfqpoint{0.544897in}{0.485883in}}%
\pgfpathlineto{\pgfqpoint{0.569918in}{0.492830in}}%
\pgfpathlineto{\pgfqpoint{0.590769in}{0.501838in}}%
\pgfpathlineto{\pgfqpoint{0.607450in}{0.512620in}}%
\pgfpathlineto{\pgfqpoint{0.619960in}{0.524025in}}%
\pgfpathlineto{\pgfqpoint{0.632470in}{0.539769in}}%
\pgfpathlineto{\pgfqpoint{0.644981in}{0.562055in}}%
\pgfpathlineto{\pgfqpoint{0.653321in}{0.582217in}}%
\pgfpathlineto{\pgfqpoint{0.661662in}{0.608346in}}%
\pgfpathlineto{\pgfqpoint{0.670002in}{0.642509in}}%
\pgfpathlineto{\pgfqpoint{0.678342in}{0.687390in}}%
\pgfpathlineto{\pgfqpoint{0.686682in}{0.746213in}}%
\pgfpathlineto{\pgfqpoint{0.695023in}{0.822233in}}%
\pgfpathlineto{\pgfqpoint{0.707533in}{0.972077in}}%
\pgfpathlineto{\pgfqpoint{0.749235in}{1.542138in}}%
\pgfpathlineto{\pgfqpoint{0.761745in}{1.645213in}}%
\pgfpathlineto{\pgfqpoint{0.774256in}{1.716086in}}%
\pgfpathlineto{\pgfqpoint{0.786766in}{1.764669in}}%
\pgfpathlineto{\pgfqpoint{0.799276in}{1.798531in}}%
\pgfpathlineto{\pgfqpoint{0.811787in}{1.822703in}}%
\pgfpathlineto{\pgfqpoint{0.824297in}{1.840399in}}%
\pgfpathlineto{\pgfqpoint{0.836808in}{1.853667in}}%
\pgfpathlineto{\pgfqpoint{0.853488in}{1.866693in}}%
\pgfpathlineto{\pgfqpoint{0.874339in}{1.878084in}}%
\pgfpathlineto{\pgfqpoint{0.899360in}{1.887326in}}%
\pgfpathlineto{\pgfqpoint{0.928551in}{1.894468in}}%
\pgfpathlineto{\pgfqpoint{0.966082in}{1.900317in}}%
\pgfpathlineto{\pgfqpoint{1.016124in}{1.904863in}}%
\pgfpathlineto{\pgfqpoint{1.087017in}{1.907893in}}%
\pgfpathlineto{\pgfqpoint{1.174590in}{1.908369in}}%
\pgfpathlineto{\pgfqpoint{1.253822in}{1.905955in}}%
\pgfpathlineto{\pgfqpoint{1.312205in}{1.901480in}}%
\pgfpathlineto{\pgfqpoint{1.353906in}{1.895610in}}%
\pgfpathlineto{\pgfqpoint{1.387267in}{1.887875in}}%
\pgfpathlineto{\pgfqpoint{1.412288in}{1.878905in}}%
\pgfpathlineto{\pgfqpoint{1.433139in}{1.867891in}}%
\pgfpathlineto{\pgfqpoint{1.449819in}{1.855341in}}%
\pgfpathlineto{\pgfqpoint{1.462330in}{1.842602in}}%
\pgfpathlineto{\pgfqpoint{1.474840in}{1.825669in}}%
\pgfpathlineto{\pgfqpoint{1.487351in}{1.802623in}}%
\pgfpathlineto{\pgfqpoint{1.499861in}{1.770459in}}%
\pgfpathlineto{\pgfqpoint{1.508201in}{1.741716in}}%
\pgfpathlineto{\pgfqpoint{1.516542in}{1.704900in}}%
\pgfpathlineto{\pgfqpoint{1.524882in}{1.657472in}}%
\pgfpathlineto{\pgfqpoint{1.533222in}{1.596417in}}%
\pgfpathlineto{\pgfqpoint{1.541563in}{1.518707in}}%
\pgfpathlineto{\pgfqpoint{1.554073in}{1.367586in}}%
\pgfpathlineto{\pgfqpoint{1.595775in}{0.799206in}}%
\pgfpathlineto{\pgfqpoint{1.608285in}{0.699118in}}%
\pgfpathlineto{\pgfqpoint{1.620795in}{0.632058in}}%
\pgfpathlineto{\pgfqpoint{1.633306in}{0.587436in}}%
\pgfpathlineto{\pgfqpoint{1.645816in}{0.557291in}}%
\pgfpathlineto{\pgfqpoint{1.658327in}{0.536436in}}%
\pgfpathlineto{\pgfqpoint{1.670837in}{0.521633in}}%
\pgfpathlineto{\pgfqpoint{1.683348in}{0.510862in}}%
\pgfpathlineto{\pgfqpoint{1.700028in}{0.500632in}}%
\pgfpathlineto{\pgfqpoint{1.720879in}{0.492042in}}%
\pgfpathlineto{\pgfqpoint{1.745900in}{0.485382in}}%
\pgfpathlineto{\pgfqpoint{1.779261in}{0.479937in}}%
\pgfpathlineto{\pgfqpoint{1.825133in}{0.475722in}}%
\pgfpathlineto{\pgfqpoint{1.839282in}{0.474856in}}%
\pgfpathlineto{\pgfqpoint{1.839282in}{0.474856in}}%
\pgfusepath{stroke}%
\end{pgfscope}%
\begin{pgfscope}%
\pgfpathrectangle{\pgfqpoint{0.369491in}{0.395708in}}{\pgfqpoint{1.455903in}{1.603770in}}%
\pgfusepath{clip}%
\pgfsetbuttcap%
\pgfsetroundjoin%
\pgfsetlinewidth{2.007500pt}%
\definecolor{currentstroke}{rgb}{1.000000,0.498039,0.054902}%
\pgfsetstrokecolor{currentstroke}%
\pgfsetdash{{7.400000pt}{3.200000pt}}{0.000000pt}%
\pgfpathmoveto{\pgfqpoint{0.355602in}{0.474856in}}%
\pgfpathlineto{\pgfqpoint{0.411453in}{0.479433in}}%
\pgfpathlineto{\pgfqpoint{0.448984in}{0.485382in}}%
\pgfpathlineto{\pgfqpoint{0.474005in}{0.492042in}}%
\pgfpathlineto{\pgfqpoint{0.494856in}{0.500632in}}%
\pgfpathlineto{\pgfqpoint{0.511536in}{0.510862in}}%
\pgfpathlineto{\pgfqpoint{0.524047in}{0.521633in}}%
\pgfpathlineto{\pgfqpoint{0.536557in}{0.536436in}}%
\pgfpathlineto{\pgfqpoint{0.549068in}{0.557291in}}%
\pgfpathlineto{\pgfqpoint{0.557408in}{0.576085in}}%
\pgfpathlineto{\pgfqpoint{0.565748in}{0.600372in}}%
\pgfpathlineto{\pgfqpoint{0.574088in}{0.632058in}}%
\pgfpathlineto{\pgfqpoint{0.582429in}{0.673653in}}%
\pgfpathlineto{\pgfqpoint{0.590769in}{0.728257in}}%
\pgfpathlineto{\pgfqpoint{0.599109in}{0.799206in}}%
\pgfpathlineto{\pgfqpoint{0.611620in}{0.941163in}}%
\pgfpathlineto{\pgfqpoint{0.628300in}{1.183865in}}%
\pgfpathlineto{\pgfqpoint{0.644981in}{1.422464in}}%
\pgfpathlineto{\pgfqpoint{0.657491in}{1.559808in}}%
\pgfpathlineto{\pgfqpoint{0.670002in}{1.657472in}}%
\pgfpathlineto{\pgfqpoint{0.682512in}{1.724462in}}%
\pgfpathlineto{\pgfqpoint{0.695023in}{1.770459in}}%
\pgfpathlineto{\pgfqpoint{0.707533in}{1.802623in}}%
\pgfpathlineto{\pgfqpoint{0.720044in}{1.825669in}}%
\pgfpathlineto{\pgfqpoint{0.732554in}{1.842602in}}%
\pgfpathlineto{\pgfqpoint{0.749235in}{1.858884in}}%
\pgfpathlineto{\pgfqpoint{0.765915in}{1.870440in}}%
\pgfpathlineto{\pgfqpoint{0.786766in}{1.880664in}}%
\pgfpathlineto{\pgfqpoint{0.811787in}{1.889058in}}%
\pgfpathlineto{\pgfqpoint{0.840978in}{1.895610in}}%
\pgfpathlineto{\pgfqpoint{0.882679in}{1.901480in}}%
\pgfpathlineto{\pgfqpoint{0.936891in}{1.905731in}}%
\pgfpathlineto{\pgfqpoint{1.011954in}{1.908254in}}%
\pgfpathlineto{\pgfqpoint{1.099527in}{1.908074in}}%
\pgfpathlineto{\pgfqpoint{1.174590in}{1.905131in}}%
\pgfpathlineto{\pgfqpoint{1.228802in}{1.900317in}}%
\pgfpathlineto{\pgfqpoint{1.270503in}{1.893620in}}%
\pgfpathlineto{\pgfqpoint{1.299694in}{1.886031in}}%
\pgfpathlineto{\pgfqpoint{1.324715in}{1.876133in}}%
\pgfpathlineto{\pgfqpoint{1.341396in}{1.866693in}}%
\pgfpathlineto{\pgfqpoint{1.358076in}{1.853667in}}%
\pgfpathlineto{\pgfqpoint{1.370587in}{1.840399in}}%
\pgfpathlineto{\pgfqpoint{1.383097in}{1.822703in}}%
\pgfpathlineto{\pgfqpoint{1.395607in}{1.798531in}}%
\pgfpathlineto{\pgfqpoint{1.408118in}{1.764669in}}%
\pgfpathlineto{\pgfqpoint{1.416458in}{1.734325in}}%
\pgfpathlineto{\pgfqpoint{1.424799in}{1.695393in}}%
\pgfpathlineto{\pgfqpoint{1.433139in}{1.645213in}}%
\pgfpathlineto{\pgfqpoint{1.441479in}{1.580708in}}%
\pgfpathlineto{\pgfqpoint{1.453990in}{1.451102in}}%
\pgfpathlineto{\pgfqpoint{1.466500in}{1.281809in}}%
\pgfpathlineto{\pgfqpoint{1.491521in}{0.917464in}}%
\pgfpathlineto{\pgfqpoint{1.504031in}{0.781897in}}%
\pgfpathlineto{\pgfqpoint{1.516542in}{0.687390in}}%
\pgfpathlineto{\pgfqpoint{1.529052in}{0.624271in}}%
\pgfpathlineto{\pgfqpoint{1.541563in}{0.582217in}}%
\pgfpathlineto{\pgfqpoint{1.554073in}{0.553716in}}%
\pgfpathlineto{\pgfqpoint{1.566584in}{0.533925in}}%
\pgfpathlineto{\pgfqpoint{1.579094in}{0.519823in}}%
\pgfpathlineto{\pgfqpoint{1.595775in}{0.506719in}}%
\pgfpathlineto{\pgfqpoint{1.612455in}{0.497765in}}%
\pgfpathlineto{\pgfqpoint{1.633306in}{0.490151in}}%
\pgfpathlineto{\pgfqpoint{1.662497in}{0.483401in}}%
\pgfpathlineto{\pgfqpoint{1.700028in}{0.478322in}}%
\pgfpathlineto{\pgfqpoint{1.754240in}{0.474398in}}%
\pgfpathlineto{\pgfqpoint{1.839282in}{0.471580in}}%
\pgfpathlineto{\pgfqpoint{1.839282in}{0.471580in}}%
\pgfusepath{stroke}%
\end{pgfscope}%
\begin{pgfscope}%
\pgfpathrectangle{\pgfqpoint{0.369491in}{0.395708in}}{\pgfqpoint{1.455903in}{1.603770in}}%
\pgfusepath{clip}%
\pgfsetbuttcap%
\pgfsetroundjoin%
\pgfsetlinewidth{1.505625pt}%
\definecolor{currentstroke}{rgb}{0.121569,0.466667,0.705882}%
\pgfsetstrokecolor{currentstroke}%
\pgfsetdash{{1.500000pt}{2.475000pt}}{0.000000pt}%
\pgfpathmoveto{\pgfqpoint{1.144096in}{0.395708in}}%
\pgfpathlineto{\pgfqpoint{1.144096in}{1.887214in}}%
\pgfusepath{stroke}%
\end{pgfscope}%
\begin{pgfscope}%
\pgfpathrectangle{\pgfqpoint{0.369491in}{0.395708in}}{\pgfqpoint{1.455903in}{1.603770in}}%
\pgfusepath{clip}%
\pgfsetbuttcap%
\pgfsetroundjoin%
\pgfsetlinewidth{1.505625pt}%
\definecolor{currentstroke}{rgb}{1.000000,0.647059,0.000000}%
\pgfsetstrokecolor{currentstroke}%
\pgfsetdash{{1.500000pt}{2.475000pt}}{0.000000pt}%
\pgfpathmoveto{\pgfqpoint{1.050788in}{0.395708in}}%
\pgfpathlineto{\pgfqpoint{1.050788in}{1.887214in}}%
\pgfusepath{stroke}%
\end{pgfscope}%
\begin{pgfscope}%
\pgfsetrectcap%
\pgfsetmiterjoin%
\pgfsetlinewidth{0.803000pt}%
\definecolor{currentstroke}{rgb}{0.000000,0.000000,0.000000}%
\pgfsetstrokecolor{currentstroke}%
\pgfsetdash{}{0pt}%
\pgfpathmoveto{\pgfqpoint{0.369491in}{0.395708in}}%
\pgfpathlineto{\pgfqpoint{0.369491in}{1.999478in}}%
\pgfusepath{stroke}%
\end{pgfscope}%
\begin{pgfscope}%
\pgfsetrectcap%
\pgfsetmiterjoin%
\pgfsetlinewidth{0.803000pt}%
\definecolor{currentstroke}{rgb}{0.000000,0.000000,0.000000}%
\pgfsetstrokecolor{currentstroke}%
\pgfsetdash{}{0pt}%
\pgfpathmoveto{\pgfqpoint{1.825393in}{0.395708in}}%
\pgfpathlineto{\pgfqpoint{1.825393in}{1.999478in}}%
\pgfusepath{stroke}%
\end{pgfscope}%
\begin{pgfscope}%
\pgfsetrectcap%
\pgfsetmiterjoin%
\pgfsetlinewidth{0.803000pt}%
\definecolor{currentstroke}{rgb}{0.000000,0.000000,0.000000}%
\pgfsetstrokecolor{currentstroke}%
\pgfsetdash{}{0pt}%
\pgfpathmoveto{\pgfqpoint{0.369491in}{0.395708in}}%
\pgfpathlineto{\pgfqpoint{1.825393in}{0.395708in}}%
\pgfusepath{stroke}%
\end{pgfscope}%
\begin{pgfscope}%
\pgfsetrectcap%
\pgfsetmiterjoin%
\pgfsetlinewidth{0.803000pt}%
\definecolor{currentstroke}{rgb}{0.000000,0.000000,0.000000}%
\pgfsetstrokecolor{currentstroke}%
\pgfsetdash{}{0pt}%
\pgfpathmoveto{\pgfqpoint{0.369491in}{1.999478in}}%
\pgfpathlineto{\pgfqpoint{1.825393in}{1.999478in}}%
\pgfusepath{stroke}%
\end{pgfscope}%
\begin{pgfscope}%
\definecolor{textcolor}{rgb}{0.000000,0.000000,0.000000}%
\pgfsetstrokecolor{textcolor}%
\pgfsetfillcolor{textcolor}%
\pgftext[x=0.413168in,y=1.839101in,left,base]{\color{textcolor}\sffamily\fontsize{12.000000}{14.400000}\selectfont (a)}%
\end{pgfscope}%
\begin{pgfscope}%
\definecolor{textcolor}{rgb}{1.000000,0.647059,0.000000}%
\pgfsetstrokecolor{textcolor}%
\pgfsetfillcolor{textcolor}%
\pgftext[x=1.305428in,y=1.343390in,left,base]{\color{textcolor}\sffamily\fontsize{17.280000}{20.736000}\selectfont \(\displaystyle \downarrow\)}%
\end{pgfscope}%
\begin{pgfscope}%
\definecolor{textcolor}{rgb}{0.290196,0.474510,0.654902}%
\pgfsetstrokecolor{textcolor}%
\pgfsetfillcolor{textcolor}%
\pgftext[x=1.596609in,y=1.343390in,left,base]{\color{textcolor}\sffamily\fontsize{17.280000}{20.736000}\selectfont \(\displaystyle \uparrow\)}%
\end{pgfscope}%
\begin{pgfscope}%
\pgfsetbuttcap%
\pgfsetmiterjoin%
\definecolor{currentfill}{rgb}{1.000000,1.000000,1.000000}%
\pgfsetfillcolor{currentfill}%
\pgfsetlinewidth{0.000000pt}%
\definecolor{currentstroke}{rgb}{0.000000,0.000000,0.000000}%
\pgfsetstrokecolor{currentstroke}%
\pgfsetstrokeopacity{0.000000}%
\pgfsetdash{}{0pt}%
\pgfpathmoveto{\pgfqpoint{2.294491in}{0.395708in}}%
\pgfpathlineto{\pgfqpoint{3.750393in}{0.395708in}}%
\pgfpathlineto{\pgfqpoint{3.750393in}{1.999478in}}%
\pgfpathlineto{\pgfqpoint{2.294491in}{1.999478in}}%
\pgfpathclose%
\pgfusepath{fill}%
\end{pgfscope}%
\begin{pgfscope}%
\pgfpathrectangle{\pgfqpoint{2.294491in}{0.395708in}}{\pgfqpoint{1.455903in}{1.603770in}}%
\pgfusepath{clip}%
\pgfsetbuttcap%
\pgfsetmiterjoin%
\definecolor{currentfill}{rgb}{1.000000,0.000000,0.000000}%
\pgfsetfillcolor{currentfill}%
\pgfsetlinewidth{1.003750pt}%
\definecolor{currentstroke}{rgb}{1.000000,0.000000,0.000000}%
\pgfsetstrokecolor{currentstroke}%
\pgfsetdash{}{0pt}%
\pgfpathmoveto{\pgfqpoint{3.116036in}{0.833100in}}%
\pgfpathlineto{\pgfqpoint{3.091077in}{0.760201in}}%
\pgfpathlineto{\pgfqpoint{3.091077in}{0.814875in}}%
\pgfpathlineto{\pgfqpoint{3.022442in}{0.814875in}}%
\pgfpathlineto{\pgfqpoint{3.022442in}{0.851325in}}%
\pgfpathlineto{\pgfqpoint{3.091077in}{0.851325in}}%
\pgfpathlineto{\pgfqpoint{3.091077in}{0.905999in}}%
\pgfpathclose%
\pgfusepath{stroke,fill}%
\end{pgfscope}%
\begin{pgfscope}%
\pgfsetbuttcap%
\pgfsetroundjoin%
\definecolor{currentfill}{rgb}{0.000000,0.000000,0.000000}%
\pgfsetfillcolor{currentfill}%
\pgfsetlinewidth{0.803000pt}%
\definecolor{currentstroke}{rgb}{0.000000,0.000000,0.000000}%
\pgfsetstrokecolor{currentstroke}%
\pgfsetdash{}{0pt}%
\pgfsys@defobject{currentmarker}{\pgfqpoint{0.000000in}{-0.048611in}}{\pgfqpoint{0.000000in}{0.000000in}}{%
\pgfpathmoveto{\pgfqpoint{0.000000in}{0.000000in}}%
\pgfpathlineto{\pgfqpoint{0.000000in}{-0.048611in}}%
\pgfusepath{stroke,fill}%
}%
\begin{pgfscope}%
\pgfsys@transformshift{2.606470in}{0.395708in}%
\pgfsys@useobject{currentmarker}{}%
\end{pgfscope}%
\end{pgfscope}%
\begin{pgfscope}%
\definecolor{textcolor}{rgb}{0.000000,0.000000,0.000000}%
\pgfsetstrokecolor{textcolor}%
\pgfsetfillcolor{textcolor}%
\pgftext[x=2.606470in,y=0.298486in,,top]{\color{textcolor}\sffamily\fontsize{10.000000}{12.000000}\selectfont −1}%
\end{pgfscope}%
\begin{pgfscope}%
\pgfsetbuttcap%
\pgfsetroundjoin%
\definecolor{currentfill}{rgb}{0.000000,0.000000,0.000000}%
\pgfsetfillcolor{currentfill}%
\pgfsetlinewidth{0.803000pt}%
\definecolor{currentstroke}{rgb}{0.000000,0.000000,0.000000}%
\pgfsetstrokecolor{currentstroke}%
\pgfsetdash{}{0pt}%
\pgfsys@defobject{currentmarker}{\pgfqpoint{0.000000in}{-0.048611in}}{\pgfqpoint{0.000000in}{0.000000in}}{%
\pgfpathmoveto{\pgfqpoint{0.000000in}{0.000000in}}%
\pgfpathlineto{\pgfqpoint{0.000000in}{-0.048611in}}%
\pgfusepath{stroke,fill}%
}%
\begin{pgfscope}%
\pgfsys@transformshift{3.022442in}{0.395708in}%
\pgfsys@useobject{currentmarker}{}%
\end{pgfscope}%
\end{pgfscope}%
\begin{pgfscope}%
\definecolor{textcolor}{rgb}{0.000000,0.000000,0.000000}%
\pgfsetstrokecolor{textcolor}%
\pgfsetfillcolor{textcolor}%
\pgftext[x=3.022442in,y=0.298486in,,top]{\color{textcolor}\sffamily\fontsize{10.000000}{12.000000}\selectfont 0}%
\end{pgfscope}%
\begin{pgfscope}%
\pgfsetbuttcap%
\pgfsetroundjoin%
\definecolor{currentfill}{rgb}{0.000000,0.000000,0.000000}%
\pgfsetfillcolor{currentfill}%
\pgfsetlinewidth{0.803000pt}%
\definecolor{currentstroke}{rgb}{0.000000,0.000000,0.000000}%
\pgfsetstrokecolor{currentstroke}%
\pgfsetdash{}{0pt}%
\pgfsys@defobject{currentmarker}{\pgfqpoint{0.000000in}{-0.048611in}}{\pgfqpoint{0.000000in}{0.000000in}}{%
\pgfpathmoveto{\pgfqpoint{0.000000in}{0.000000in}}%
\pgfpathlineto{\pgfqpoint{0.000000in}{-0.048611in}}%
\pgfusepath{stroke,fill}%
}%
\begin{pgfscope}%
\pgfsys@transformshift{3.438414in}{0.395708in}%
\pgfsys@useobject{currentmarker}{}%
\end{pgfscope}%
\end{pgfscope}%
\begin{pgfscope}%
\definecolor{textcolor}{rgb}{0.000000,0.000000,0.000000}%
\pgfsetstrokecolor{textcolor}%
\pgfsetfillcolor{textcolor}%
\pgftext[x=3.438414in,y=0.298486in,,top]{\color{textcolor}\sffamily\fontsize{10.000000}{12.000000}\selectfont 1}%
\end{pgfscope}%
\begin{pgfscope}%
\pgfsetbuttcap%
\pgfsetroundjoin%
\definecolor{currentfill}{rgb}{0.000000,0.000000,0.000000}%
\pgfsetfillcolor{currentfill}%
\pgfsetlinewidth{0.602250pt}%
\definecolor{currentstroke}{rgb}{0.000000,0.000000,0.000000}%
\pgfsetstrokecolor{currentstroke}%
\pgfsetdash{}{0pt}%
\pgfsys@defobject{currentmarker}{\pgfqpoint{0.000000in}{-0.027778in}}{\pgfqpoint{0.000000in}{0.000000in}}{%
\pgfpathmoveto{\pgfqpoint{0.000000in}{0.000000in}}%
\pgfpathlineto{\pgfqpoint{0.000000in}{-0.027778in}}%
\pgfusepath{stroke,fill}%
}%
\begin{pgfscope}%
\pgfsys@transformshift{2.398484in}{0.395708in}%
\pgfsys@useobject{currentmarker}{}%
\end{pgfscope}%
\end{pgfscope}%
\begin{pgfscope}%
\pgfsetbuttcap%
\pgfsetroundjoin%
\definecolor{currentfill}{rgb}{0.000000,0.000000,0.000000}%
\pgfsetfillcolor{currentfill}%
\pgfsetlinewidth{0.602250pt}%
\definecolor{currentstroke}{rgb}{0.000000,0.000000,0.000000}%
\pgfsetstrokecolor{currentstroke}%
\pgfsetdash{}{0pt}%
\pgfsys@defobject{currentmarker}{\pgfqpoint{0.000000in}{-0.027778in}}{\pgfqpoint{0.000000in}{0.000000in}}{%
\pgfpathmoveto{\pgfqpoint{0.000000in}{0.000000in}}%
\pgfpathlineto{\pgfqpoint{0.000000in}{-0.027778in}}%
\pgfusepath{stroke,fill}%
}%
\begin{pgfscope}%
\pgfsys@transformshift{2.814456in}{0.395708in}%
\pgfsys@useobject{currentmarker}{}%
\end{pgfscope}%
\end{pgfscope}%
\begin{pgfscope}%
\pgfsetbuttcap%
\pgfsetroundjoin%
\definecolor{currentfill}{rgb}{0.000000,0.000000,0.000000}%
\pgfsetfillcolor{currentfill}%
\pgfsetlinewidth{0.602250pt}%
\definecolor{currentstroke}{rgb}{0.000000,0.000000,0.000000}%
\pgfsetstrokecolor{currentstroke}%
\pgfsetdash{}{0pt}%
\pgfsys@defobject{currentmarker}{\pgfqpoint{0.000000in}{-0.027778in}}{\pgfqpoint{0.000000in}{0.000000in}}{%
\pgfpathmoveto{\pgfqpoint{0.000000in}{0.000000in}}%
\pgfpathlineto{\pgfqpoint{0.000000in}{-0.027778in}}%
\pgfusepath{stroke,fill}%
}%
\begin{pgfscope}%
\pgfsys@transformshift{3.230428in}{0.395708in}%
\pgfsys@useobject{currentmarker}{}%
\end{pgfscope}%
\end{pgfscope}%
\begin{pgfscope}%
\pgfsetbuttcap%
\pgfsetroundjoin%
\definecolor{currentfill}{rgb}{0.000000,0.000000,0.000000}%
\pgfsetfillcolor{currentfill}%
\pgfsetlinewidth{0.602250pt}%
\definecolor{currentstroke}{rgb}{0.000000,0.000000,0.000000}%
\pgfsetstrokecolor{currentstroke}%
\pgfsetdash{}{0pt}%
\pgfsys@defobject{currentmarker}{\pgfqpoint{0.000000in}{-0.027778in}}{\pgfqpoint{0.000000in}{0.000000in}}{%
\pgfpathmoveto{\pgfqpoint{0.000000in}{0.000000in}}%
\pgfpathlineto{\pgfqpoint{0.000000in}{-0.027778in}}%
\pgfusepath{stroke,fill}%
}%
\begin{pgfscope}%
\pgfsys@transformshift{3.646400in}{0.395708in}%
\pgfsys@useobject{currentmarker}{}%
\end{pgfscope}%
\end{pgfscope}%
\begin{pgfscope}%
\definecolor{textcolor}{rgb}{0.000000,0.000000,0.000000}%
\pgfsetstrokecolor{textcolor}%
\pgfsetfillcolor{textcolor}%
\pgftext[x=3.284504in,y=0.261295in,,top]{\color{textcolor}\sffamily\fontsize{12.000000}{14.400000}\selectfont \(\displaystyle k\)}%
\end{pgfscope}%
\begin{pgfscope}%
\pgfsetbuttcap%
\pgfsetroundjoin%
\definecolor{currentfill}{rgb}{0.000000,0.000000,0.000000}%
\pgfsetfillcolor{currentfill}%
\pgfsetlinewidth{0.803000pt}%
\definecolor{currentstroke}{rgb}{0.000000,0.000000,0.000000}%
\pgfsetstrokecolor{currentstroke}%
\pgfsetdash{}{0pt}%
\pgfsys@defobject{currentmarker}{\pgfqpoint{-0.048611in}{0.000000in}}{\pgfqpoint{0.000000in}{0.000000in}}{%
\pgfpathmoveto{\pgfqpoint{0.000000in}{0.000000in}}%
\pgfpathlineto{\pgfqpoint{-0.048611in}{0.000000in}}%
\pgfusepath{stroke,fill}%
}%
\begin{pgfscope}%
\pgfsys@transformshift{2.294491in}{0.468607in}%
\pgfsys@useobject{currentmarker}{}%
\end{pgfscope}%
\end{pgfscope}%
\begin{pgfscope}%
\definecolor{textcolor}{rgb}{0.000000,0.000000,0.000000}%
\pgfsetstrokecolor{textcolor}%
\pgfsetfillcolor{textcolor}%
\pgftext[x=2.025000in,y=0.415845in,left,base]{\color{textcolor}\sffamily\fontsize{10.000000}{12.000000}\selectfont 0.0}%
\end{pgfscope}%
\begin{pgfscope}%
\pgfsetbuttcap%
\pgfsetroundjoin%
\definecolor{currentfill}{rgb}{0.000000,0.000000,0.000000}%
\pgfsetfillcolor{currentfill}%
\pgfsetlinewidth{0.803000pt}%
\definecolor{currentstroke}{rgb}{0.000000,0.000000,0.000000}%
\pgfsetstrokecolor{currentstroke}%
\pgfsetdash{}{0pt}%
\pgfsys@defobject{currentmarker}{\pgfqpoint{-0.048611in}{0.000000in}}{\pgfqpoint{0.000000in}{0.000000in}}{%
\pgfpathmoveto{\pgfqpoint{0.000000in}{0.000000in}}%
\pgfpathlineto{\pgfqpoint{-0.048611in}{0.000000in}}%
\pgfusepath{stroke,fill}%
}%
\begin{pgfscope}%
\pgfsys@transformshift{2.294491in}{1.197593in}%
\pgfsys@useobject{currentmarker}{}%
\end{pgfscope}%
\end{pgfscope}%
\begin{pgfscope}%
\definecolor{textcolor}{rgb}{0.000000,0.000000,0.000000}%
\pgfsetstrokecolor{textcolor}%
\pgfsetfillcolor{textcolor}%
\pgftext[x=2.025000in,y=1.144832in,left,base]{\color{textcolor}\sffamily\fontsize{10.000000}{12.000000}\selectfont 0.5}%
\end{pgfscope}%
\begin{pgfscope}%
\pgfsetbuttcap%
\pgfsetroundjoin%
\definecolor{currentfill}{rgb}{0.000000,0.000000,0.000000}%
\pgfsetfillcolor{currentfill}%
\pgfsetlinewidth{0.803000pt}%
\definecolor{currentstroke}{rgb}{0.000000,0.000000,0.000000}%
\pgfsetstrokecolor{currentstroke}%
\pgfsetdash{}{0pt}%
\pgfsys@defobject{currentmarker}{\pgfqpoint{-0.048611in}{0.000000in}}{\pgfqpoint{0.000000in}{0.000000in}}{%
\pgfpathmoveto{\pgfqpoint{0.000000in}{0.000000in}}%
\pgfpathlineto{\pgfqpoint{-0.048611in}{0.000000in}}%
\pgfusepath{stroke,fill}%
}%
\begin{pgfscope}%
\pgfsys@transformshift{2.294491in}{1.926579in}%
\pgfsys@useobject{currentmarker}{}%
\end{pgfscope}%
\end{pgfscope}%
\begin{pgfscope}%
\definecolor{textcolor}{rgb}{0.000000,0.000000,0.000000}%
\pgfsetstrokecolor{textcolor}%
\pgfsetfillcolor{textcolor}%
\pgftext[x=2.025000in,y=1.873818in,left,base]{\color{textcolor}\sffamily\fontsize{10.000000}{12.000000}\selectfont 1.0}%
\end{pgfscope}%
\begin{pgfscope}%
\pgfsetbuttcap%
\pgfsetroundjoin%
\definecolor{currentfill}{rgb}{0.000000,0.000000,0.000000}%
\pgfsetfillcolor{currentfill}%
\pgfsetlinewidth{0.602250pt}%
\definecolor{currentstroke}{rgb}{0.000000,0.000000,0.000000}%
\pgfsetstrokecolor{currentstroke}%
\pgfsetdash{}{0pt}%
\pgfsys@defobject{currentmarker}{\pgfqpoint{-0.027778in}{0.000000in}}{\pgfqpoint{0.000000in}{0.000000in}}{%
\pgfpathmoveto{\pgfqpoint{0.000000in}{0.000000in}}%
\pgfpathlineto{\pgfqpoint{-0.027778in}{0.000000in}}%
\pgfusepath{stroke,fill}%
}%
\begin{pgfscope}%
\pgfsys@transformshift{2.294491in}{0.833100in}%
\pgfsys@useobject{currentmarker}{}%
\end{pgfscope}%
\end{pgfscope}%
\begin{pgfscope}%
\pgfsetbuttcap%
\pgfsetroundjoin%
\definecolor{currentfill}{rgb}{0.000000,0.000000,0.000000}%
\pgfsetfillcolor{currentfill}%
\pgfsetlinewidth{0.602250pt}%
\definecolor{currentstroke}{rgb}{0.000000,0.000000,0.000000}%
\pgfsetstrokecolor{currentstroke}%
\pgfsetdash{}{0pt}%
\pgfsys@defobject{currentmarker}{\pgfqpoint{-0.027778in}{0.000000in}}{\pgfqpoint{0.000000in}{0.000000in}}{%
\pgfpathmoveto{\pgfqpoint{0.000000in}{0.000000in}}%
\pgfpathlineto{\pgfqpoint{-0.027778in}{0.000000in}}%
\pgfusepath{stroke,fill}%
}%
\begin{pgfscope}%
\pgfsys@transformshift{2.294491in}{1.562086in}%
\pgfsys@useobject{currentmarker}{}%
\end{pgfscope}%
\end{pgfscope}%
\begin{pgfscope}%
\definecolor{textcolor}{rgb}{0.000000,0.000000,0.000000}%
\pgfsetstrokecolor{textcolor}%
\pgfsetfillcolor{textcolor}%
\pgftext[x=2.205556in,y=1.518347in,,bottom,rotate=90.000000]{\color{textcolor}\sffamily\fontsize{12.000000}{14.400000}\selectfont \(\displaystyle n_k\)}%
\end{pgfscope}%
\begin{pgfscope}%
\pgfpathrectangle{\pgfqpoint{2.294491in}{0.395708in}}{\pgfqpoint{1.455903in}{1.603770in}}%
\pgfusepath{clip}%
\pgfsetrectcap%
\pgfsetroundjoin%
\pgfsetlinewidth{2.007500pt}%
\definecolor{currentstroke}{rgb}{0.121569,0.466667,0.705882}%
\pgfsetstrokecolor{currentstroke}%
\pgfsetdash{}{0pt}%
\pgfpathmoveto{\pgfqpoint{2.280602in}{0.470779in}}%
\pgfpathlineto{\pgfqpoint{2.403175in}{0.473953in}}%
\pgfpathlineto{\pgfqpoint{2.465727in}{0.478229in}}%
\pgfpathlineto{\pgfqpoint{2.507429in}{0.483996in}}%
\pgfpathlineto{\pgfqpoint{2.536620in}{0.491143in}}%
\pgfpathlineto{\pgfqpoint{2.557470in}{0.499265in}}%
\pgfpathlineto{\pgfqpoint{2.574151in}{0.508880in}}%
\pgfpathlineto{\pgfqpoint{2.586662in}{0.518949in}}%
\pgfpathlineto{\pgfqpoint{2.599172in}{0.532715in}}%
\pgfpathlineto{\pgfqpoint{2.611682in}{0.552000in}}%
\pgfpathlineto{\pgfqpoint{2.624193in}{0.579718in}}%
\pgfpathlineto{\pgfqpoint{2.632533in}{0.605094in}}%
\pgfpathlineto{\pgfqpoint{2.640873in}{0.638245in}}%
\pgfpathlineto{\pgfqpoint{2.649214in}{0.681785in}}%
\pgfpathlineto{\pgfqpoint{2.657554in}{0.738894in}}%
\pgfpathlineto{\pgfqpoint{2.665894in}{0.812869in}}%
\pgfpathlineto{\pgfqpoint{2.678405in}{0.959591in}}%
\pgfpathlineto{\pgfqpoint{2.699256in}{1.268093in}}%
\pgfpathlineto{\pgfqpoint{2.715936in}{1.488759in}}%
\pgfpathlineto{\pgfqpoint{2.728447in}{1.607708in}}%
\pgfpathlineto{\pgfqpoint{2.740957in}{1.690434in}}%
\pgfpathlineto{\pgfqpoint{2.753467in}{1.747023in}}%
\pgfpathlineto{\pgfqpoint{2.765978in}{1.786144in}}%
\pgfpathlineto{\pgfqpoint{2.778488in}{1.813790in}}%
\pgfpathlineto{\pgfqpoint{2.790999in}{1.833822in}}%
\pgfpathlineto{\pgfqpoint{2.803509in}{1.848700in}}%
\pgfpathlineto{\pgfqpoint{2.820190in}{1.863166in}}%
\pgfpathlineto{\pgfqpoint{2.836870in}{1.873548in}}%
\pgfpathlineto{\pgfqpoint{2.857721in}{1.882830in}}%
\pgfpathlineto{\pgfqpoint{2.882742in}{1.890526in}}%
\pgfpathlineto{\pgfqpoint{2.916103in}{1.897273in}}%
\pgfpathlineto{\pgfqpoint{2.957805in}{1.902463in}}%
\pgfpathlineto{\pgfqpoint{3.016187in}{1.906433in}}%
\pgfpathlineto{\pgfqpoint{3.095420in}{1.908467in}}%
\pgfpathlineto{\pgfqpoint{3.182993in}{1.907641in}}%
\pgfpathlineto{\pgfqpoint{3.253885in}{1.904181in}}%
\pgfpathlineto{\pgfqpoint{3.303927in}{1.899059in}}%
\pgfpathlineto{\pgfqpoint{3.341458in}{1.892410in}}%
\pgfpathlineto{\pgfqpoint{3.370649in}{1.884165in}}%
\pgfpathlineto{\pgfqpoint{3.391500in}{1.875449in}}%
\pgfpathlineto{\pgfqpoint{3.408181in}{1.865762in}}%
\pgfpathlineto{\pgfqpoint{3.424861in}{1.852361in}}%
\pgfpathlineto{\pgfqpoint{3.437372in}{1.838676in}}%
\pgfpathlineto{\pgfqpoint{3.449882in}{1.820377in}}%
\pgfpathlineto{\pgfqpoint{3.462393in}{1.795310in}}%
\pgfpathlineto{\pgfqpoint{3.474903in}{1.760098in}}%
\pgfpathlineto{\pgfqpoint{3.483243in}{1.728479in}}%
\pgfpathlineto{\pgfqpoint{3.491584in}{1.687866in}}%
\pgfpathlineto{\pgfqpoint{3.499924in}{1.635512in}}%
\pgfpathlineto{\pgfqpoint{3.508264in}{1.568309in}}%
\pgfpathlineto{\pgfqpoint{3.520775in}{1.434056in}}%
\pgfpathlineto{\pgfqpoint{3.533285in}{1.261096in}}%
\pgfpathlineto{\pgfqpoint{3.558306in}{0.900203in}}%
\pgfpathlineto{\pgfqpoint{3.570816in}{0.769462in}}%
\pgfpathlineto{\pgfqpoint{3.583327in}{0.679008in}}%
\pgfpathlineto{\pgfqpoint{3.595837in}{0.618705in}}%
\pgfpathlineto{\pgfqpoint{3.608348in}{0.578478in}}%
\pgfpathlineto{\pgfqpoint{3.620858in}{0.551147in}}%
\pgfpathlineto{\pgfqpoint{3.633369in}{0.532114in}}%
\pgfpathlineto{\pgfqpoint{3.645879in}{0.518513in}}%
\pgfpathlineto{\pgfqpoint{3.662560in}{0.505838in}}%
\pgfpathlineto{\pgfqpoint{3.679240in}{0.497151in}}%
\pgfpathlineto{\pgfqpoint{3.700091in}{0.489743in}}%
\pgfpathlineto{\pgfqpoint{3.729282in}{0.483155in}}%
\pgfpathlineto{\pgfqpoint{3.764282in}{0.478442in}}%
\pgfpathlineto{\pgfqpoint{3.764282in}{0.478442in}}%
\pgfusepath{stroke}%
\end{pgfscope}%
\begin{pgfscope}%
\pgfpathrectangle{\pgfqpoint{2.294491in}{0.395708in}}{\pgfqpoint{1.455903in}{1.603770in}}%
\pgfusepath{clip}%
\pgfsetbuttcap%
\pgfsetroundjoin%
\pgfsetlinewidth{2.007500pt}%
\definecolor{currentstroke}{rgb}{1.000000,0.498039,0.054902}%
\pgfsetstrokecolor{currentstroke}%
\pgfsetdash{{7.400000pt}{3.200000pt}}{0.000000pt}%
\pgfpathmoveto{\pgfqpoint{2.280602in}{0.472822in}}%
\pgfpathlineto{\pgfqpoint{2.357303in}{0.476887in}}%
\pgfpathlineto{\pgfqpoint{2.403175in}{0.482154in}}%
\pgfpathlineto{\pgfqpoint{2.432366in}{0.488095in}}%
\pgfpathlineto{\pgfqpoint{2.457387in}{0.496363in}}%
\pgfpathlineto{\pgfqpoint{2.474068in}{0.504711in}}%
\pgfpathlineto{\pgfqpoint{2.490748in}{0.516844in}}%
\pgfpathlineto{\pgfqpoint{2.503259in}{0.529812in}}%
\pgfpathlineto{\pgfqpoint{2.515769in}{0.547894in}}%
\pgfpathlineto{\pgfqpoint{2.528279in}{0.573759in}}%
\pgfpathlineto{\pgfqpoint{2.536620in}{0.597353in}}%
\pgfpathlineto{\pgfqpoint{2.544960in}{0.628105in}}%
\pgfpathlineto{\pgfqpoint{2.553300in}{0.668458in}}%
\pgfpathlineto{\pgfqpoint{2.561641in}{0.721453in}}%
\pgfpathlineto{\pgfqpoint{2.569981in}{0.790435in}}%
\pgfpathlineto{\pgfqpoint{2.582491in}{0.929203in}}%
\pgfpathlineto{\pgfqpoint{2.599172in}{1.169714in}}%
\pgfpathlineto{\pgfqpoint{2.615853in}{1.410654in}}%
\pgfpathlineto{\pgfqpoint{2.628363in}{1.551085in}}%
\pgfpathlineto{\pgfqpoint{2.640873in}{1.651429in}}%
\pgfpathlineto{\pgfqpoint{2.653384in}{1.720332in}}%
\pgfpathlineto{\pgfqpoint{2.665894in}{1.767602in}}%
\pgfpathlineto{\pgfqpoint{2.678405in}{1.800602in}}%
\pgfpathlineto{\pgfqpoint{2.690915in}{1.824203in}}%
\pgfpathlineto{\pgfqpoint{2.703426in}{1.841512in}}%
\pgfpathlineto{\pgfqpoint{2.715936in}{1.854512in}}%
\pgfpathlineto{\pgfqpoint{2.732617in}{1.867297in}}%
\pgfpathlineto{\pgfqpoint{2.753467in}{1.878498in}}%
\pgfpathlineto{\pgfqpoint{2.778488in}{1.887603in}}%
\pgfpathlineto{\pgfqpoint{2.807679in}{1.894650in}}%
\pgfpathlineto{\pgfqpoint{2.845211in}{1.900428in}}%
\pgfpathlineto{\pgfqpoint{2.895252in}{1.904924in}}%
\pgfpathlineto{\pgfqpoint{2.966145in}{1.907914in}}%
\pgfpathlineto{\pgfqpoint{3.053718in}{1.908358in}}%
\pgfpathlineto{\pgfqpoint{3.132951in}{1.905906in}}%
\pgfpathlineto{\pgfqpoint{3.191333in}{1.901380in}}%
\pgfpathlineto{\pgfqpoint{3.233034in}{1.895441in}}%
\pgfpathlineto{\pgfqpoint{3.266396in}{1.887603in}}%
\pgfpathlineto{\pgfqpoint{3.291416in}{1.878498in}}%
\pgfpathlineto{\pgfqpoint{3.312267in}{1.867297in}}%
\pgfpathlineto{\pgfqpoint{3.328948in}{1.854512in}}%
\pgfpathlineto{\pgfqpoint{3.341458in}{1.841512in}}%
\pgfpathlineto{\pgfqpoint{3.353969in}{1.824203in}}%
\pgfpathlineto{\pgfqpoint{3.366479in}{1.800602in}}%
\pgfpathlineto{\pgfqpoint{3.378990in}{1.767602in}}%
\pgfpathlineto{\pgfqpoint{3.387330in}{1.738071in}}%
\pgfpathlineto{\pgfqpoint{3.395670in}{1.700213in}}%
\pgfpathlineto{\pgfqpoint{3.404010in}{1.651429in}}%
\pgfpathlineto{\pgfqpoint{3.412351in}{1.588667in}}%
\pgfpathlineto{\pgfqpoint{3.424861in}{1.462129in}}%
\pgfpathlineto{\pgfqpoint{3.437372in}{1.295405in}}%
\pgfpathlineto{\pgfqpoint{3.466563in}{0.878112in}}%
\pgfpathlineto{\pgfqpoint{3.479073in}{0.753744in}}%
\pgfpathlineto{\pgfqpoint{3.491584in}{0.668458in}}%
\pgfpathlineto{\pgfqpoint{3.504094in}{0.611697in}}%
\pgfpathlineto{\pgfqpoint{3.516604in}{0.573759in}}%
\pgfpathlineto{\pgfqpoint{3.529115in}{0.547894in}}%
\pgfpathlineto{\pgfqpoint{3.541625in}{0.529812in}}%
\pgfpathlineto{\pgfqpoint{3.554136in}{0.516844in}}%
\pgfpathlineto{\pgfqpoint{3.570816in}{0.504711in}}%
\pgfpathlineto{\pgfqpoint{3.587497in}{0.496363in}}%
\pgfpathlineto{\pgfqpoint{3.608348in}{0.489217in}}%
\pgfpathlineto{\pgfqpoint{3.637539in}{0.482837in}}%
\pgfpathlineto{\pgfqpoint{3.675070in}{0.478001in}}%
\pgfpathlineto{\pgfqpoint{3.733452in}{0.474038in}}%
\pgfpathlineto{\pgfqpoint{3.764282in}{0.472822in}}%
\pgfpathlineto{\pgfqpoint{3.764282in}{0.472822in}}%
\pgfusepath{stroke}%
\end{pgfscope}%
\begin{pgfscope}%
\pgfpathrectangle{\pgfqpoint{2.294491in}{0.395708in}}{\pgfqpoint{1.455903in}{1.603770in}}%
\pgfusepath{clip}%
\pgfsetbuttcap%
\pgfsetroundjoin%
\pgfsetlinewidth{1.505625pt}%
\definecolor{currentstroke}{rgb}{0.121569,0.466667,0.705882}%
\pgfsetstrokecolor{currentstroke}%
\pgfsetdash{{1.500000pt}{2.475000pt}}{0.000000pt}%
\pgfpathmoveto{\pgfqpoint{3.115737in}{0.395708in}}%
\pgfpathlineto{\pgfqpoint{3.115737in}{1.887214in}}%
\pgfusepath{stroke}%
\end{pgfscope}%
\begin{pgfscope}%
\pgfpathrectangle{\pgfqpoint{2.294491in}{0.395708in}}{\pgfqpoint{1.455903in}{1.603770in}}%
\pgfusepath{clip}%
\pgfsetbuttcap%
\pgfsetroundjoin%
\pgfsetlinewidth{1.505625pt}%
\definecolor{currentstroke}{rgb}{1.000000,0.647059,0.000000}%
\pgfsetstrokecolor{currentstroke}%
\pgfsetdash{{1.500000pt}{2.475000pt}}{0.000000pt}%
\pgfpathmoveto{\pgfqpoint{3.022442in}{0.395708in}}%
\pgfpathlineto{\pgfqpoint{3.022442in}{1.887214in}}%
\pgfusepath{stroke}%
\end{pgfscope}%
\begin{pgfscope}%
\pgfsetrectcap%
\pgfsetmiterjoin%
\pgfsetlinewidth{0.803000pt}%
\definecolor{currentstroke}{rgb}{0.000000,0.000000,0.000000}%
\pgfsetstrokecolor{currentstroke}%
\pgfsetdash{}{0pt}%
\pgfpathmoveto{\pgfqpoint{2.294491in}{0.395708in}}%
\pgfpathlineto{\pgfqpoint{2.294491in}{1.999478in}}%
\pgfusepath{stroke}%
\end{pgfscope}%
\begin{pgfscope}%
\pgfsetrectcap%
\pgfsetmiterjoin%
\pgfsetlinewidth{0.803000pt}%
\definecolor{currentstroke}{rgb}{0.000000,0.000000,0.000000}%
\pgfsetstrokecolor{currentstroke}%
\pgfsetdash{}{0pt}%
\pgfpathmoveto{\pgfqpoint{3.750393in}{0.395708in}}%
\pgfpathlineto{\pgfqpoint{3.750393in}{1.999478in}}%
\pgfusepath{stroke}%
\end{pgfscope}%
\begin{pgfscope}%
\pgfsetrectcap%
\pgfsetmiterjoin%
\pgfsetlinewidth{0.803000pt}%
\definecolor{currentstroke}{rgb}{0.000000,0.000000,0.000000}%
\pgfsetstrokecolor{currentstroke}%
\pgfsetdash{}{0pt}%
\pgfpathmoveto{\pgfqpoint{2.294491in}{0.395708in}}%
\pgfpathlineto{\pgfqpoint{3.750393in}{0.395708in}}%
\pgfusepath{stroke}%
\end{pgfscope}%
\begin{pgfscope}%
\pgfsetrectcap%
\pgfsetmiterjoin%
\pgfsetlinewidth{0.803000pt}%
\definecolor{currentstroke}{rgb}{0.000000,0.000000,0.000000}%
\pgfsetstrokecolor{currentstroke}%
\pgfsetdash{}{0pt}%
\pgfpathmoveto{\pgfqpoint{2.294491in}{1.999478in}}%
\pgfpathlineto{\pgfqpoint{3.750393in}{1.999478in}}%
\pgfusepath{stroke}%
\end{pgfscope}%
\begin{pgfscope}%
\definecolor{textcolor}{rgb}{0.000000,0.000000,0.000000}%
\pgfsetstrokecolor{textcolor}%
\pgfsetfillcolor{textcolor}%
\pgftext[x=2.338168in,y=1.839101in,left,base]{\color{textcolor}\sffamily\fontsize{12.000000}{14.400000}\selectfont (b)}%
\end{pgfscope}%
\begin{pgfscope}%
\definecolor{textcolor}{rgb}{1.000000,0.647059,0.000000}%
\pgfsetstrokecolor{textcolor}%
\pgfsetfillcolor{textcolor}%
\pgftext[x=2.481678in,y=1.343390in,left,base]{\color{textcolor}\sffamily\fontsize{17.280000}{20.736000}\selectfont \(\displaystyle \downarrow\)}%
\end{pgfscope}%
\begin{pgfscope}%
\definecolor{textcolor}{rgb}{0.290196,0.474510,0.654902}%
\pgfsetstrokecolor{textcolor}%
\pgfsetfillcolor{textcolor}%
\pgftext[x=2.731261in,y=1.343390in,left,base]{\color{textcolor}\sffamily\fontsize{17.280000}{20.736000}\selectfont \(\displaystyle \uparrow\)}%
\end{pgfscope}%
\begin{pgfscope}%
\definecolor{textcolor}{rgb}{1.000000,0.000000,0.000000}%
\pgfsetstrokecolor{textcolor}%
\pgfsetfillcolor{textcolor}%
\pgftext[x=3.064039in,y=0.935158in,,base]{\color{textcolor}\sffamily\fontsize{10.000000}{12.000000}\selectfont \(\displaystyle \kappa\)}%
\end{pgfscope}%
\end{pgfpicture}%
\makeatother%
\endgroup%